\documentclass{emulateapj}
\graphicspath{{figures/}}

\begin{document}

\title{Time-Series Photometry of Stars in and around the Lagoon Nebula.\ 
I.\ Rotation Periods of 290 Low-Mass Pre--Main-Sequence Stars in NGC~6530}
\author{
{Calen B.~Henderson}\altaffilmark{1} and
{Keivan G.~Stassun}\altaffilmark{2,3,4}
}
\altaffiltext{1}{Department of Astronomy, The Ohio State University, 140 W.\ 18th Ave., Columbus, OH 43210, USA; henderson@astronomy.ohio-state.edu}
\altaffiltext{2}{Department of Physics \& Astronomy, Vanderbilt University, VU Station B 1807, Nashville, TN 37235, USA}
\altaffiltext{3}{Department of Physics, Fisk University, 1000 17th Ave.\ N., Nashville, TN 37208, USA}
\altaffiltext{4}{Department of Physics, Massachusetts Institute of Technology, 77 Massachusetts Ave., Cambridge, MA 02139, USA}

\shorttitle{Rotation Periods in NGC~6530}
\shortauthors{Henderson \& Stassun}

% -----------------------------------------------------------------
\begin{abstract}

We have conducted a long-term, wide-field, high-cadence photometric 
monitoring survey of $\sim$50,000 stars in the Lagoon Nebula \ion{H}{2} region. 
This first paper presents rotation periods for 290 low-mass stars in 
NGC 6530, the young cluster illuminating the nebula, and for which we 
assemble a catalog of infrared and spectroscopic disk indicators, 
estimated masses and ages, and X-ray luminosities. 
The distribution of rotation periods we measure is broadly uniform for 
$0.5 < P < 10$~d; the short-period cutoff corresponds to breakup.
We observe no obvious bimodality in the period distribution, but 
we do find that stars with disk signatures rotate more slowly on average.
The stars' X-ray luminosities are roughly flat with rotation period,
at the saturation level ($\log L_X / L_{\rm bol} \approx -3.3$).
However, we find a significant positive correlation between $L_X / L_{\rm bol}$ 
and co-rotation radius, suggesting that the observed X-ray luminosities are
regulated by centrifugal stripping of the stellar coronae. 
The period--mass relationship in NGC 6530 is broadly similar
to that of the Orion Nebula Cluster (ONC), but the slope of the relationship
among the slowest rotators differs from that in the ONC and other young clusters. 
We show that the slope of the period--mass relationship for the slowest rotators can be used 
as a proxy for the age of a young cluster, and we argue that NGC~6530 may be slightly 
younger than the ONC, making it a particularly important 
touchstone for models of angular momentum evolution in young, low-mass stars.

\end{abstract}

% -----------------------------------------------------------------


%%%%%%%%%%%%%%%%%%%%%%%%%%%%%%%%%%%%%%%%%%%%%%%%%%%%%%%%%%%%%%%%%%%%%%%%%%%%
\section{Introduction}
\label{sec:intro}
%%%%%%%%%%%%%%%%%%%%%%%%%%%%%%%%%%%%%%%%%%%%%%%%%%%%%%%%%%%%%%%%%%%%%%%%%%%%

Time-domain photometric monitoring surveys of young stars have been
crucial to our understanding of a variety of fundamental questions
related to low-mass stars in the pre--main-sequence (PMS) phase of
evolution. Indeed, our empirical understanding of time-variable
accretion processes
\citep[e.g.][]{gullbring96,bouvier99,stassun99b,bouvier04},
of magnetic activity
\citep[e.g.][]{walter87,feigelson02,feigelson05,feigelson07,walter04,stassun04b,stassun06,stassun07b}, 
of the early evolution of stellar angular momentum
\citep[e.g.][]{attridge92,bouvier93,choi96,stassun99a,rebull01,herbst02,makidon04,lamm05,irwin08a,irwin08b},
of the inner architectures of protoplanetary disks
\citep[e.g.][]{bouvier03,winn06,bouvier07,herbst08,herbst10},
of the formation of binary stars
\citep[e.g.][]{mathieu97,basri97,jensen07,stassun08}, 
of outbursts
\citep[e.g.][]{walter04,briceno04,kastner04,grosso05,aspin06,aspin11,covey11,bastien11}, 
of pulsations
\citep[e.g.][]{zwintz06,guenther07}, 
and of the fundamental masses and radii of PMS stars
\citep[e.g.][]{casey98,covino00,stassun04a,stassun06a,stassun07a,irwin07,stempels08,stassun08,hebb10}, 
has relied upon detailed light curves of
T~Tauri stars (TTSs) in a variety of young clusters and star-forming
regions spanning a range of ages and star-forming environments. 

The most extensively monitored regions include the Taurus-Auriga association
(age $\sim$1--3 Myr), the Orion Nebula Cluster ($\sim$1--2 Myr), NGC 2264
($\sim$3 Myr), NGC 2362 ($\sim$3--4 Myr), IC 348 ($\sim$4--5 Myr), and
NGC 2547 ($\sim$40 Myr). At a distance of $\sim$400 pc
\citep[e.g.][]{menten07}, the Orion Nebula
Cluster (ONC) and the larger star-forming region surrounding it is the
nearest and perhaps the single best-studied massive star-forming region. 
This region contributes photometrically-determined rotation periods
for hundreds of TTSs \citep[see][and references therein]{herbst07}, 
nearly all of the known PMS eclipsing binary stars
\citep[see][and references therein]{stassun09},
including the only brown-dwarf eclipsing binary system yet discovered
\citep{stassun06a,stassun07a,gomez09,mohanty09,mohanty10},
and the most well-studied FUor/EXor type eruptive star discovered in recent
times \citep[V1647 Ori; see, e.g.,][and references therein]{bastien11}. 
Thanks in large part to the broad array of discoveries
enabled by the extensive photometric monitoring surveys of this region,
the ONC has become a crucial testbed for star formation theory, from PMS
angular momentum evolution \citep[see][]{stassun03,herbst07}
to the nature of the initial mass function
\citep[see][]{hillenbrand97,dario10}.
As wide-area photometric monitoring campaigns begin to survey a larger
number of star-forming regions, particularly rich young clusters like the ONC,
the discovery space is becoming enlarged for larger samples of benchmark
PMS objects. For example, the recent PTF surveys of the 25~Ori and North
America/Pelican nebulae have already resulted in the discovery of numerous 
new candidate PMS eclipsing binaries and new FUor/EXor outburst systems
\citep{vaneyken11,covey11}.

In this paper we report the first results of a large-scale, multi-year,
high-cadence photometric monitoring survey of the bright Lagoon
Nebula (Messier 8) \ion{H}{2} star-forming region. The massive star
cluster illuminating the nebula, NGC 6530, includes a rich population of
$\gtrsim$1100 stellar members spanning the full IMF,
from massive OB type stars down to at least the hydrogen burning limit
\citep{damiani04,prisinzano05,damiani06,prisinzano07}. With
a nominal age of $\sim$1 Myr \citep[e.g.][]{mayne07}, NGC 6530 is thus in
many respects an analog of the ONC, but at a distance of $\sim$1.25 kpc
NGC 6530 provides an ONC-like stellar laboratory beyond the immediate 
solar neighborhood.

An extensive literature has emerged over the past few years,
characterizing the PMS population of NGC 6530 from X-ray to infrared
wavelengths.  \citet{damiani04} conducted the first large-scale study of
the stellar population of NGC 6530. Using {\it Chandra} they detected 884
X-ray point sources, finding that 90--95\% of them constitute cluster
members. \citet{prisinzano05} subsequently performed a complementary
deep optical survey of the region, obtaining $BVI_C$ photometry down to
$V\approx 23$ using the Wide Field Imager at the MPG/ESO 2.2m telescope. They
matched their catalog to that of \citet{damiani04} and found 828 common
stars, the vast majority of which are cluster members. 
From their deep color-magnitude diagrams (CMDs) and X-ray membership selection, 
they determined a cluster distance of 1.25 kpc and a modest cluster 
extinction of $A_V = 1.1$ mag, and they moreover estimated masses and
ages for the stars through comparison with the PMS evolutionary tracks of \citet{siess00}.
\citet{damiani06} performed a near-infrared (NIR) survey of the cluster
to identify stars with NIR excess emission indicative of objects 
bearing massive protoplanetary disks, and in the process
increased the number of PMS cluster members to more than 1100. 
\citet{prisinzano07} spectroscopically
studied a subsample of 332 cluster members, using H$\alpha$ emission
to classify the stars as classical T~Tauri stars (CTTSs) or weak-lined
T~Tauri stars (WTTSs).

Our synoptic survey of the photometric variability properties of NGC 6530
builds on these studies. Our light curves of a $40\arcmin \times 40\arcmin$
region centered on NGC 6530 include the known cluster members as well as a
total of $\sim$50,000 other stars within and surrounding the larger Lagoon
Nebula star-forming region. These light curves in concert with the extant
literature enable an array of variability studies for the region,
including measurement of stellar rotation periods, identification of PMS
eclipsing binaries, characterization of accretion induced variations and
stellar occultations due to disk obscuration, and discovery of eruptive
variables. This first paper presents the results of our systematic search
for rotation periods among the members of NGC 6530, the first such survey
for stellar rotation periods yet reported for this important cluster.

Understanding the evolution of stellar angular momentum, particularly during
the first few Myr of a star's life, remains one of the longest outstanding
questions in star formation research
\citep[e.g.][]{vogel81,hartmann86,bouvier86}.
As PMS stars contract
toward the main sequence, they would be expected to rapidly spin up to near
breakup velocity as a consequence of angular momentum conservation. However,
numerous surveys of rotation periods among low-mass PMS stars clearly
show that these stars typically rotate at a small fraction of breakup,
despite significant contraction in stellar radius
\citep[e.g.][]{stauffer87}.
A variety of mechanisms for efficiently removing angular momentum from
PMS stars during the first $\sim$10 Myr has been proposed, including
magnetic star-disk interaction 
\citep[i.e., ``disk-locking";][]{konigl91,shu94,najita95,ostriker95},
scaled-up 
solar-type magnetized winds perhaps driven by accretion
\citep[e.g.][]{matt04,matt05a,matt05b,matt08a,matt08b}, 
and scaled-up solar-type coronal mass ejections
\citep[e.g.][]{aarnio09,aarnio10,aarnio11}. However, the observational support 
for and theoretical efficacy of these mechanisms remains debated.
There is not yet a consensus on the dominant mechanism(s) responsible
for governing the angular momentum evolution of low-mass PMS stars.

In contrast, by the time a young cluster of stars reaches the age of the
Pleiades ($\sim$125 Myr), the observational picture is
much more clear. By Pleiades' age, a cluster of coeval stars develops two
distinct populations in period versus mass, now commonly referred to as the
``I'' and ``C'' sequences \citep{barnes03,barnes07,barnes10a,barnes10b}. 
These respectively correspond
to curves that trace the upper and lower envelopes of stellar rotation
periods. At an age of $\sim$600 Myr, about the age of the Hyades, nearly
all stars inhabiting the ``C'' sequence will have spun down and transitioned
onto the ``I'' sequence. This general behavior is attributed principally to 
changes in the stars' internal structures as a function of mass, and
has been found to hold true
for clusters at a variety of ages \citep{patten96, barnes99, allain96,
krishnamurthi98, radick87, radick90}.  
While this ``gyrochronology" paradigm has yet to be extended back
to the PMS stage, it is clear that young low-mass stars must at some stage
prior to the main sequence develop specific relationships between rotation
period and stellar mass, and that these encode the stellar age.
Exploiting the very young age of NGC 6530, we are in a position to better
define the initial conditions of PMS angular momentum evolution, and to 
identify when and how the period--mass--age relationship begins to take shape.

In this paper we report rotation periods for 290 members of NGC 6530.
We match these to the literature to produce a catalog with which we
investigate correlations between rotation period and other stellar
properties, including disk presence, mass, age, spatial distribution
within the cluster, and X-ray activity. In \S\ref{sec:data} we describe the
photometric observations and their reduction, data from the literature,
and the construction of our catalog. Our period search and statistical
methods are discussed in \S\ref{sec:analysis}. 
In \S\ref{sec:results} we examine the distribution of rotation periods, 
and investigate correlations between rotation period and the other stellar
properties in our catalog. 
We discuss the implications of our results in \S\ref{sec:disc}, 
and in \S\ref{sec:summary} we summarize the main conclusions.

%%%%%%%%%%%%%%%%%%%%%%%%%%%%%%%%%%%%%%%%%%%%%%%%%%%%%%%%%%%%%%%%%%%%%%%%%%%%%%%
\section{Data}
\label{sec:data}
%%%%%%%%%%%%%%%%%%%%%%%%%%%%%%%%%%%%%%%%%%%%%%%%%%%%%%%%%%%%%%%%%%%%%%%%%%%%%%%

% -------------------------
\subsection{Observations}
\label{sec:obs}
% -------------------------

We have photometrically monitored the Lagoon Nebula over a period of 
several years using the SMARTS 1.0-meter and 0.9-meter telescopes at
the Cerro Tololo Inter-American Observatory (CTIO). Table~\ref{tab:obsdata} summarizes
the observing runs, the telescope used, and the number of nights with
useable data for each. For this paper we use the data from the longest
of the observing runs, in June--July 2006.

We repeatedly imaged the Lagoon Nebula on 27 clear nights over a time 
baseline spanning the 35 nights from June 15 to July 19. 
We used the Y4Kcam on the SMARTS 1.0-meter telescope at CTIO. The observations consist of
720~s exposures taken through the Cousins $I$ filter. The Y4Kcam
has a field of view (FOV) of $20\arcmin \times 20\arcmin$ square,
and we imaged four fields, alternating imaging each one. This gives
us a full FOV of $40\arcmin \times 40\arcmin$ square which we imaged with
a sampling cadence of $\sim$1~hr$^{-1}$. The field was centered on 
the NGC 6530 cluster,
$(\alpha,\delta)=(18^{\rm h}04^{\rm m}24\fs2, -24\arcdeg21\arcmin06\farcs0)$
(J2000.0), the same as the {\it Chandra} observation of \citet{damiani04}. 

\begin{figure}[ht]
\plotone{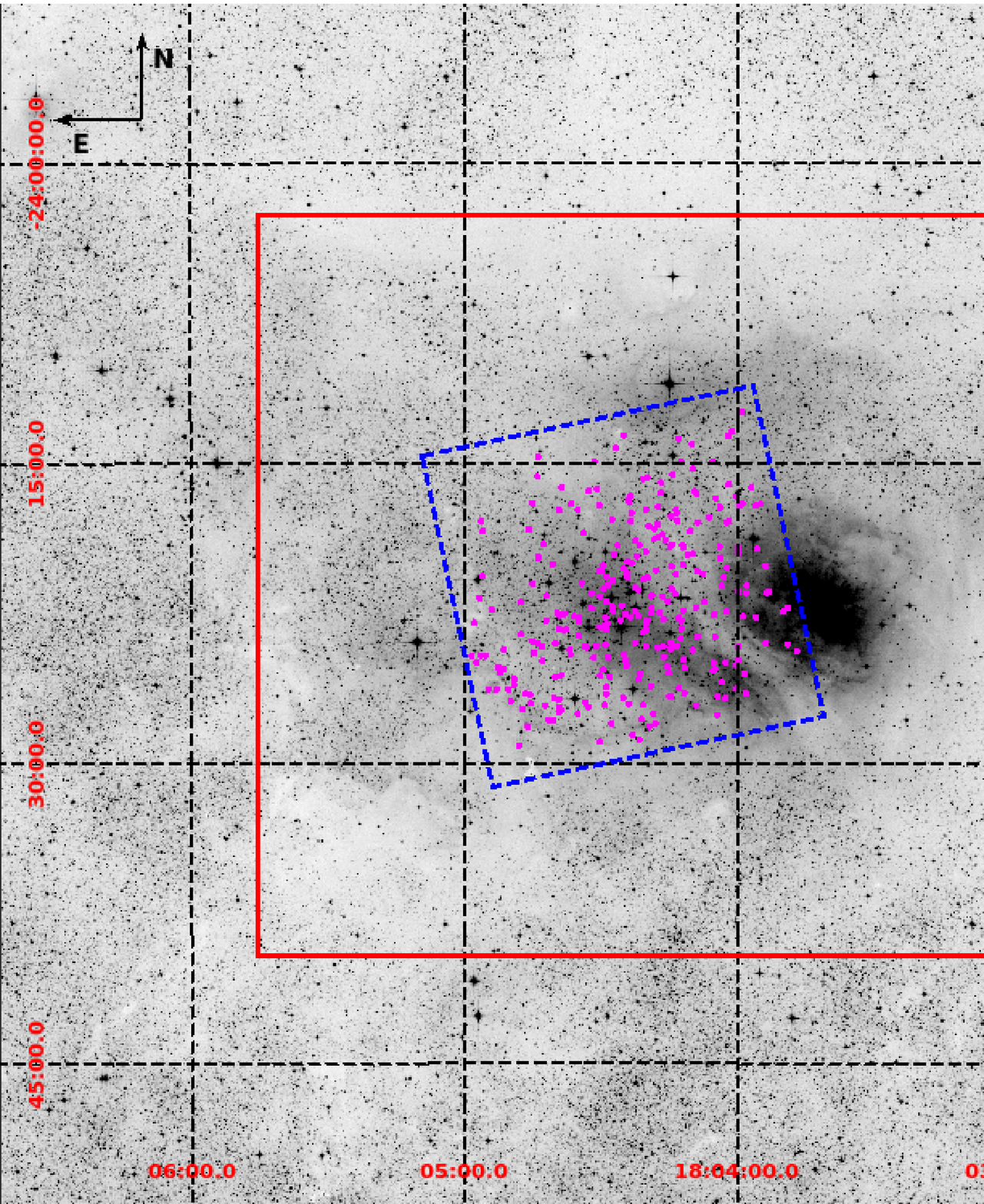}
\caption{A NASA SkyView image of the region around NGC 6530. The image
size is $1\times1$ deg$^{2}$ and centered on $(\alpha,\delta)=(18^{\rm
h}04^{\rm m}24\fs4, -24\arcdeg21\arcmin06\farcs0)$ with north up and east
to the left.  The area covered by this study is shown in red while the
X-ray study of \citet{damiani04} is shown in dashed blue. Our 290 stars
with rotation periods are shown in magenta. 
See the electronic journal for a color version of this figure. 
\label{fig:fichart}
}
\end{figure}

Table~\ref{tab:2006data} gives an overview of the observations used in 
this paper.
Figure~\ref{fig:fichart} shows a $1\times1$
deg$^{2}$ image of the region with overlays of this study and the X-ray
study of \citet{damiani04}.

% -------------------------
\subsection{Data Reduction}
\label{sec:reduc}
% -------------------------

The images were reduced, and instrumental magnitudes for all point
sources extracted, using standard IRAF\footnote{IRAF is distributed 
by the National Optical Astronomy Observatory, which is operated by 
the Association of Universities for Research in Astronomy under 
cooperative agreement with the National Science Foundation.}
procedures. Differential light curves were determined from PSF photometry
using an algorithm for inhomogeneous ensemble photometry \citep{honeycutt92}
as implemented in \citet{stassun99a, stassun02} for observations of
high-nebulosity regions such as M8. We used a point-source detection 
threshold of 7$\sigma$ above the sky background noise, and we kept only 
sources detected in at least 50 frames.

\begin{figure}[ht]
\plotone{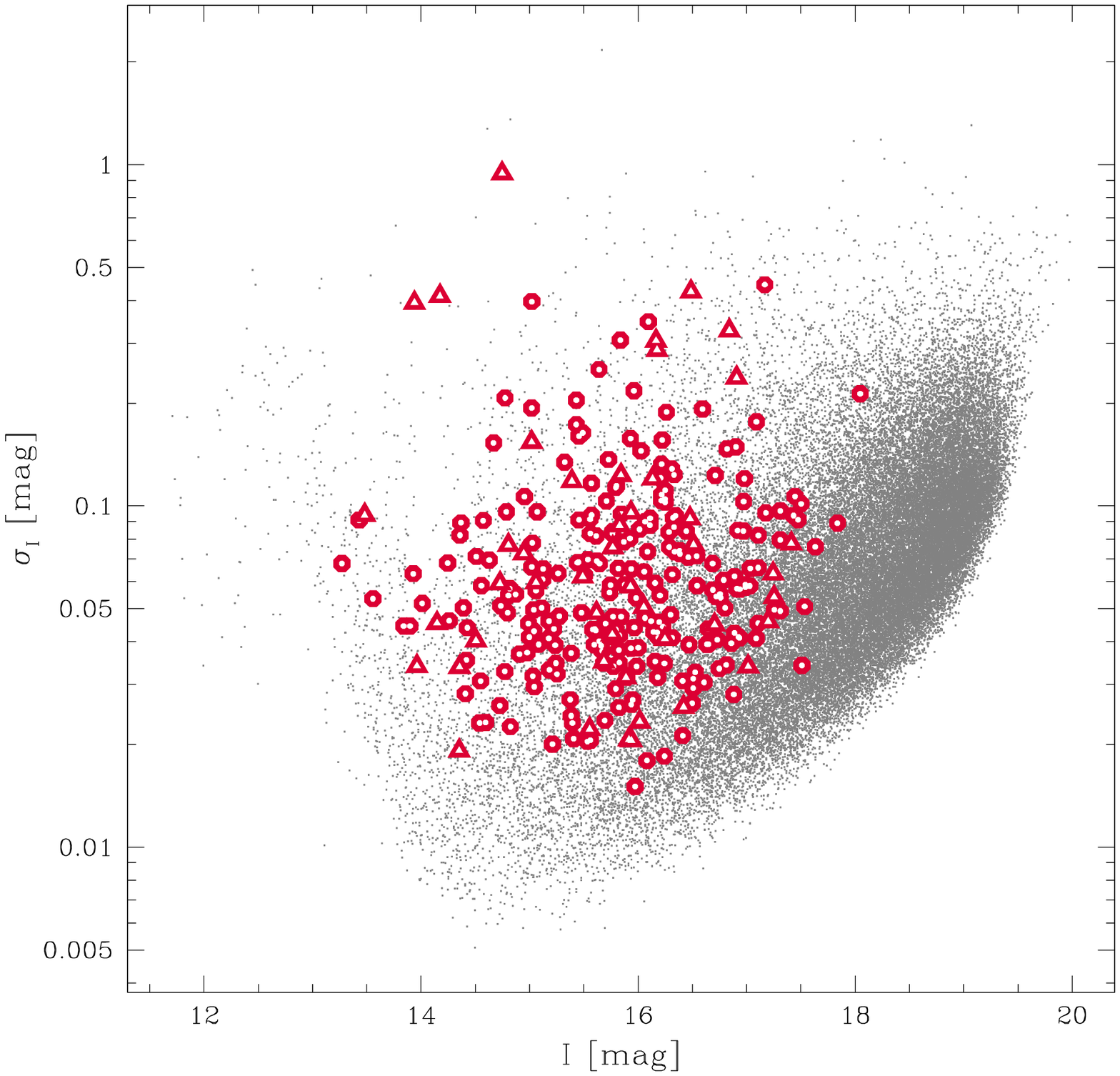}
\caption{RMS of the light curves as a function of calibrated $I_{C}$
magnitude for the 53,500 stars in our four fields. The red circles denote
the 244 stars with ``definite'' rotation periods (FAP~$\leq$~0.001) while
the red triangles mark the 46 stars with ``possible'' rotation periods
(0.001~$<$~FAP~$\leq$~0.01).
See the electronic journal for a color version of this figure. 
\label{fig:sigma}
}
\end{figure}

Figure~\ref{fig:sigma} shows the r.m.s.\ of the light curves as a function 
of $I_{C}$ magnitude 
\citep[calibrated using the absolute photometry of][]{prisinzano05}
for each of the 53,500 stars in our images. In the figure, the lower 
envelope of points with declining r.m.s.\ toward brighter $I_C$ magnitude
represents intrinsically non-variable stars to within the precision of our
photometry, which is $\sim$0.008 mag at $I_{C}\approx 14$ and rising to 
$\sim$0.04 mag at $I_C\approx 18.0$ (the faint limit of our period search;
see below). The rising envelope of r.m.s.\ for bright stars with 
$I_{C}\lesssim 14$ is due to CCD non-linearity effects for stars approaching
saturation ($I_C\approx 12.5$), and so we limit our period search to stars
with $I_C\ge 13.0$ (see below).

We determined astrometric positions for each star in our catalog using the
{\tt astrometry.net} tool suite \citep{lang10}. The absolute positions of our
sample stars are expected to be accurate to $\lesssim$1$\arcsec$.

%------------------------------------
\subsection{Data from the Literature}
\label{sec:litdata}
%------------------------------------

Large-scale X-ray observations of young clusters have demonstrated that
X-ray emission is a highly efficient means for separating low-mass PMS
stars from field contaminants 
\citep[e.g.][]{getman05a,getman05b,guedel07,feigelson11}. For example,
the \textit{Chandra} Orion Ultradeep Project (COUP) found the rate of contaminants
(due to foreground/background field stars and extragalactic sources) to be
$<$10\% \citep{getman05b}.
Thus we begin with the X-ray catalog of NGC 6530 from \citet{damiani04} to
identify the most likely cluster members.
They expect contamination in their population from non-member field stars
to be $\sim$5$\%$.

First we match each of the stars in our photometric database to the 
\citet{damiani04} X-ray source list using a positional 
tolerance of 2$\arcsec$. From this we determined the mean offsets between 
the two astrometric systems for each of our fields, which are listed in Table~\ref{tab:xrayoffsets}
(where offsets are positionally calculated as Damiani$-$ours), and corrected
our astrometry to place all of the stars onto the 
\citet{damiani04} system. Then we re-matched our stars to the 
\citet{damiani04} catalog employing a 1.5$\arcsec$ tolerance.
This results in 662 unique cluster members which form the master sample for
the present study.
For these stars we also derive X-ray luminosities from the X-ray count rates
reported by \citet{damiani04}.

Stellar masses and ages come from the optical catalog of \citet{prisinzano05}, 
and we use the isochrones of \citet{siess00} to infer stellar bolometric
luminosities. 
We identify stars likely possessing warm, massive circumstellar disks using
two different indicators---the reddening-free index of NIR-excess, $Q_{VIJK}$,
reported by \citet{damiani06}, and the CTTS/WTTS classifications of 
\citet{prisinzano07} and \citet{arias07} based on their spectroscopic survey
of H$\alpha$ emitting stars in the cluster.
Finally, known spectroscopic binaries are identified from the catalog of
\citet{prisinzano07}.

\begin{figure}[ht]
\plotone{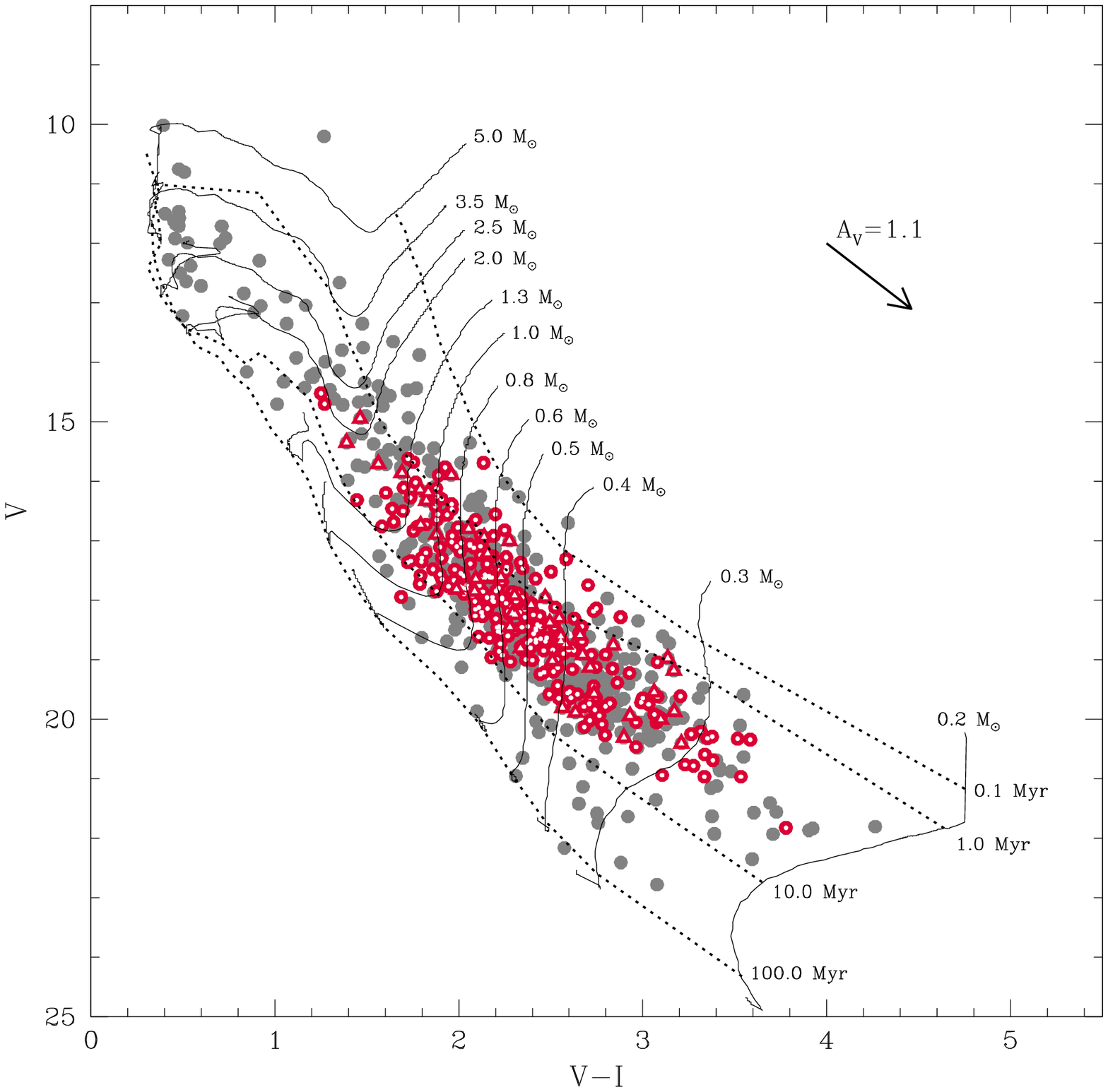}
\caption{$V$ versus $V-I_C$ color-magnitude diagram for our master sample
drawn from the catalog of X-ray members of \citet{damiani04} (filled symbols). 
Stars for which we report rotation periods are highlighted. Overplotted are
the PMS evolutionary tracks of \citet{siess00} assuming a distance of
1.25 kpc and an extinction of $A_V=1.1$ mag \citep{prisinzano05}.
The reddening vector shown uses the above $A_V$ and the color excess
$E(V-I)=0.46$ value from \citet{prisinzano05}.
See the electronic journal for a color version of this figure.
\label{fig:cmd}
}
\end{figure}

Figure \ref{fig:cmd} shows the $V$ versus $V-I_C$ CMD for our 
X-ray selected master study sample. The PMS evolutionary tracks
of \citet{siess00} are overlaid for context \citep[we transformed the tracks
from effective temperature and bolometric luminosity to the CMD plane using the
main-sequence relations of][]{kenyon95}. The sample stars span a range of
inferred masses $0.2 \lesssim M/{\rm M}_\odot \lesssim 5$.

%%%%%%%%%%%%%%%%%%%%%%%%%%%%%%%%%%%%%%%%%%%%%%%%%%%%%%%%%%%%%%%%%%%%%%%%%%%%%%%
\section{Analysis}
\label{sec:analysis}
%%%%%%%%%%%%%%%%%%%%%%%%%%%%%%%%%%%%%%%%%%%%%%%%%%%%%%%%%%%%%%%%%%%%%%%%%%%%%%%

%-------------------------
\subsection{Period Search}
\label{sec:persearch}
%-------------------------

We use the VARTOOLS light curve analysis program \citep{hartman08} to search
for periods in our light curves, employing the Lomb-Scargle (LS) period
search algorithm \citep{lomb76, scargle82, press89, press92}. Because the
temporal baseline of our light curves is 35 days, we restrict the search
to periods shorter than 20~d, i.e.\ just over 50\% of the baseline. We also
limit the search to periods longer than 0.1~d, corresponding to the Nyquist
limit given our typical sampling frequency of $\approx$0.05~d 
(see \S\ref{sec:data}). We first clipped the light curve data 
with iterative 3-sigma outlier rejection and then selected
only those periods whose peaks in the resulting LS periodogram 
satisfied a signal-to-noise ratio, $\mid$\textit{SNR}$\mid$ $>4.0$.

We perform a Monte Carlo simulation with our observed light curves to 
determine the false-alarm probability (FAP) of the periods we detect. 
Following the procedures described in \citet{stassun99a}, for each star we 
generate 10,000 synthetic light curves from which we empirically determine 
the distribution of peak heights in the LS power spectrum that would arise 
from noise. We then compare the height of the peak in the star's observed LS
power spectrum to this distribution of peak heights to determine the FAP.
Each of the 10,000 synthetic light curves consists of two sources of
noise. The first is the point-to-point scatter in the photometry which
we simulate by scrambling the star's actual light curve data. The second
is a correlated noise with a timescale of 1~d, whose amplitude we estimate 
from the standard deviation of nightly means from the star's light curve. 
The former preserves the specific noise distribution and the time sampling 
pattern of the actual data for each star, while the
latter gives the synthetic light curves the freedom to vary on timescales 
that are long compared to our sampling interval, allowing them to mimic any 
slow variability of stellar origin (such as accretion activity) that could 
produce spurious periodic behavior that would be misinterpreted as a 
rotation period.

We consider ``definite'' rotation periods to be those with FAP~$\leq$~0.001.
In other words, these stars' LS power spectra evince peaks whose strengths
occur by chance in 10 or fewer of the 10,000 noise light curves. As 
our master sample includes 662 stars, we therefore expect at most $\sim$1
false positive period. In addition, we consider ``possible'' rotation periods
to be those with 0.001~$<$~FAP~$\leq$~0.01. 
Throughout our analysis we generally only utilize the definite rotation 
period stars, but we include the possible rotation period stars here for 
the benefit of future follow-up studies. 

\begin{figure*}[ht]
\plotone{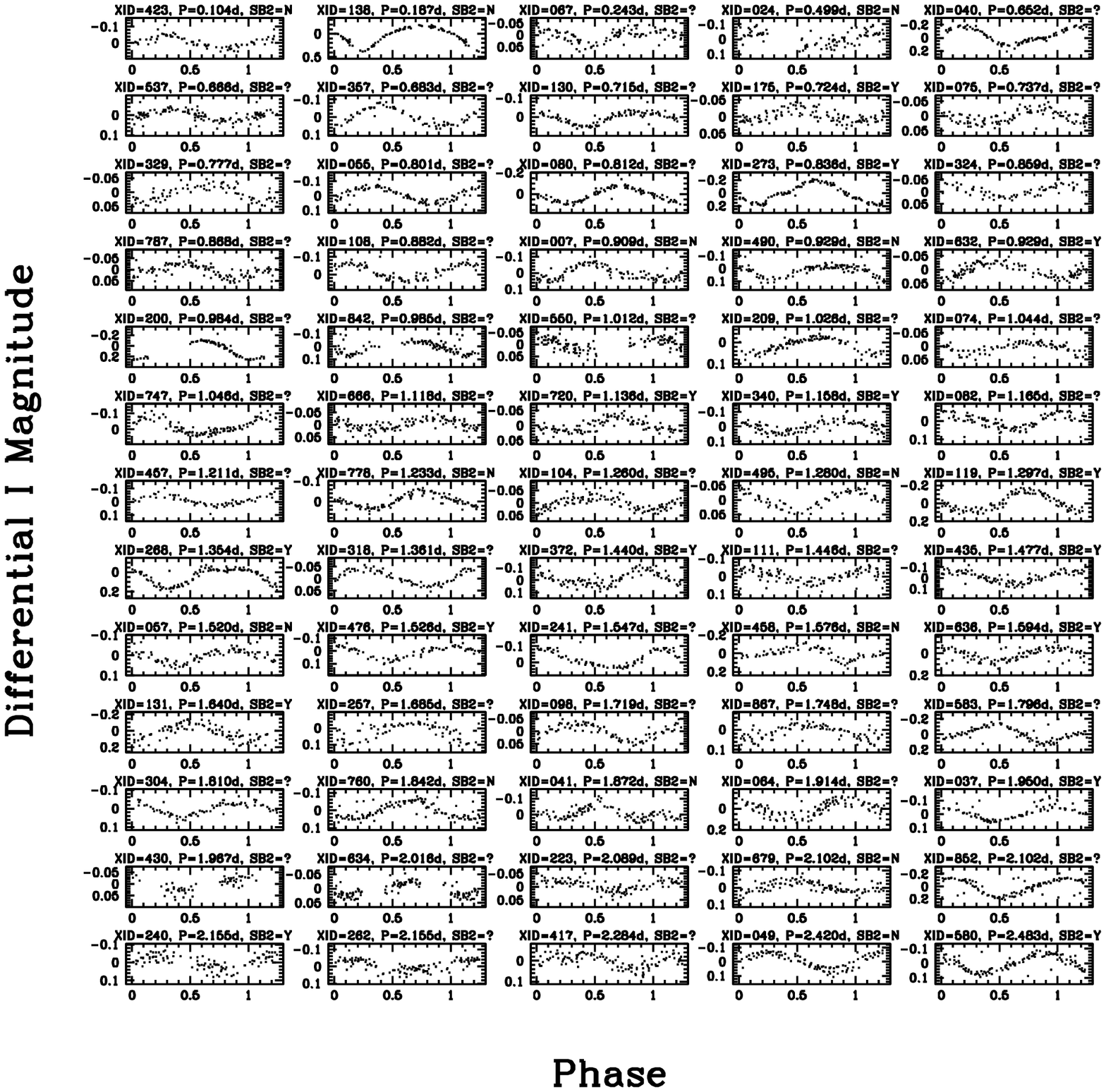}
\caption{Light curves of NGC 6530 cluster members with ``definite''
rotation periods (FAP~$\leq$~0.001). The light curves are folded on
the derived rotation period (shown above each light curve) and
replicated over an additional 0.25 phase for clarity. Also shown above
each light curve is the ID number from the X-ray study of 
\citet{damiani04}, as well as a flag indicating whether previous
spectroscopic observations suggest the star is a spectroscopic binary
(SB2). The stars are shown ordered by increasing period.
\label{fig:lc.perdef.01}}
\end{figure*}

From our master sample of 662 cluster members 
we find 256 definite periods and 47 possible periods.
Finally, for the remainder of our analysis, we include only those stars 
included in the optical catalog of \citet{prisinzano05}, for which we have 
estimated stellar masses and ages. This
yields a final catalog of 244 cluster members with definite periods
and 46 with possible periods. 
The sample of 244 definite periods represents the successful detection of
periods for one-third of the 
\citet{damiani04} X-ray catalog of NGC 6530 cluster members.
The full catalogs of cluster members with definite and possible rotation periods
are in Tables~\ref{tab:defparms} and~\ref{tab:possparms}, respectively, along with all of the associated data that we
have gleaned from the literature (see \S\ref{sec:litdata}).
Figure~\ref{fig:lc.perdef.01} shows the phase-folded light curves of the 
244 definite rotators, ordered by increasing period. 
Figure~\ref{fig:lc.perposs.01} shows the same for the 46 possible rotators.

The definite and possible rotators are also highlighted in the CMD shown
in Figure~\ref{fig:cmd}.
Our sample of rotators span the range $13.0 \lesssim I_{C} \lesssim 18.0$.
Note that many stars with high r.m.s.\ are not identified here as 
rotation period detections because they were not identified as cluster 
members in the X-ray study of \citet{damiani04}; future investigations
of cluster membership would enable an even larger sample of rotation
period determinations for NGC 6530 from our light curves.

We quantify our period detection sensitivity and any biases as
functions of period and stellar brightness.
We use 1000 of the non-variable stars in our full data set, and that are in
in the same magnitude range as our rotators, as a control sample. 
For each star we inject sinusoids with amplitude in the range $0.02 \leq \Delta \textit{I} \leq 0.2$, 
typical for our sample of rotators (see Fig.~\ref{fig:lc.perdef.01}),
into the light curves with periods in the range 0.1--20~d and run the VARTOOLS 
LS period search algorithm using the same criteria as above. We consider the
period successfully recovered if it agrees with the input period to within
10\%, which is an acceptable margin of error in the rotation periods for the
purposes of our analysis below. Some of the observed light curves for our
sample of rotators show modest departures from sinusoidal shapes (see
Fig.\ref{fig:lc.perdef.01}), but we expect any resultant errors in the
periods to be within the 10\% tolerance that we adopt for this test.

Figure~\ref{fig:detecteff}a shows the fraction of correctly recovered 
periods as a function of input period, while Figure~\ref{fig:detecteff}b
shows the fraction of correctly recovered periods as a function of $I_C$ 
magnitude. We find that our period detection efficiency is roughly constant
at $\approx$90\% for the full range of rotation periods tested. Thus, while
this simulation suggests we are missing $\sim$10\% of the true underlying
population of rotators, with variability amplitudes larger than 0.02 mag,
we infer no strong biases in the rotation period distribution as a function
of period. At the same time, there is a strong bias against period detection
for the bright stars with $I_C\lesssim 14$, clearly the result of the 
higher r.m.s.\ in our light curves for the brightest stars 
(see \S\ref{sec:obs}). For fainter stars the detection efficiency is 
approximately 100\%, and thus the $\sim$10\% loss of efficiency seen in
Fig.~\ref{fig:detecteff}a is entirely a consequence of the brighter stars.
However, again, Fig.~\ref{fig:detecteff}a indicates that this loss of 
efficiency is largely independent of period, and thus we expect no strong 
biases in the period distribution of our sample.
Figure~\ref{fig:detecteff}c shows that our detection efficiency is also
independent of amplitude over this range. We also performed a simulation
with injected amplitudes as low as 0.002 mag and found that our sensitivity remained high
down to $\Delta \textit{I} \sim 0.006$ mag. However, given the simple nature of these
simulations (e.g.~we inject perfectly sinusoidal signals), we conservatively
assume that we are not sensitive to periodic variations below $\sim$0.02 mag
given the $\sim$0.01 mag precision in the photometry.

Finally, we note that 21 of the 244 stars in the definite rotator group 
have been previously identified as candidate spectroscopic binaries 
(SB2s; see Table~\ref{tab:defparms}). In these cases it is possible that 
the periodicity we observe is related to the binary orbit but not to the
rotation of the star(s), or that the rotation of the star dominating
the light has been affected by the presence of a close faint companion
star. We do not attempt to correct for these possibilities, but note here
that these SB2s constitute less than 10\% of our sample and are not
concentrated at specific rotation periods, so we do not expect these to
alter our results. We do comment on specific interesting cases below.
We note that those SB2s with shorter periods could potentially 
be non-member field contaminants. Our catalog of members with rotation periods is
X-ray-selected, and tidally-locked short-period binaries in the foreground might
display enhanced X-ray emission that would mimic that of cluster members.

%-----------------------------
\subsection{Correlation Tests}
\label{sec:stats}
%-----------------------------

We employ two standard statistical tests to examine possible differences
or trends in the rotation periods as a function of various stellar properties:
the Kolmogorov-Smirnov (KS) test, which determines the probability that two
samples were drawn from the same parent distribution, and the Student's
$t$ test, which determines the probability that two samples possess identical
means. For a given stellar property (e.g.\ mass), we divide the period 
distribution into two bins (e.g.\ low-mass and high-mass) and apply
these statistical tests on the rotation periods of the two bins. In all of
the statistical comparisons below we include only stars with definite 
rotation periods as defined above (see \S\ref{sec:persearch}).

%%%%%%%%%%%%%%%%%%%%%%%%%%%%%%%%%%%%%%%%%%%%%%%%%%%%%%%%%%%%%%%%%%%%%%%%%%%%%%%
\section{Results}
\label{sec:results}
%%%%%%%%%%%%%%%%%%%%%%%%%%%%%%%%%%%%%%%%%%%%%%%%%%%%%%%%%%%%%%%%%%%%%%%%%%%%%%%

% -----------------------------
\subsection{Period Distribution}
\label{sec:perdist}
% -----------------------------

Figure~\ref{fig:perhist} shows the rotation period distribution of our 
entire sample of 244 cluster members with definite periods. 
The masses and ages for these stars from \citet{prisinzano05}
using the PMS evolutionary tracks of \citet{siess00} are shown in 
Figure~\ref{fig:masshist}. The typical star in our sample 
is inferred to have $M_\star$ $\sim$0.6~M$_\odot$ and age $\sim$2~Myr
according to these tracks (but see \S\ref{sec:disc} for a detailed 
discussion of the likely age of the cluster). 
Correlations between the rotation periods
and these masses and ages are discussed below.

The distribution is roughly flat for $P \le 10$ d. A peak is apparent near
$P = 1$ d, which may suggest that some aliasing effects at the diurnal 
sampling frequency of the light curves is still present. Despite this,
a one-sided KS test comparing the observed distribution with a uniform 
distribution for $P \le 10$ d yields a probability of 18\% that the two
distributions represent the same parent population. Thus, the null 
hypothesis---a uniform distribution in this case---is not rejected by the
observed period distribution.

We observe two clear cutoffs in the period distribution, despite our
good sensitivity to periodic signals for periods both longer and shorter 
than the observed cutoffs. At the long
period end, the distribution tapers off strongly for $P\gtrsim 10$ d. This
is very similar to the observed long-period cutoff in the distributions of
other young clusters, such as the ONC \citep[e.g.][]{stassun99a}.
At the short-period end, the distribution drops dramatically for 
$P\lesssim 0.5$ d.
For the typical star in our sample, with $M_\star \approx 0.6~M_\odot$ and
$R_\star \approx 2.2~R_\odot$ 
\citep[according to the PMS tracks of][]{siess00}, 
the rotation period corresponding to breakup velocity is $\approx$0.5 d. 
Therefore, we ascribe the short-period cutoff to a real physical limit 
on the minimum rotation period for the stars in our sample. 

%---------------------------
\subsection{Disk and Accretion Indicators}
\label{sec:disk}
%---------------------------

We use the reddening-free index of NIR excess, $Q_{VIJK}$ \citep{damiani06}, to 
segregate our sample according to the likelihood that they possess massive 
circumstellar disks. Those stars that are identified in \citet{damiani06} 
as having a high $Q_{VIJK}$ value display significant IR emission relative to 
their optical color, and we adopt their same classifications to indicate
which stars harbor a circumstellar disk.
We note that $Q_{VIJK}$ is a fairly crude measure of NIR excess, 
and certainly does not yield information on disk structure.
Stars without large $Q_{VIJK}$ index values may still possess disks, for
example, if the disk has an evacuated inner hole.
We perform a two-sided KS test as well as a $t$ test to compare the period
distributions of the ``disked" and ``non-disked" stars. 
Table~\ref{tab:perstats} gives the number
of stars in each group and the details of the statistical results. We
find that the period distribution for stars with NIR excess
is statistically different from those without NIR excess, with a $\sim$1\%
probability that they were drawn from the same distribution. We also find 
that the means of the distributions (6.3~d and 3.7~d for the disked and
non-disked stars, respectively) have only a $\sim$0.1\% probability of 
being the same (i.e.\ the difference in means is statistically significant). 
Figure~\ref{fig:perhist.qvijk}a
offers a visual comparison of this result. While those stars with no NIR
excess are concentrated at faster periods, those with NIR excess are more
uniformly distributed and exhibit a significant long-period component.

We also segregate the sample into accretors and non-accretors based on
their classification as a CTTS or WTTS, as determined from their H$\alpha$ 
emission strength. The distributions are shown in 
Figure~\ref{fig:perhist.qvijk}b.
The number of classified CTTS and WTTS stars is small, however the WTTSs 
appear to be concentrated at faster rotation periods while the distribution 
for CTTSs is shifted toward longer periods.
To quantify the comparison we again perform a two-sided KS tests as well as 
a $t$ test. The sample sizes and 
statistical results are in Table~\ref{tab:perstats}. The probability that 
the two period distributions were drawn from the same parent distribution 
is only $\sim$2\%, and the probability that their means (7.3~d and 3.4~d for 
the CTTS and WTTS samples, respectively) are identical is only 0.6\%.

% ---------------
\subsection{Stellar Mass}
\label{sec:mass}
% ---------------

Figure~\ref{fig:permass.qvijk} shows rotation period as a function of 
stellar mass \citep[inferred from the PMS evolutionary tracks of][]{siess00}
for the sample of NGC 6530 rotators.
There is an apparent trend of decreasing rotation period (faster
rotation) toward the higher stellar masses.
We also show in the figure the stars for which we have NIR excess
information (left panel) and the stars for which we have CTTS/WTTS
status information (right panel).
There is a clear bias present such that almost none of the stars with
either NIR excess information or CTTS/WTTS status are present for
$M_{\star} < 0.5~M_{\odot}$. This is the result of the observational
limits of the NIR and spectroscopic surveys of the cluster, which were
not sensitive to the fainter, low-mass members of the cluster
\citep{prisinzano07, arias07}.

As an initial quantitative measure of the dependence of rotation 
period on stellar mass, we divide the sample of rotators into two groups 
of comparable size based on the overall distribution of the stellar masses 
(see Fig.~\ref{fig:masshist}): a ``low mass" group with 
$M_\star \leq 0.5~M_{\odot}$, and a ``high mass" group with 
$M_\star > 0.5~M_{\odot}$.
We chose the mass cut to be at 0.5~M$_{\odot}$ because previous studies
have suggested a change in the behavior of the period distribution at around
0.4~M$_{\odot}$ (for the \citet{siess00} tracks used here), but dividing the
sample at 0.4~M$_{\odot}$ would have created imbalanced groups (63 ``low-mass"
stars to 181 ``high-mass" stars) for definite rotators in our catalog. We have
checked that all statistical results reported below based on this mass division
are not changed qualitatively if we instead cut on 0.4~M$_{\odot}$.
Figure~\ref{fig:perhist.mass} compares the period distributions of these 
two mass groupings, which are clearly different. 
A KS test gives that the probability 
of the two period distributions being drawn from the same parent 
distribution is only 0.1\%, and a $t$ test gives a probability of only 
0.5\% that the means of the two period distributions are identical
(see Table~\ref{tab:perstats}). 
The high-mass stars rotate faster than the low-mass stars; their mean
rotation periods are 4.9~d and 6.4~d, respectively 
(Table~\ref{tab:perstats}).

% --------------
\subsection{Stellar Age}
\label{sec:age}
% --------------

Figure~\ref{fig:perage.qvijk} shows rotation period as a function of
age \citep[inferred from the isochrones of][]{siess00}
for our sample of NGC 6530 rotators.
Here any trends between the rotation periods of the stars and their
inferred ages are more subtle than is the case with stellar mass (see 
above). The stars with available NIR excess and CTTS/WTTS status indicators
are not strongly biased with respect to inferred stellar age.

To quantify any relationship between rotation period and inferred stellar 
age, we divide the sample into two age groups.
Based on the distribution of inferred stellar ages 
(see Fig.~\ref{fig:masshist}b), we divide the stars into
a ``young'' group with log(age/yr)~$\leq$~6.25 and an ``old'' group 
with log(age/yr)~$>$~6.25. 
As was suggested visually in Fig.~\ref{fig:perage.qvijk}, a $t$ test
does not indicate strong evidence for a statistically significant 
difference in the mean periods of the two groups
(Table~\ref{tab:perstats}). However, a KS test does indicate that
the young and old stars have only a 1.4\% chance of being drawn from 
the same parent rotation period distribution. We conclude that there
is weak evidence for a difference in the rotation period distributions
of the stars as a function of inferred stellar age, with the older
stars rotating slightly more slowly on average (mean rotation periods
5.9~d and 4.9~d, respectively; Table~\ref{tab:perstats}).

% -------------------------------
\subsection{Spatial Distribution}
\label{sec:spatial}
% -------------------------------

Previous works have suggested evidence for sequential star formation in 
NGC 6530 and the larger Lagoon Nebula region. For example, \citet{lada76}
suggested that star formation has progressed from NGC 6530 
to Herschel~36, nearby and to the west. 
Similarly, the X-ray study of \citet{damiani04} found evidence for
an age gradient in NGC 6530, wherein the younger stars are more 
concentrated in the southeast and older stars in the northwest 
(cf.\ Fig.~12 in that paper).

Figure~\ref{fig:spatial.qvijk} shows the spatial distribution of 
our sample of rotators in NGC 6530. 
Following \citet{damiani04} we divide the cluster into
quadrants (northwest, southeast, etc.) and perform our statistical 
tests on the rotation period distributions of the stars in different 
pairs of quadrants. 
As shown in Table~\ref{tab:perstats}, the Student's $t$ test does not
indicate any statistically significant difference in the mean rotation
periods as a function of spatial position. However a KS test does show
a modestly 
significant difference in the period distributions when the southeast
quadrant is compared to the northwest quadrant, with a 
probability of $\sim$1\% that the rotation period distributions
of the two groups were drawn from the same parent distribution. 
The mean rotation periods in the southeast and northwest quadrants
are 6.2~d and 5.2~d, respectively (Table~\ref{tab:perstats}).
Together with the age gradient found above, this difference in periods 
would imply that the younger stars
(to the southeast) rotate more slowly on average than the older stars
(to the northwest). 

However, this is in the opposite sense of the (weak) period-age trend 
found above, in which the isochronally younger stars rotate more 
rapidly on average. Thus we conclude that any trends in rotation with
isochronal age or with age inferred from spatial location are weak at best
and inconsistent in sense.

% -------------------------
\subsection{X-ray Activity}
\label{sec:xray}
% -------------------------

To investigate the relationship between the rotation period of
NGC 6530 cluster members and their X-ray activity, we require estimates
of the X-ray luminosities ($L_X$) and the bolometric luminosities 
($L_{\rm bol}$) of the stars 
in our sample. Neither of these quantities was tabulated by the 
previous studies of the cluster, so here we adopt a procedure to 
provide estimates of both quantities. 

To estimate $L_X$ for each star, we use 
the PIMMS software\footnote{Distributed by 
NASA's High Energy Astrophysics Science Research Center; 
http://heasarc.gsfc.nasa.gov/docs/software/tools/pimms.html}
we convert the X-ray count rates of \citet{damiani04} into X-ray fluxes,
using the PIMMS {\sc mekal} model. 
The model requires as input the temperature of the emitting
coronal gas ($kT$) and the hydrogen column density toward the source.
For $kT$, we adopt the median of the $kT$ distribution found by COUP \citep{getman05a}. 
We adopt an extinction of $A_{V}$=1.1 to the cluster 
\citep{prisinzano05}, which yields a hydrogen column 
density of $N_\mathrm{H}=2.431\times10^{21}$ cm$^{-2}$ \citep{guver09}.
Finally, adopting a cluster distance of 1.25 kpc \citep{prisinzano05}, 
we convert the PIMMS X-ray fluxes into $L_X$.
This approach obviously does not take into account potential differences
in $kT$ or $A_V$ to individual stars, however as only the X-ray count
rates are available from \citet{damiani04}, a
more sophisticated approach is not warranted.

To estimate $L_{\rm bol}$, we interpolate on the PMS evolutionary 
tracks of \citet{siess00} to obtain the predicted $L_{\rm bol}$ for
each star, given the mass and age estimates from \citet{prisinzano05}
using these same evolutionary tracks.

Figure~\ref{fig:lxlbol} shows the resulting $L_X / L_{\rm bol}$ of the
NGC 6530 rotators as a function of rotation period. 
As a whole the sample shows a roughly constant $L_X / L_{\rm bol}$
at approximately the ``saturation" value of 
$\log L_X / L_{\rm bol} \approx -3.3$ \citep[e.g.][]{pizzolato03}.
However, the most rapidly rotating stars appear to exhibit a
systematically reduced $L_X / L_{\rm bol}$. To quantify this, we
perform both a KS test and a Student's $t$ test comparing the 
$L_X / L_{\rm bol}$ for rapid rotators with $P < 2.5$~d versus more 
slowly rotating stars with $P > 2.5$~d (see Table~\ref{tab:xraystats}).
We find a stastisically significant difference from both tests, with
the mean $\log L_X/L_{\rm bol}$ for the rapid rotators ($-3.56$)
being lower than that for the slower rotators ($-3.37$) with a 
statistical significance of 99.97\%.

In addition, 
we have checked whether our sample of rotators is representative of 
the underlying population of NGC 6530 members in $L_X / L_{\rm bol}$.
Figure~\ref{fig:lxlbolhist} compares the distribution of $L_X / L_{\rm bol}$
for stars with and without a measured rotation period.
We again find a very statistically significant difference in the two 
distributions from both a KS test and a $t$ test. The mean 
$\log L_X / L_{\rm bol}$ for the sample of rotators ($-3.43$) is 
significantly higher than that for the sample without detected rotation 
periods ($-3.63$). The probability that there is no difference in the 
mean $\log L_X / L_{\rm bol}$ of the two samples is $6\times10^{-6}$
(see Table~\ref{tab:xraystats}).

%%%%%%%%%%%%%%%%%%%%%%%%%%%%%%%%%%%%%%%%%%%%%%%%%%%%%%%%%%%%%%%%%%%%%%%%%%%%%%%
\section{Discussion}
%%%%%%%%%%%%%%%%%%%%%%%%%%%%%%%%%%%%%%%%%%%%%%%%%%%%%%%%%%%%%%%%%%%%%%%%%%%%%%%
\label{sec:disc}

Ever since the pioneering efforts of Bouvier and collaborators to measure
rotation periods of Tau-Aur stars
\citep[e.g.][]{bouvier86,bouvier93,bouvier97}, 
and of Herbst and collaborators to measure rotation periods of ONC stars 
\citep[e.g.][]{attridge92,herbst94,choi96}, 
a fundamental goal has been to characterize the morphology of the 
rotation period distribution for young low-mass stars.
Early works on the ONC emphasized the apparent bimodality of the period
distribution, with peaks around $\sim$2~d and $\sim$8~d and a deep gap
in the distribution around $\sim$4--5~d
\citep[e.g.][]{attridge92}. In contrast, \citet{stassun99a} 
found a unimodal distribution in the ONC. These differing results were
subsequently argued to be a manifestation of the mass dependence of the 
period distribution \citep[e.g.][]{herbst01}: 
a bimodal period distribution for solar-mass stars,
a unimodal distribution for lower mass stars, and with the lower mass stars
rotating faster than the higher mass stars.
Several studies have confirmed these trends in the ONC and in other, 
slightly older clusters, including NGC 2264, NGC 2362, and IC 348
\citep[e.g.][]{kearns97,kearns98,herbst00b,lamm05,cieza06,irwin08a}.
Thus a picture has emerged in which young stars at $\gtrsim$1~Myr exhibit 
a mass-dependent period distribution that is bimodal at higher masses, 
in which lower mass stars rotate faster on average, and in which older
stars tend to spin faster, presumably due to spin-up as the stars contract
toward the main sequence.

The distribution of rotation periods we have measured for NGC~6530 is 
consistent with a uniform distribution for $0.5 < P < 10$~d;
we do not observe obvious bimodality in the period distribution. 
When we subdivide the sample stars into groups by mass, we do not
observe bimodality among the higher mass stars (nor for the lower mass stars),
and moreover the lower mass stars rotate more slowly on average. These
features of the NGC~6530 period distribution and its dependence on stellar
mass differ strongly from the trends discussed above for numerous other
clusters with ages $\gtrsim$1~Myr.

We find that NGC~6530 stars with older isochrone ages 
rotate more slowly on average than their younger counterparts.
The statistical significance of this trend is not strong,
but as with the other rotational properties of NGC 6530 noted above, such a
trend is in contrast with that expected from the longer-term evolution observed
between other extensively-studied young clusters, which show a tendency for stars
to spin up modestly between the age of the ONC and that of the Pleiades.

These differences in rotational characteristics between NGC 6530 and the 
other young clusters might be understood if NGC 6530 represents a
population of PMS stars that is significantly younger than the other 
clusters, such that in particular the higher mass stars in the cluster
are having their period distribution shaped from a unimodal one into
a bimodal one, and the lowest mass stars in the cluster are actively 
spinning up so that they will end up spinning faster than their higher
mass counterparts. That is, the NGC 6530 stars are perhaps currently 
evolving toward an evolutionary state when their rotational properties 
would presumably resemble those of the ONC.
Indeed, the age of NGC 6530 has been estimated by \citet{mayne07} to be 
similar to, and possibly slightly younger than, the ONC. 

In an attempt to more firmly place NGC 6530 in an evolutionary context
relative to other well studied young clusters, 
and motivated by the findings of \citet{irwin08a} who suggested patterns 
with age in the mass--period relationship of young clusters,
we show in Figure~\ref{fig:permassclusters} the rotation periods as a 
function of stellar mass for NGC 6530 and five other young clusters
with extant rotation period measurements. NGC 6530 is shown first, and
the other clusters ordered chronologically with ages from 
\citet{mayne07} and \citet{mayne08}.
Finally, we also include the zero-age main sequence cluster NGC 2516.
The six clusters from ONC to NGC 2516 thus span a range of ages from 
$\sim$2~Myr to $\sim$150~Myr.
The period--mass relationship for NGC 6530 as expected appears most similar
to that seen in the ONC. Broadly speaking, whereas the older clusters exhibit 
an increasing tendency for the upper envelope of rotation periods to slope
downward at low stellar masses, the upper envelope of rotation periods in 
NGC 6530 is, like the ONC's, roughly flat with stellar mass. However, 
whereas the ONC does exhibit a modest downward slope toward decreasing
stellar masses (i.e.\ the lowest mass stars in the ONC rotate on average
faster than the higher mass stars), in NGC 6530 the trend is in the opposite
sense (i.e.\ the lowest mass stars rotate on average more slowly; see
\S\ref{sec:mass}), and this appears in Fig.~\ref{fig:permassclusters} as a
slightly upward slope in the upper envelope of NGC 6530 rotation periods.

To better quantify these trends of rotation period with mass, we fit a
linear trend line to the upper envelopes of the rotation periods versus
mass for each of the clusters in Fig.~\ref{fig:permassclusters}
(shown as red lines) as follows.
For stars with masses in the range $0.5 < M/{\rm M}_\odot < 0.1$, 
we grouped the stars into mass bins 0.1~M$_\odot$ wide, and within each
of these mass bins we calculated the rotation period corresponding to the 
75\%-ile of the rotation periods in that bin. (We chose the 75\%-ile 
because it is a more robust measure of the upper envelope of the 
distribution than, e.g., taking the upper-most data point in the bin.)
We calculated the uncertainty on the 75\%-ile periods as the difference
between the 75\%-ile and the 50\%-ile (the median) divided by the 
square-root of the number of data points in the bin. Finally, we fit a
linear least-squares relationship to the binned points of the form
$\log P = a \times M + b$, where $a$ and $b$ are free parameters of the fit.
The resulting slopes $a$ and their formal uncertainties are shown plotted 
versus the age of each cluster in Figure~\ref{fig:ageslopes}, in which we 
see a clear relationship of increasing slope with increasing age for the
younger clusters with ages $\lesssim$10~Myr and a flattening of the 
relationship for the two oldest clusters at $\gtrsim$40~Myr.
For specificity, the ages we assigned to each cluster are from 
\citet[][cf.\ their Table 9]{mayne08}, except that for NGC 2362 we adopted
an age of 3.5~Myr because its age was estimated as 3~Myr \citep{mayne08}
and 4~Myr \citep{mayne07}, and for IC348 we adopted 4.5~Myr as its age
was estimated by \citet{mayne08} as 4--5~Myr. 
These cluster ages are summarized in Table~\ref{tab:clusterages}.

For our linear fit in 
Fig.~\ref{fig:ageslopes} (solid line) we did not include NGC 6530;
rather, we placed the point corresponding to NGC 6530 at the age 
at which the 1-sigma upper limit for its rotation period versus mass slope
exactly lies on the linear trend fitted to the other clusters.
The maximum age inferred for NGC 6530 by this procedure
is 1.65~Myr, as compared to the 2~Myr age assigned to the ONC. 
In addition, the fit to the younger clusters was extended to only 6~Myr,
as the measurements for the two oldest clusters (NGC 2547 and NGC 2516)
clearly indicate that the trend of increasing slope in the period-mass
plane with age must flatten at approximately this age (represented by the
dotted line in Fig.~\ref{fig:ageslopes}). 
The linear relationship fitted to the younger clusters (solid line in 
Fig.~\ref{fig:ageslopes}) has the form
\begin{equation}
a = 5.98(\pm1.11) \times \tau - 1.50(\pm0.61) ,
\end{equation}
where $a$ is the slope of the linear relationship between $\log P$ 
(in days) and $M$ (in M$_\odot$)
for the upper 75\%-ile of rotation periods in each cluster over the mass
range 0.1--0.5~M$_\odot$, and $\tau$ is the cluster age (in Myr).
This relationship may be useful for assigning relative ages to PMS 
stars on the basis of the observed slope in the period--mass relationship.

If NGC 6530 is indeed younger than the ONC, then this implies that the
distance to the cluster of 1.25~kpc determined by \citet{prisinzano05}
must be slightly underestimated. The median age of the stars in our 
sample inferred from the PMS isochrones of \citet{siess00} is $\sim$2~Myr 
(see Figs.~\ref{fig:cmd} and \ref{fig:masshist}b), the 
same as the ONC using these isochrones. 
However, if the distance to NGC 6530 is taken to be just 15\% larger,
the median age of the NGC 6530 sample comes in line with our estimate 
of $\sim$1.5~Myr above. \citet{prisinzano05} do not quote an 
uncertainty on their distance determination, but most other recent
distance estimates for the cluster are $\sim$25\% larger than 1.25~kpc
\citep[e.g.][]{vandenancker97,loktin97,sung00},
so a 15\% revision would not appear to be unreasonable. Indeed, the
original X-ray study of \citet{damiani04} adopted a distance of 1.8~kpc,
and consequently determined a median age for the cluster of just 0.8~Myr
from the same PMS isochrones used here \citep{siess00}.

The NGC~6530 rotation period distribution shows a strong cutoff for
fast rotation periods, $P < 0.5$~d, and we have found that this
short-period cutoff corresponds to breakup speed for these stars.
A similar short-period cutoff associated with rotation at breakup
was observed in the ONC \citep{stassun99a}.
A few stars in our sample are found with $P < 0.5$~d. 
While rotation at breakup is a possibility for these stars, we note that 
their light curves are strikingly similar to those of contact binaries. 
Some of these stars do have prior spectroscopic data in the literature
that did not clearly identify them as spectroscopic binaries 
(see Table~\ref{tab:defparms}). However, we note that the spectra of contact
binaries can appear highly broadened, and the line splitting might not be 
readily recognized as such. 
The discovery of PMS contact binaries would be very significant in the 
context of binary formation and evolution, and we suggest that these 
stars be monitored further for indications of radial velocity variations.
A few PMS contact binary candidates have also been identified in 
Orion \citep{stassun99a,rebull01,vaneyken11}.

The X-ray luminosities of the NGC~6530 stars are flat with rotation period,
at the saturation level 
\citep[$\log L_X / L_{\rm bol} \approx -3.3$;][]{pizzolato03}, however the 
most rapidly rotating stars show significantly lower $\log L_X / L_{\rm bol}$
suggestive of so-called ``super-saturation" 
\citep[e.g.][]{james00}. A similar result was found
in the ONC by \citet{stassun04b}. 
Recent studies of rotation and X-ray activity in low-mass stars at
a variety of ages \citep[e.g.][]{wright11,jeffries11} have argued that super-saturation
may be the result of the X-ray coronae in very rapidly rotating stars extending
beyond the Keplerian co-rotation radius, causing the coronae to be centrifugally
stripped.
To examine this idea in the context of our NGC 6530 sample, in 
Figure~\ref{fig:corotation} we plot $\log L_X / L_{\rm bol}$ versus the 
Keplerian co-rotation radius ($R_{\rm co}$) for our sample, where 
$R_{\rm co}$ is determined for each star from the measured rotation 
period and from the mass and radius inferred from the \citet{siess00} 
evolutionary tracks.
We find strong evidence for a correlation between these quantities,
(shown as the line in Fig.~\ref{fig:corotation}),
similar to that suggested by \citet{wright11}. A non-parametric 
Kendall's $\tau$ rank correlation test yields a positive correlation 
coefficient of 0.21, and the probability that the two quantities are
not correlated is $<$10$^{-6}$.  The fastest rotators in our NGC 6530
sample---and those with the lowest $L_X / L_{\rm bol}$ on average---have 
$R_{\rm co}$ in the range of 1--3 $R_\star$, whereas the slower rotators
in our sample have $R_{\rm co}$ up to $\sim$15 $R_\star$.
The COUP survey \citep{getman05a} found that the coronae of
low-mass PMS stars in that study, as inferred from the lengths of the 
magnetic loops driving the observed powerful X-ray flares, can 
have extents of up to $\sim$10 $R_\star$ \citep{favata05,aarnio10}. 
Such coronal radii can be accomodated within $R_{\rm co}$ for the 
slower rotators in our NGC 6530 sample, but for the faster rotators would
extend beyond $R_{\rm co}$ and would thus be unlikely to remain stable
against centrifugal forces.
Thus, it appears plausible that the correlation we observe between
$L_X / L_{\rm bol}$ and $R_{\rm co}$ is the result of the outer coronae
of the rapidly rotating stars being increasingly centrifugally stripped,
as suggested by \citet{wright11} and \citet{jeffries11}.

\citet{stassun04b} also found in the ONC that stars with measured rotation
periods exhibit significantly higher $L_X / L_{\rm bol}$ on average than
ONC stars without rotation periods, and suggested that this
might indicate a population of stars rotating more slowly than the 
observed long-period cutoff in the ONC ($P \gtrsim 10$ d). Our NGC~6530
sample exhibits very similar properties. In particular, we observe a 
long-period cutoff for $P \gtrsim 10$ d, and moreover we find that 
NGC~6530 stars that do not exhibit a rotation period signal in our data
are less X-ray luminous on average. Perhaps there is a population of more
slowly rotating stars in NGC~6530 than our rotation period measurements
can reveal. These low L$_{X}$ stars could still have periodic signals
below 0.02 mag amplitudes that we would not detect given the precision
of our data (see \S\ref{sec:persearch}). Sensitive $v\sin i$ measurements in NGC~6530
and the ONC are needed to explore this possibilty further.

We find evidence that stars in our NGC 6530 sample with NIR excess 
emission and/or strong H$\alpha$ emission rotate more slowly on average. 
In other young clusters, an association between NIR excess and
slow rotation has been taken as evidence of the braking of stellar
rotation through a magnetic star-disk interaction (so-called ``disk-locking").
However, it is not clear that disk-locking models actually predict such a
correlation of increased NIR excess for slow rotators. A central prediction
of most disk-locking models is that the location of the inner truncation
radius of the circumstellar disk relative to the co-rotation radius determines
the magnitude and sign of the torque experienced by the star. Thus 
for slow rotators, whose co-rotation radii are large and for which a braking
torque would therefore require an even larger inner truncation radius, one
might predict less NIR emission for the slow rotators due to the large inner
hole in the disk \citep[e.g.][]{stassun01}. 

\citet{leblanc11} performed detailed modeling of the spectral energy 
distributions of stars in IC~348 with measured rotation periods in order 
to assess in detail for each star the location of the inner disk edge
relative to co-rotation. Those authors found that the slow rotators in
IC~348 tended to possess disks with inner truncation radii at or beyond
co-rotation, whereas the rapidly rotating stars tended to posses disks
with inner truncation radii within co-rotation, implying that if star-disk
interaction is important for the stars then it must be operating with a
tendency to torque down the slow rotators and torque up the rapid rotators.
In other words, if disks are important for angular momentum evolution in
that cluster, then they must be important for stars at all rotation periods,
spinning down some stars while spinning up others.
Thus, inferring the nature of any star-disk interaction among our sample
of rotators in NGC 6530 awaits detailed modeling of the spectral energy 
distributions of the stars in order to establish whether and how disks
may be sculpting the NGC 6530 rotation period distribution.

%%%%%%%%%%%%%%%%%%%%%%%%%%%%%%%%%%%%%%%%%%%%%%%%%%%%%%%%%%%%%%%%%%%%%%%%%%%%%%%
\section{Summary and Conclusions}
\label{sec:summary}
%%%%%%%%%%%%%%%%%%%%%%%%%%%%%%%%%%%%%%%%%%%%%%%%%%%%%%%%%%%%%%%%%%%%%%%%%%%%%%%

We have photometrically monitored $\sim$50,000 stars in a 
$40\arcmin\times40\arcmin$ field centered on NGC 6530, the young massive 
star-forming cluster illuminating the Lagoon Nebula, over 35 nights in the 
$I_C$-band with a cadence of 1~hr$^{-1}$. 
These observations are intended to complement recent optical, X-ray, and 
NIR surveys of the region \citep{damiani04,prisinzano05,damiani06,prisinzano07},
permitting a comprehensive characterization of the young stellar population 
in NGC 6530. 

From an analysis of periodic variations in our light curves, we measured 
rotation periods for 290 X-ray selected cluster members of NGC 6530, with 
masses in the range $0.2 < M/{\rm M}_\odot < 2.0$.
From the findings of \citet{damiani04}, we expect $\sim5\%$ of our catalog to
be contaminated by non-members, or only $\sim$15 stars in our full photometric catalog.
We investigated correlations between rotation period and other stellar
properties, including mass, age, spatial distribution within the cluster,
the presence of circumstellar disks, and X-ray activity. 
The major findings of this work are as follows:
\begin{enumerate}

\item 
The distribution of rotation periods in NGC 6530 is approximately uniform
over the range $0.5 < P < 10$~d; 
we do not observe obvious bimodality in the period distribution,
regardless of whether the distribution is considered in its entirety
or limited to narrower ranges of stellar mass.
The sharp cutoff in the period distribution at
$P \approx 0.5$~d likely results from the breakup limit for the stars in 
our sample.  A small number of stars with $P<0.5$~d are present, which
should be investigated further as possible pre--main-sequence contact
binary systems.

\item 
The X-ray luminosities of the stars are roughly flat with rotation period,
at approximately the saturation level ($\log L_X / L_{\rm bol} \approx -3.3$).
However, the fastest rotators show lower average X-ray luminosities, at
a highly statistically significant level, suggestive of so-called
``super saturation." 
At the same time, X-ray luminosity correlates most strongly with the 
stars' co-rotation radii, suggesting that centrifugal stripping of the
coronae may be the fundmental driver of the super saturation phenomenon.

\item 
Stars with NIR excesses and H$\alpha$ emission
indicative of warm circumstellar material
rotate more slowly on average than stars lacking disk signatures. 
Disked stars might be presumed to be younger on average, and indeed
we find evidence that stars with younger ages as inferred from 
spatial location within the cluster rotate more slowly on average.
However, the statistical significance is low, and indeed we find the
opposite association between rotation and age when the ages are inferred
from PMS isochrones. 

\item 
The rotation periods are a function of stellar mass: the lower mass stars 
rotate more slowly on average than the higher mass stars. This is in the
opposite sense of the period-mass relationship observed in the ONC 
and in all other slightly older clusters. 

\item
We show that the slope of the mass-period relationship among slow 
rotators (defined as the 75\%-ile rotation periods) in the mass range 
$0.1 < M/{\rm M}_\odot < 0.5$ is a good proxy for the age of a young
cluster. Calibrating this empirical mass-period-age relation to the ONC,
NGC 2264, NGC 2362, IC348, NGC 2547, and NGC 2516, we find that NGC 6530 
is the youngest of all, with a maximum age of 1.65~Myr on an age scale 
where the ONC is 2~Myr.

\end{enumerate}

The evidence points strongly to NGC 6530 being in a very early stage of 
rotational evolution in which the stars are currently evolving toward a 
state that will presumably resemble the ONC within the next 
$\lesssim$1~Myr. Thus NGC 6530 becomes an important new touchstone for
theoretical models of angular momentum evolution in young, low-mass stars.

An important question that remains to be resolved is the role of 
circumstellar disks in the rotational evolution of these stars. The 
observed correlation between NIR excess and slow rotation has been taken
in previous studies as evidence for rotational braking via star-disk
interaction. However, it is not clear that theories of star-disk 
interaction in fact predict this correlation. Additionally, recent
detailed modeling of the full spectral energy distributions of young
stars with rotation periods in IC348 indicate that, if disks do affect 
the spin rates of the stars, they must act both to spin down some stars 
and to spin up others \citep{leblanc11}. 
A similarly detailed assessment of the disk torques likely being
experienced by the stars in NGC 6530 will be important to determine
whether and how disks may yet be acting to shape the mass-period 
relationship in this very young cluster.

Finally, it remains an important challenge to empirically connect the 
rotational properties of PMS stars to 
those of main-sequence stars, and to theoretically connect 
the dominant mechanisms thought to govern the evolution of angular 
momentum in the PMS to those on the main sequence.
Main sequence angular momentum evolution is principally understood
through intrinsic structural changes in the stars that, 
through stellar winds, lead to distinct period-mass relationships that 
evolve predictably with time and thus permit reliable age-dating of stars
(gyrochronology). In contrast, in the PMS phase the dominant angular
momentum evolution mechanisms have generally been thought to be 
extrinsic to the stars (e.g., disk-locking). 
Yet it is now clear that low-mass stars already evince clear period-mass
relationships in the PMS stage. Evidently, as early as the very young
age of NGC 6530, a period-mass relationship that can be projected 
forward to the main sequence is already taking form,
and this relationship already encodes stellar age.

%%%%%%%%%%%%%%%%%%%%%%%%%%%%%%%%%%%%%%%%%%%%%%%%%%%%%%%%%%%%%%%%%%%%%%%%%%%%%%%
\acknowledgments
We thank Robert Siverd for technical assistance and Soeren Meibom
for enlightening discussions. CBH acknowledges the support of the NSF 
Graduate Research Fellowship \#2011082275 and also the NSF-funded Research Experiences
for Undergraduates program in Physics \& Astronomy at Vanderbilt University. 
KGS acknowledges support through NSF grant AST-0808072, as well as the
generous support and hospitality of a Martin Luther King Visiting
Professorship at the Massachusetts Institute of Technology.

%%%%%%%%%%%%%%%%%%%%%%%%%%%%%%%%%%%%%%%%%%%%%%%%%%%%%%%%%%%%%%%%%%%%%%%%%%%%%%%

%==============================================================================
%                                FIGURES
%==============================================================================

\clearpage

\begin{figure*}[ht]
\plotone{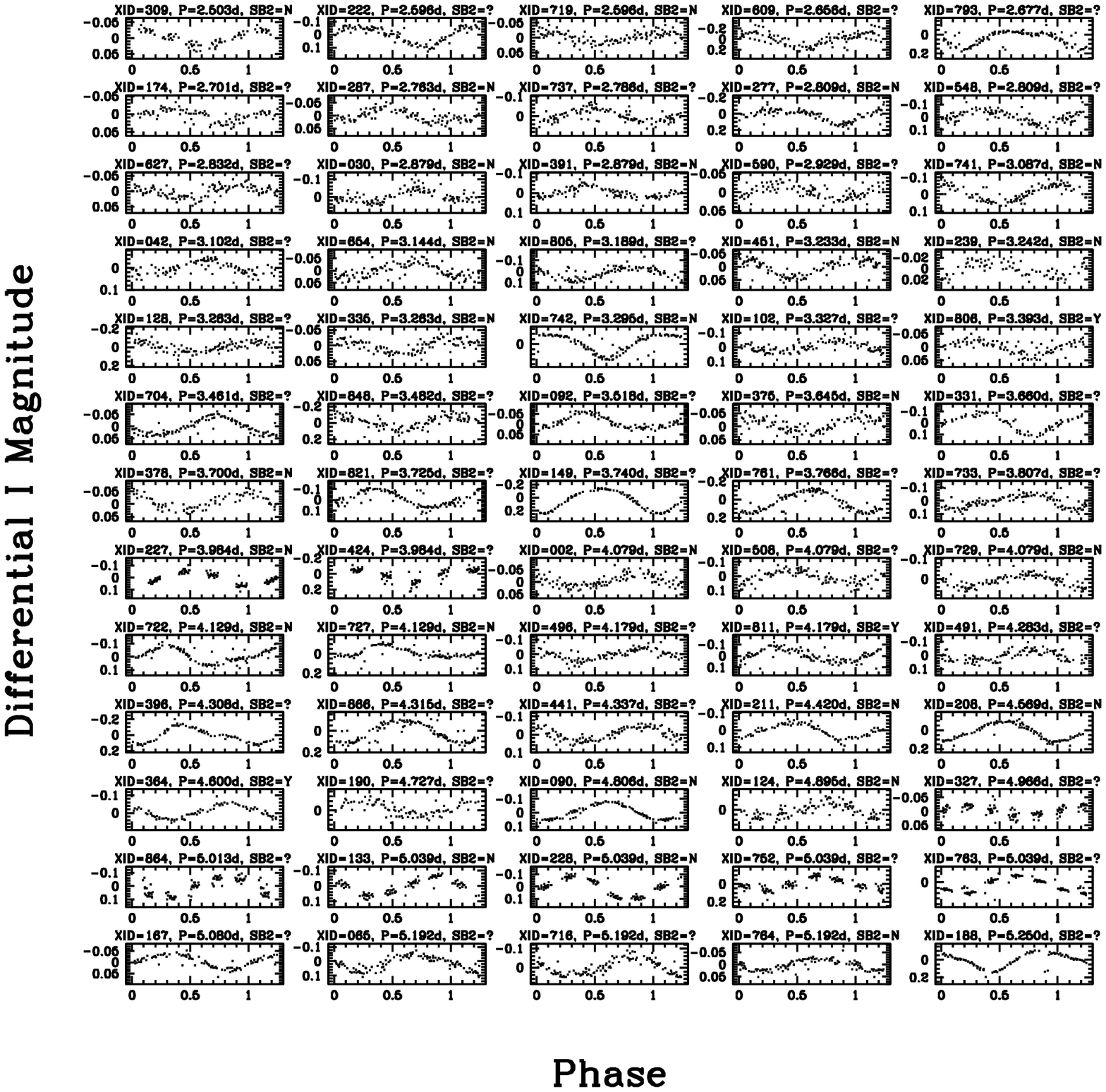}
\end{figure*}

\begin{figure*}[ht]
\plotone{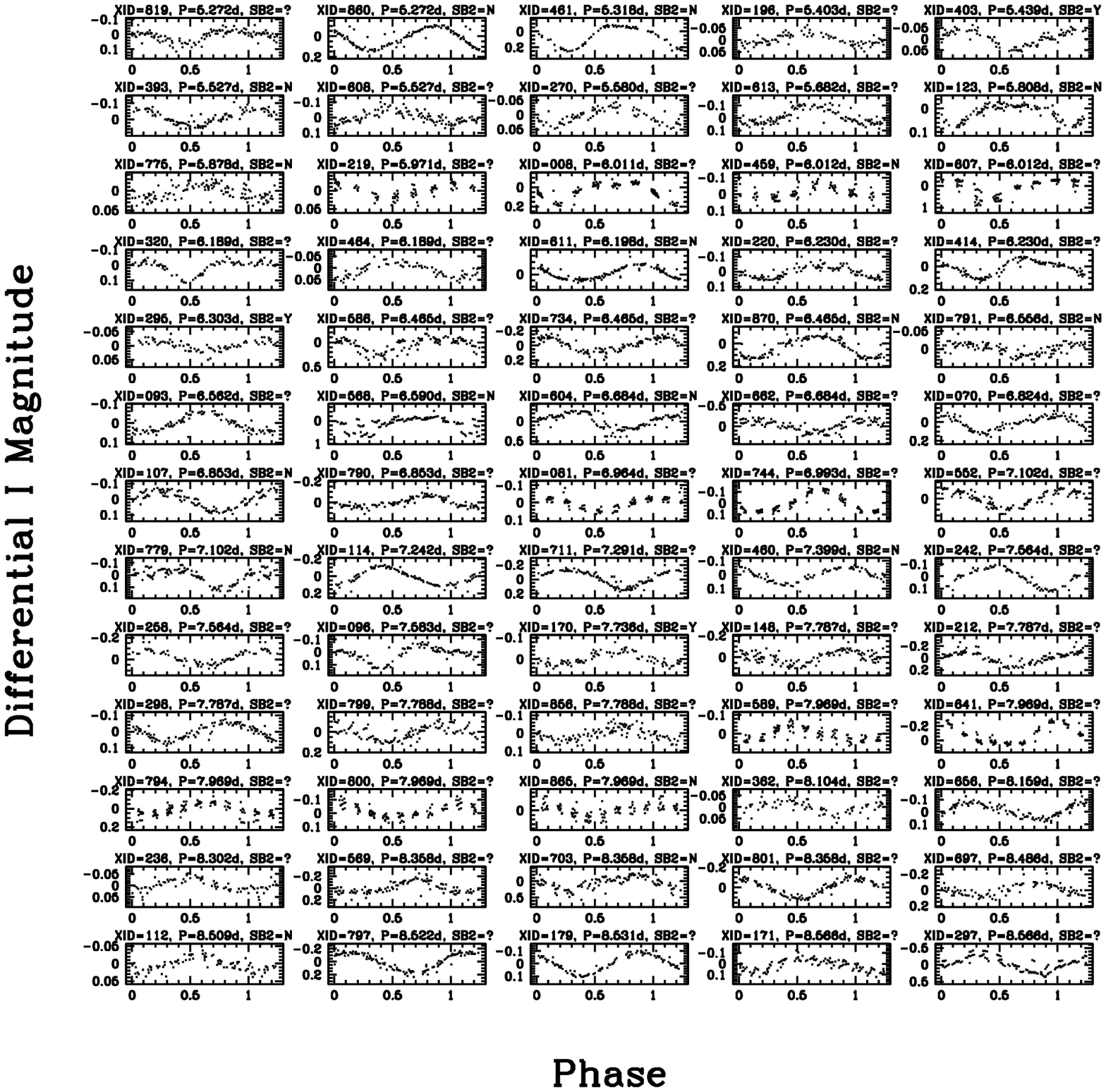}
\end{figure*}

\begin{figure*}[ht]
\plotone{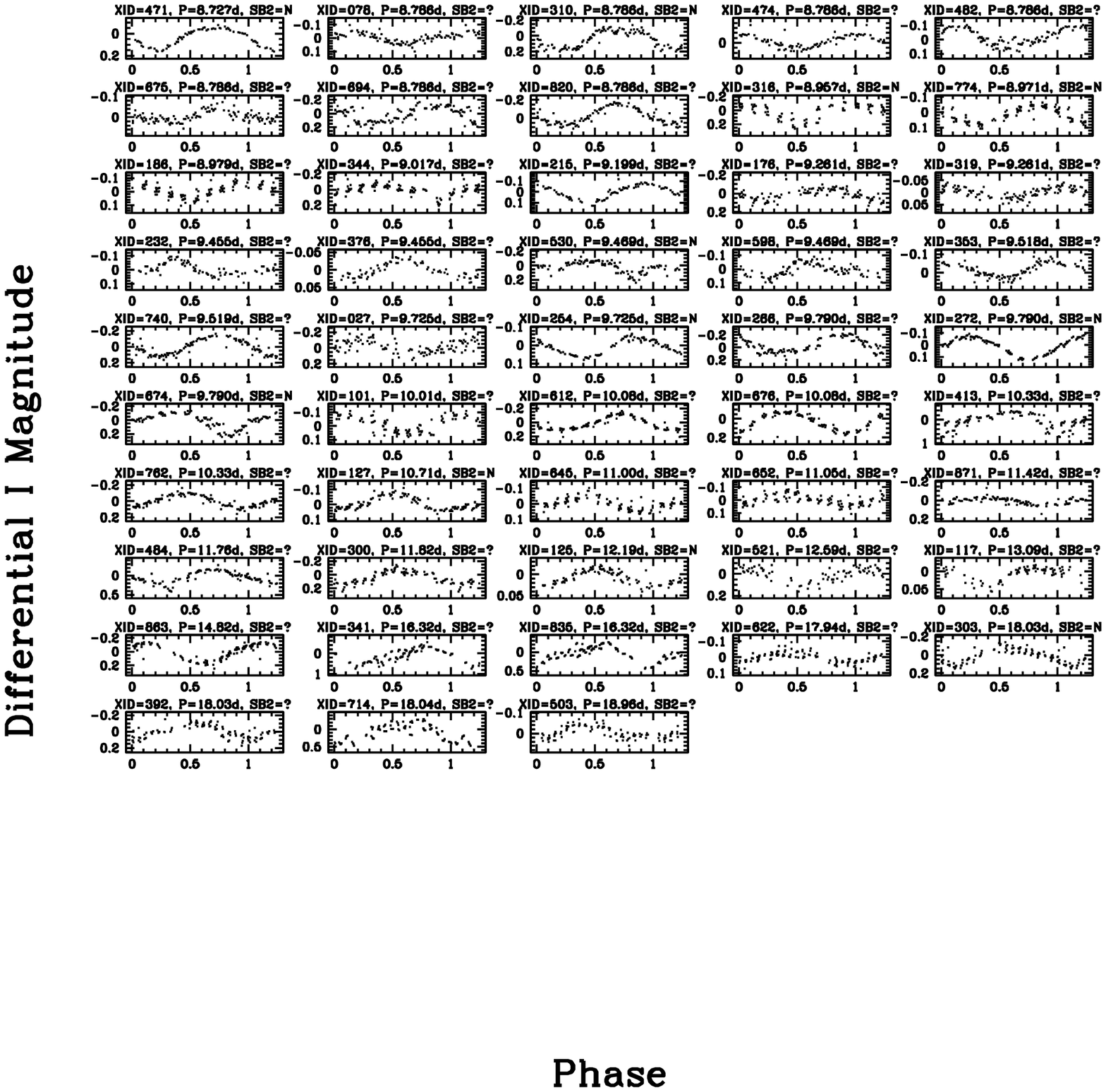}
\end{figure*}

\begin{figure*}[ht]
\plotone{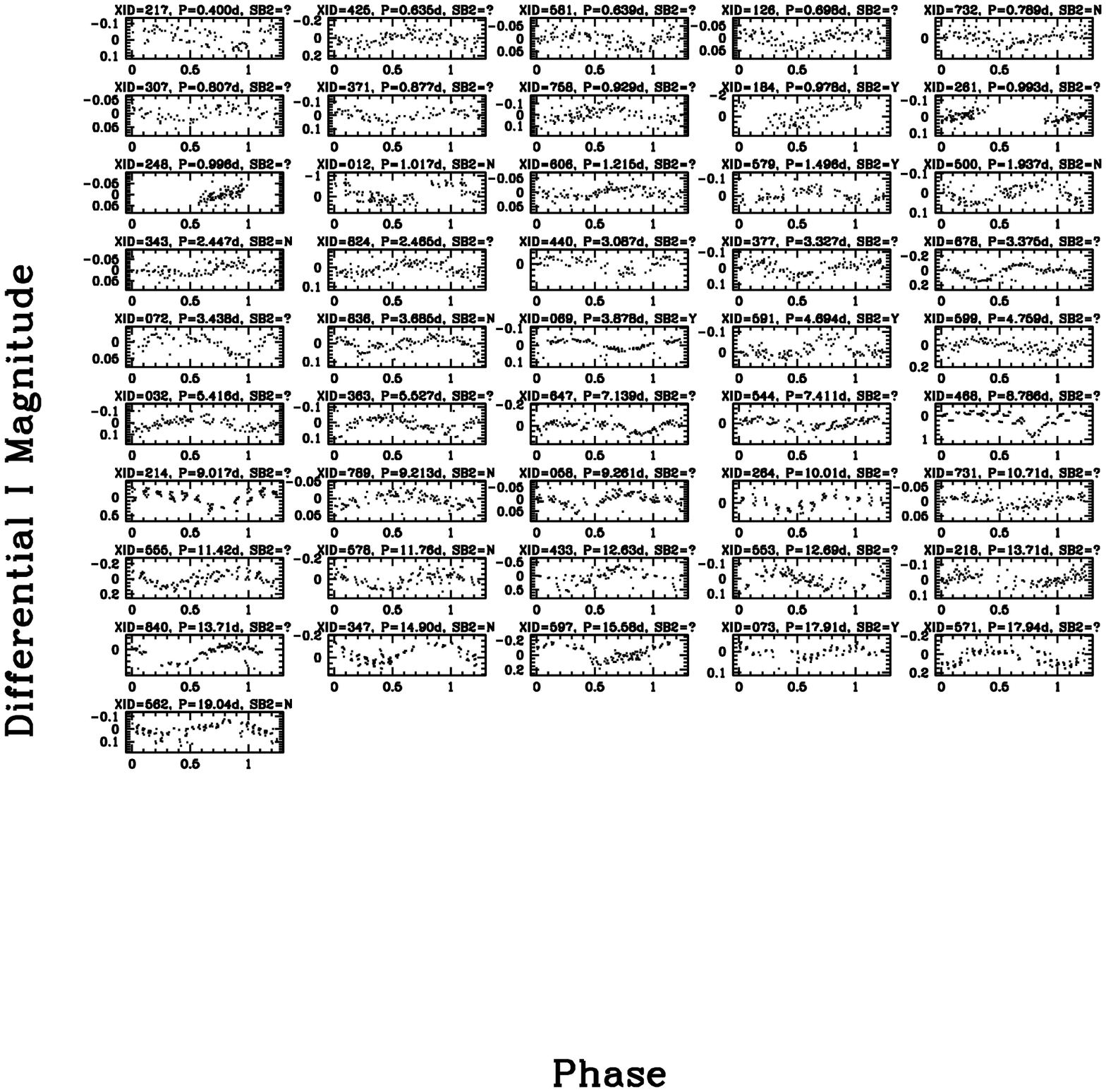}
\caption{Same as Fig.~\ref{fig:lc.perdef.01} but for
NGC 6530 cluster members with ``possible'' rotation
periods (0.001~$<$~FAP~$\leq$~0.01).
\label{fig:lc.perposs.01}}
\end{figure*}

\clearpage
\begin{figure}[ht]
\plotone{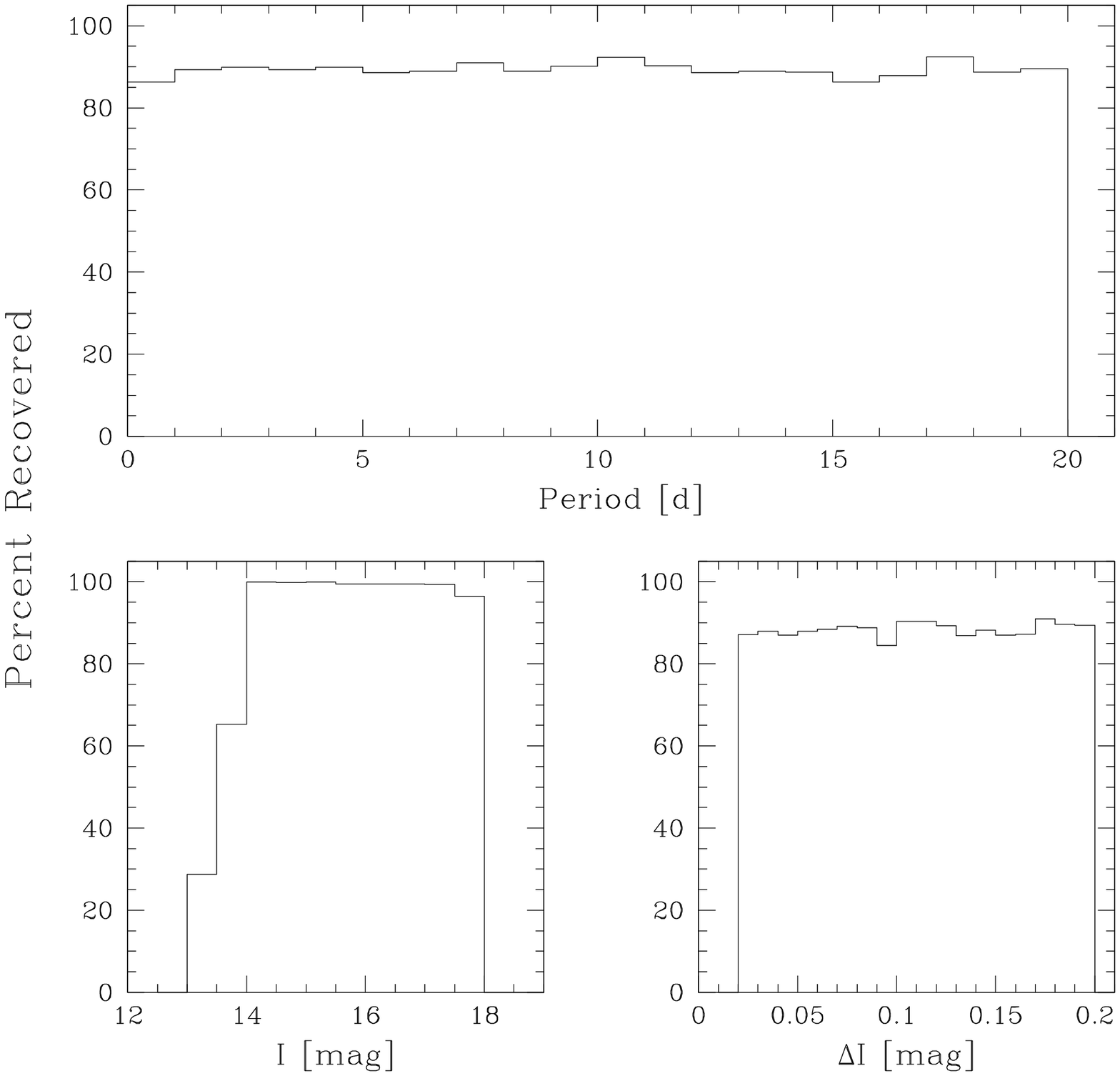}
\caption{Efficiency of period detection. 
{\it (Top:} This figure shows the fraction of
correctly detected periods as a function of input period, using as a 
control our light curves of non-variable stars spanning the same range of 
$I_C$ magnitudes as our sample of rotators. We inject sinusoids with 
amplitudes in the range $0.02 \leq \Delta \textit{I} \leq 0.2$ mag
into the light curves, and consider the period
successfully recovered if it matches the input period to within 10\%.
{\it (Bottom left:)} Same as top panel, but showing period recovery fraction
as a function of $I_C$ magnitude.
{\it (Bottom right:)} Same as other panels, but showing period recovery fraction
as a function of injected amplitude.
\label{fig:detecteff}}
\end{figure}

\begin{figure}[ht]
\plotone{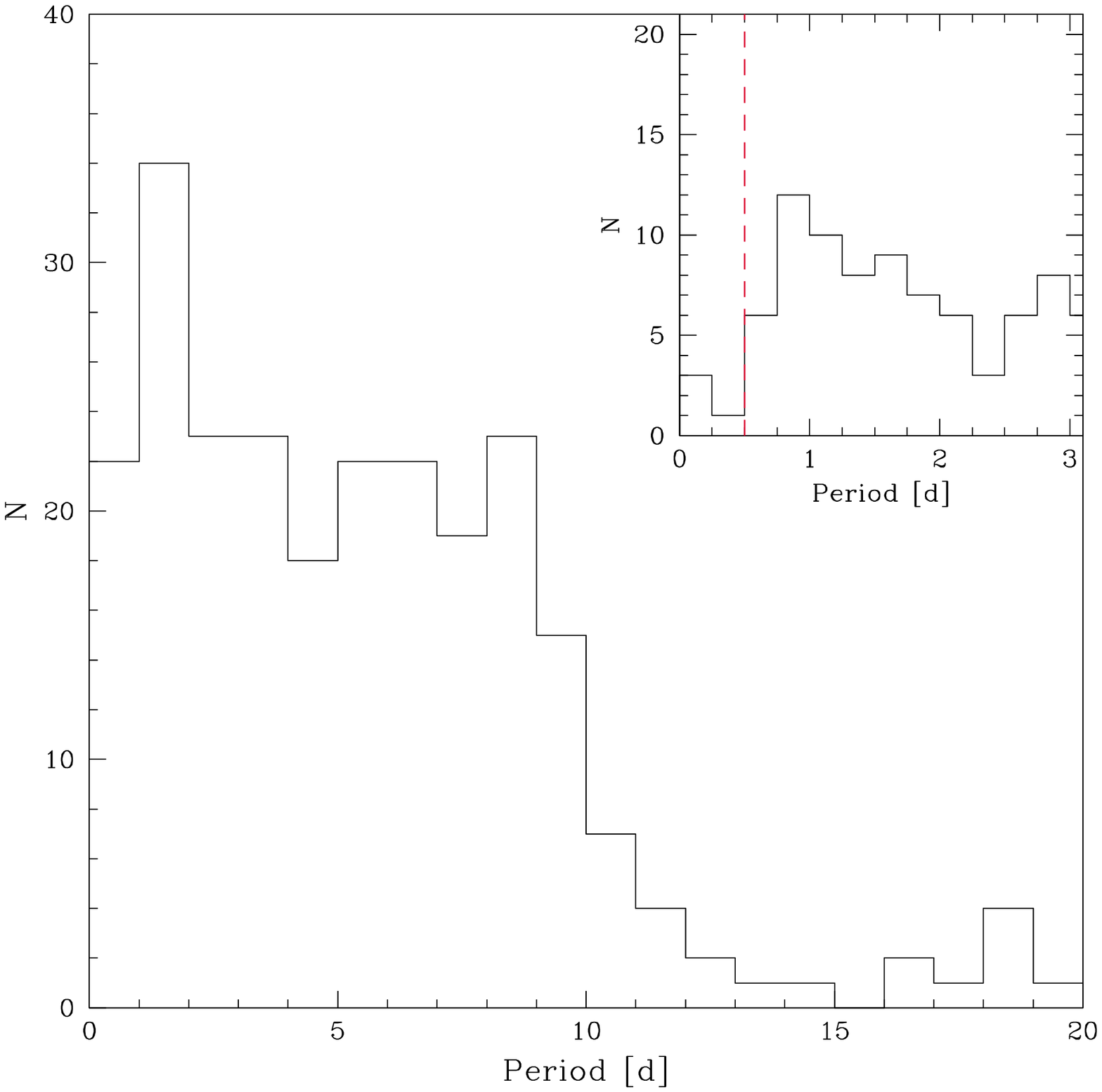}
\caption{The period distribution of our sample of 244 definite
NGC 6530 rotators, with bins of 1~d. The inset shows the same distribution, 
but with 0.25~d bins at the short-period end of the distribution.
The vertical line in the inset shows approximately the rotation period
corresponding to breakup speed for the typical rotator in our sample.
\label{fig:perhist}}
\end{figure}

\begin{figure}[ht]
\plottwo{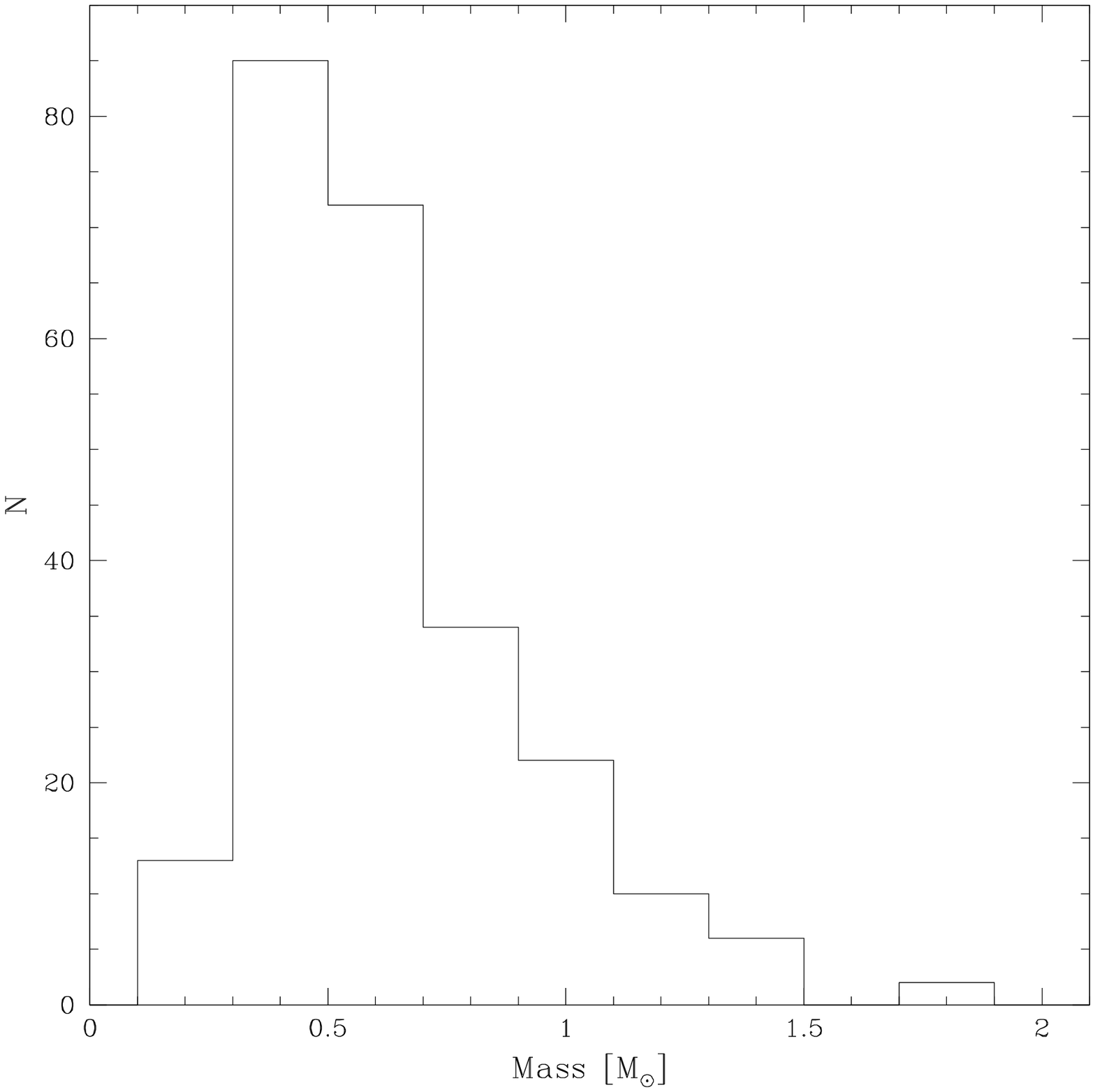}{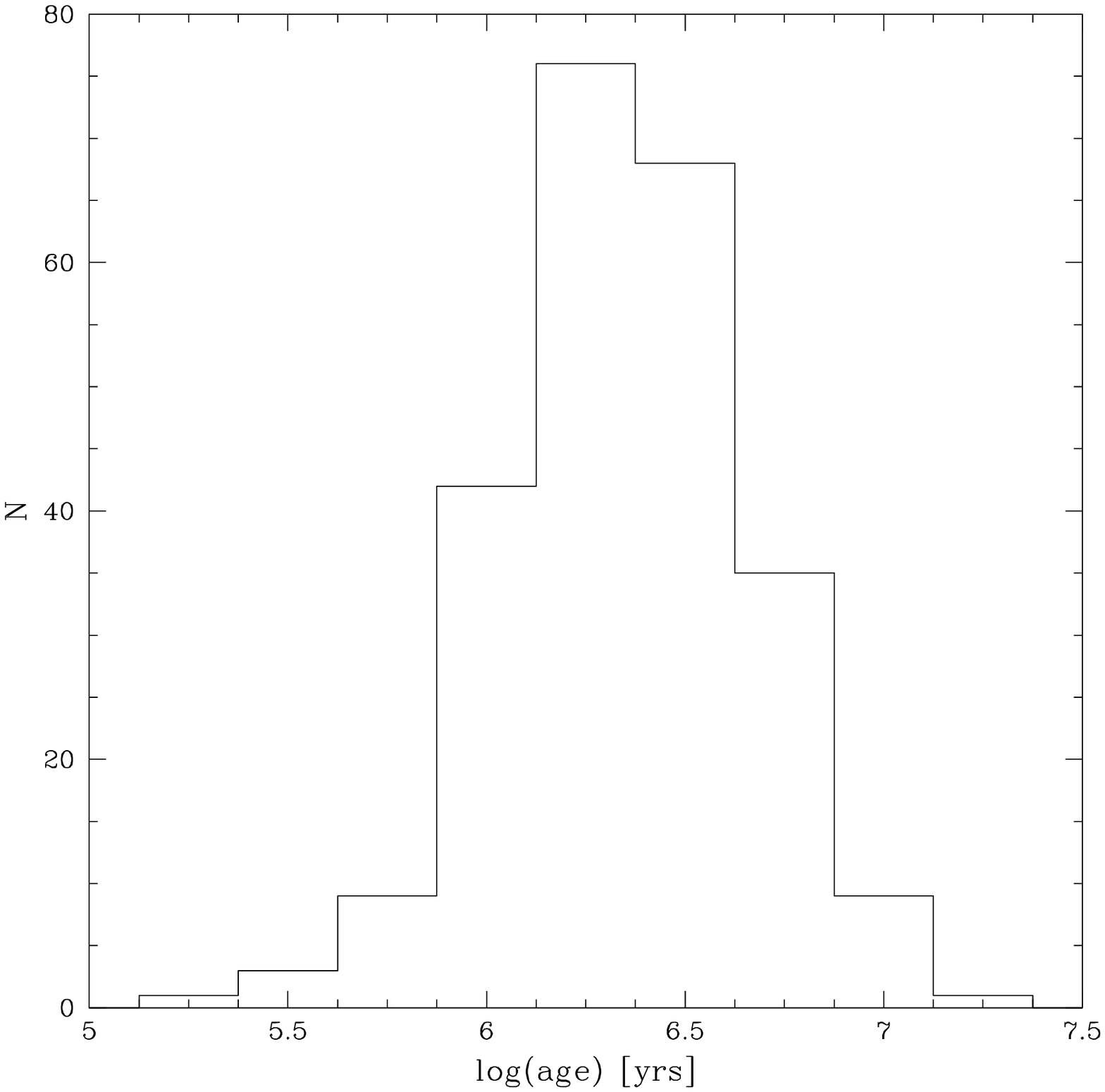}
\caption{
Distributions of masses (left) and ages (right) for our sample of NGC 6530 
cluster members with rotation periods (see Fig.~\ref{fig:perhist}). 
Masses and ages are inferred from the PMS evolutionary tracks of 
\citet{siess00}.
\label{fig:masshist}}
\end{figure}

\begin{figure}[ht]
\plottwo{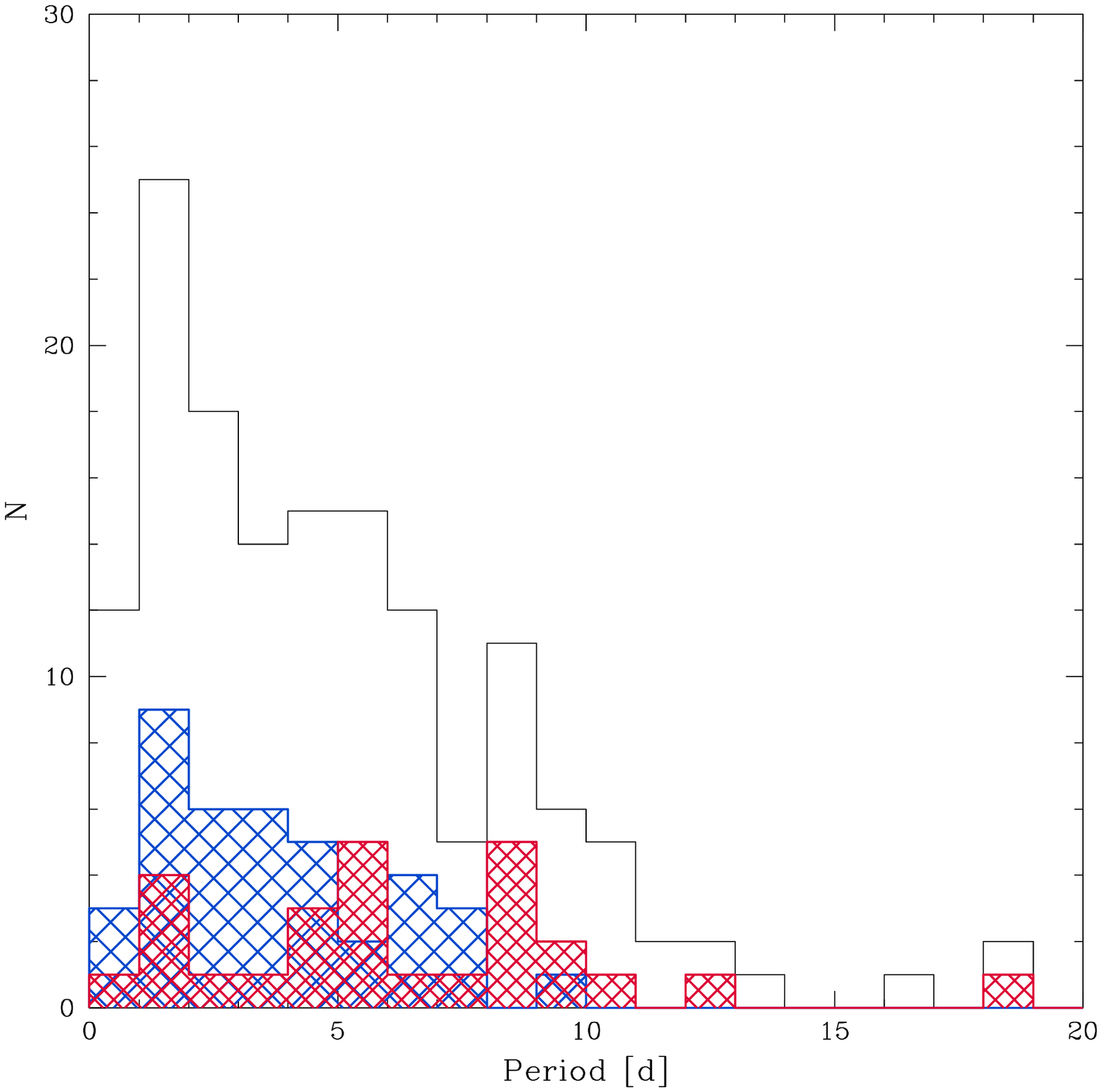}{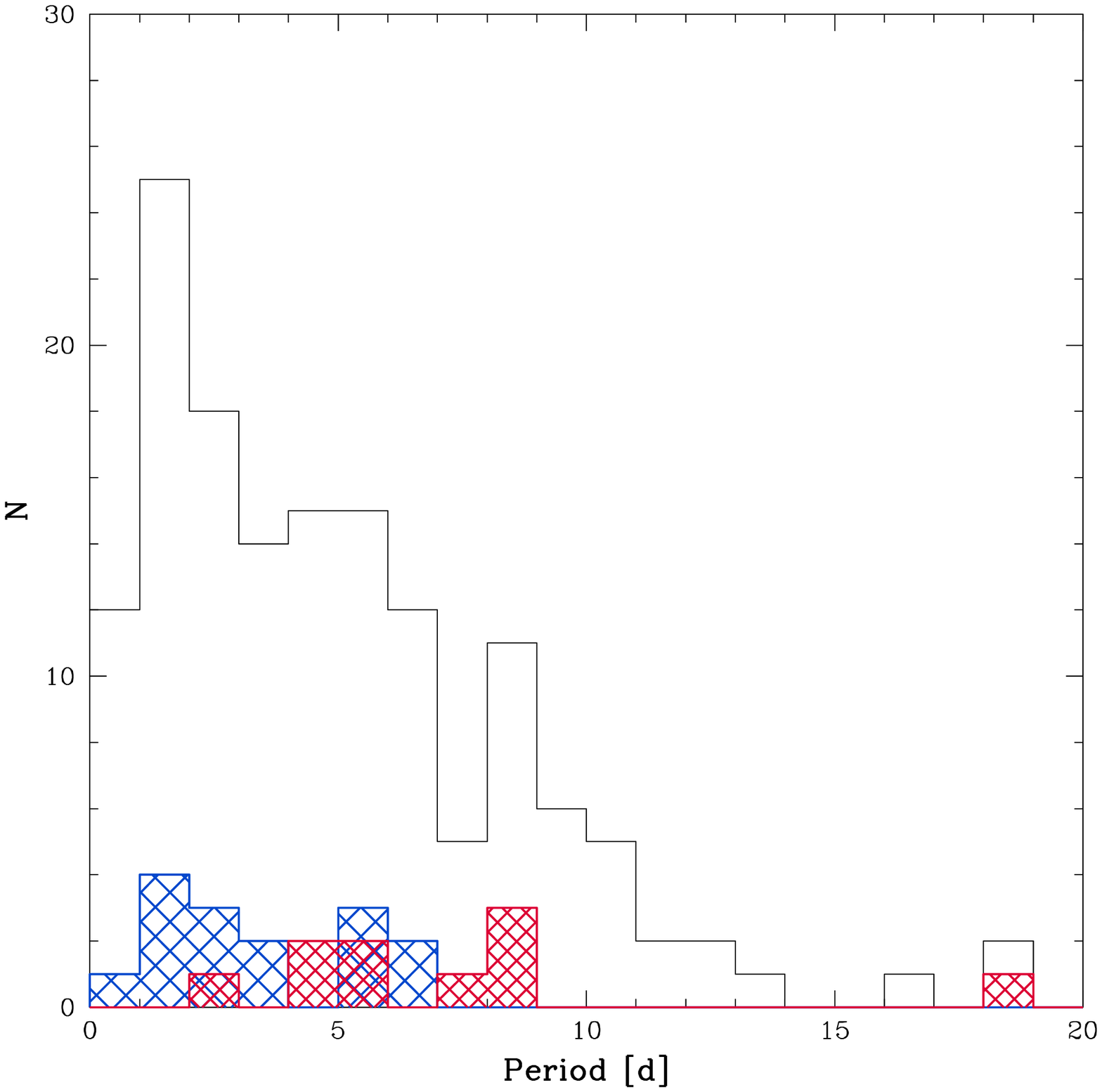}
\caption{
{\it Left:} Rotation period distributions of NGC~6530 cluster members 
exhibiting NIR
excess emission indicative of disks (shaded red histogram) and those that do not exhibit
NIR excess emission (hatched blue histogram), as determined via the $Q_{VIJK}$ index
\citep{damiani06}. 
A KS test and a Student's $t$ test both indicate that distributions of
NIR excess stars vs.\ non-excess stars are statistically different, with
the NIR excess stars rotating more slowly on average
(see Table~\ref{tab:perstats}).
{\it Right:} 
Rotation period distributions of NGC~6530 cluster members classified as
CTTSs (red) and WTTSs (blue) based on strength of H$\alpha$ emission
\citep{prisinzano07, arias07}.
A KS test and a Student's $t$ test both indicate that distributions of
CTTSs vs.\ WTTSs are statistically different, with the CTTSs
rotating more slowly on average (see Table~\ref{tab:perstats}).
In both panels the open (black) histogram shows the rotation period 
distribution for the entire sample (see Fig.~\ref{fig:perhist}), but
including only ``high mass" stars with $M_\star>0.5~M_\odot$
because the NIR excess and T~Tauri star samples are observationally
biased against low masses (see text).
\label{fig:perhist.qvijk}}
\end{figure}

\begin{figure}[ht]
\plottwo{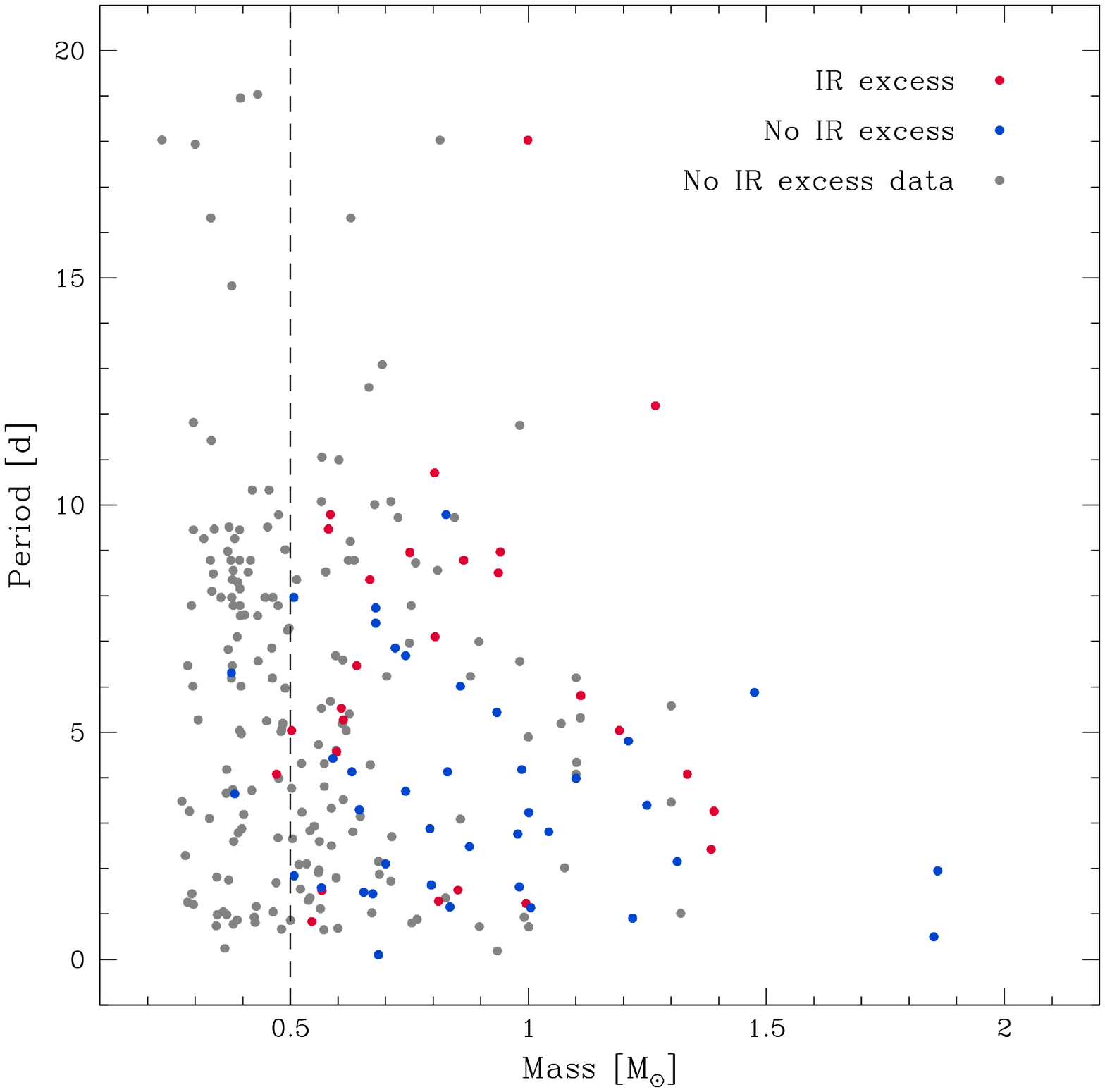}{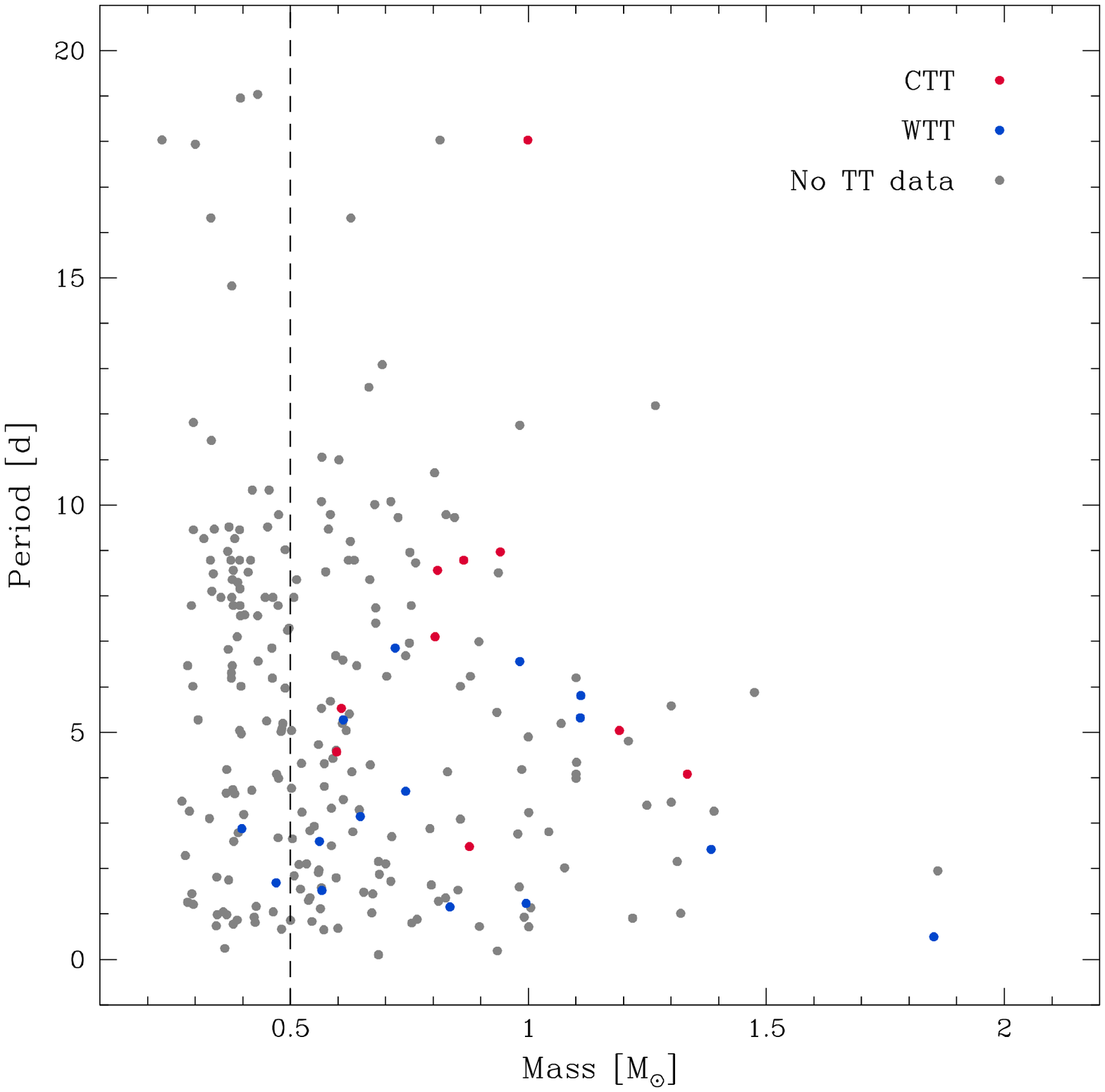}
\caption{Period vs.\ mass for our sample of NGC 6530 rotators. 
{\it (Left:)} Red points are those stars identified as having NIR 
excess as measured via the $Q_{VIJK}$ index \citep{damiani06}, 
blue points are those with no NIR excess, and gray points are the 
remainder of the sample (no NIR classification).
{\it (Right:)}
Red points are those stars identified as CTTSs, blue points are WTTSs
\citep{prisinzano07, arias07}, and gray points are the remainder of
the sample (no H$\alpha$ spectrum available).
In both panels, the dashed vertical line demarcates the two mass bins
we use for statistically comparing the period distribution as a function
of stellar mass.
A KS test and a Student's $t$ test both reveal a stastistically 
significant tendency for the lower mass stars to rotate more slowly
on average than the higher mass stars (see Table~\ref{tab:perstats}).
\label{fig:permass.qvijk}}
\end{figure}

\begin{figure}[ht]
\plotone{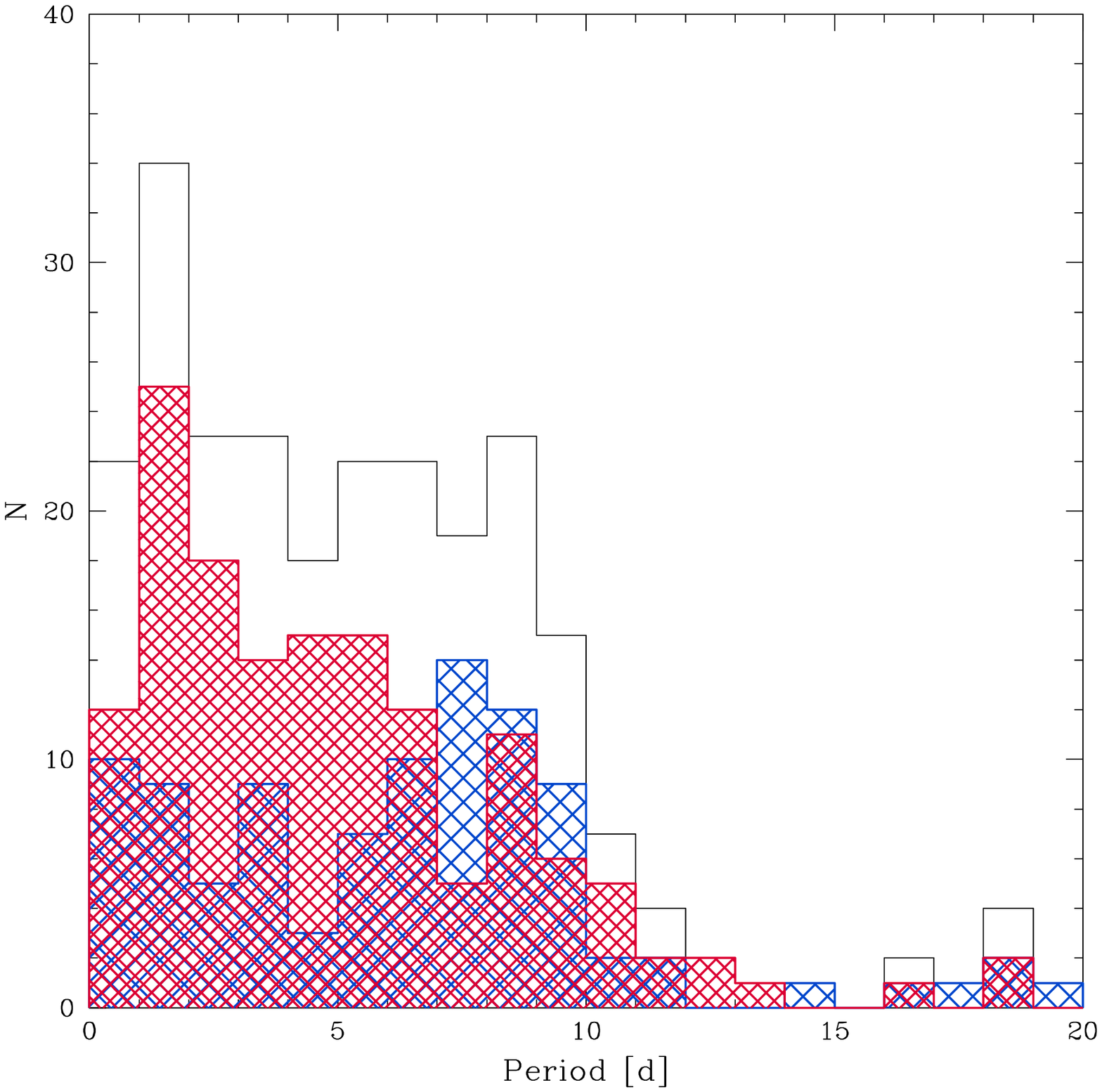}
\caption{Comparison of the rotation period distributions for
``high mass" and ``low mass" stars in our sample of NGC 6530 rotators. 
The red shaded histogram represents stars with
$M_\star > 0.5~M_{\odot}$ and the blue hatched histogram represents 
stars $M_\star \leq 0.5~M_{\odot}$.
A KS test and a Student's $t$ test both indicate strong statistical
differences between the rotation periods of the high-mass and low-mass
stars, with the high-mass stars rotating faster on average
(see Table~\ref{tab:perstats}).
\label{fig:perhist.mass}}
\end{figure}

\begin{figure}[ht]
\plottwo{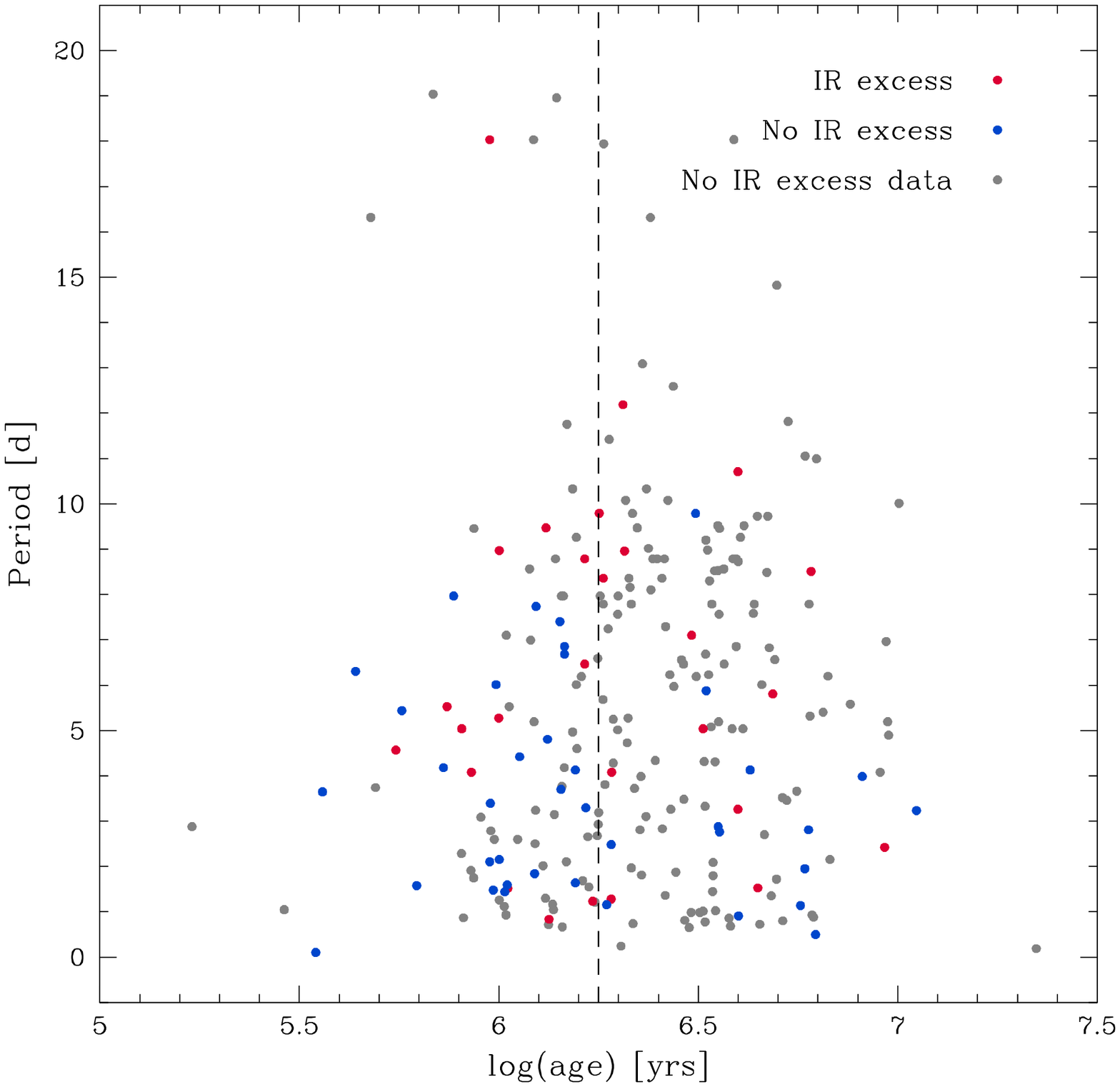}{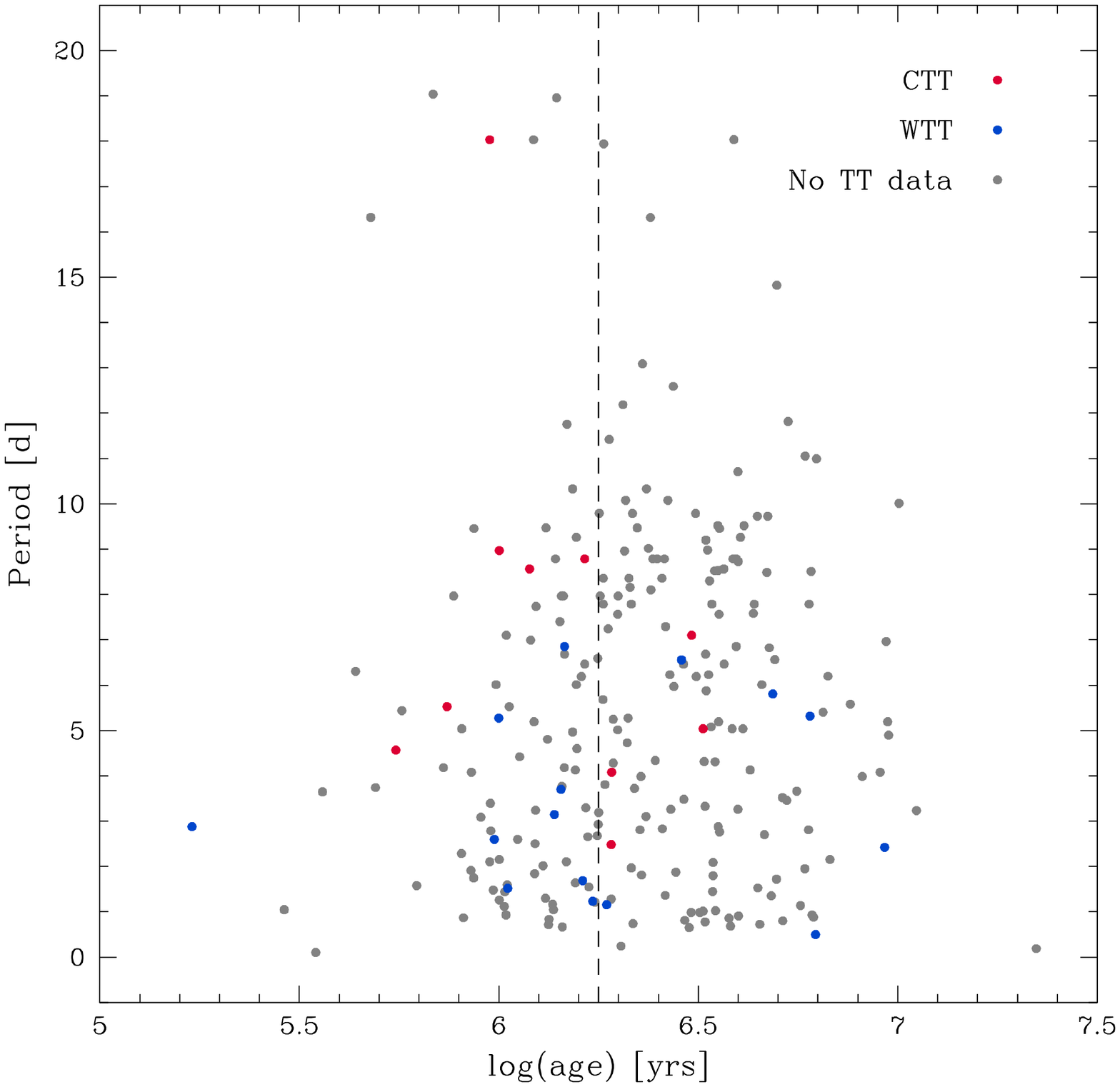}
\caption{Period vs.\ age for our sample of NGC 6530 rotators. 
{\it (Left:)} Red points are those stars identified as having NIR excess 
as measured via the $Q_{VIJK}$ index \citep{damiani06}, blue points are 
those with no NIR excess, and gray points are those with no 
NIR classification. 
{\it (Right:)} 
Red points are those stars identified as CTTSs, blue 
points are WTTSs \citep{prisinzano07, arias07}, and gray 
points are those with no H$\alpha$ spectrum. 
In both panels, the dashed vertical line demarcates the two age bins
used in our comparison of ``old" vs.\ ``young" stars.
A KS test suggests a mild tendency for the younger stars to rotate 
more rapidly on average (see Table~\ref{tab:perstats}).
\label{fig:perage.qvijk}}
\end{figure}

\begin{figure}[ht]
\plottwo{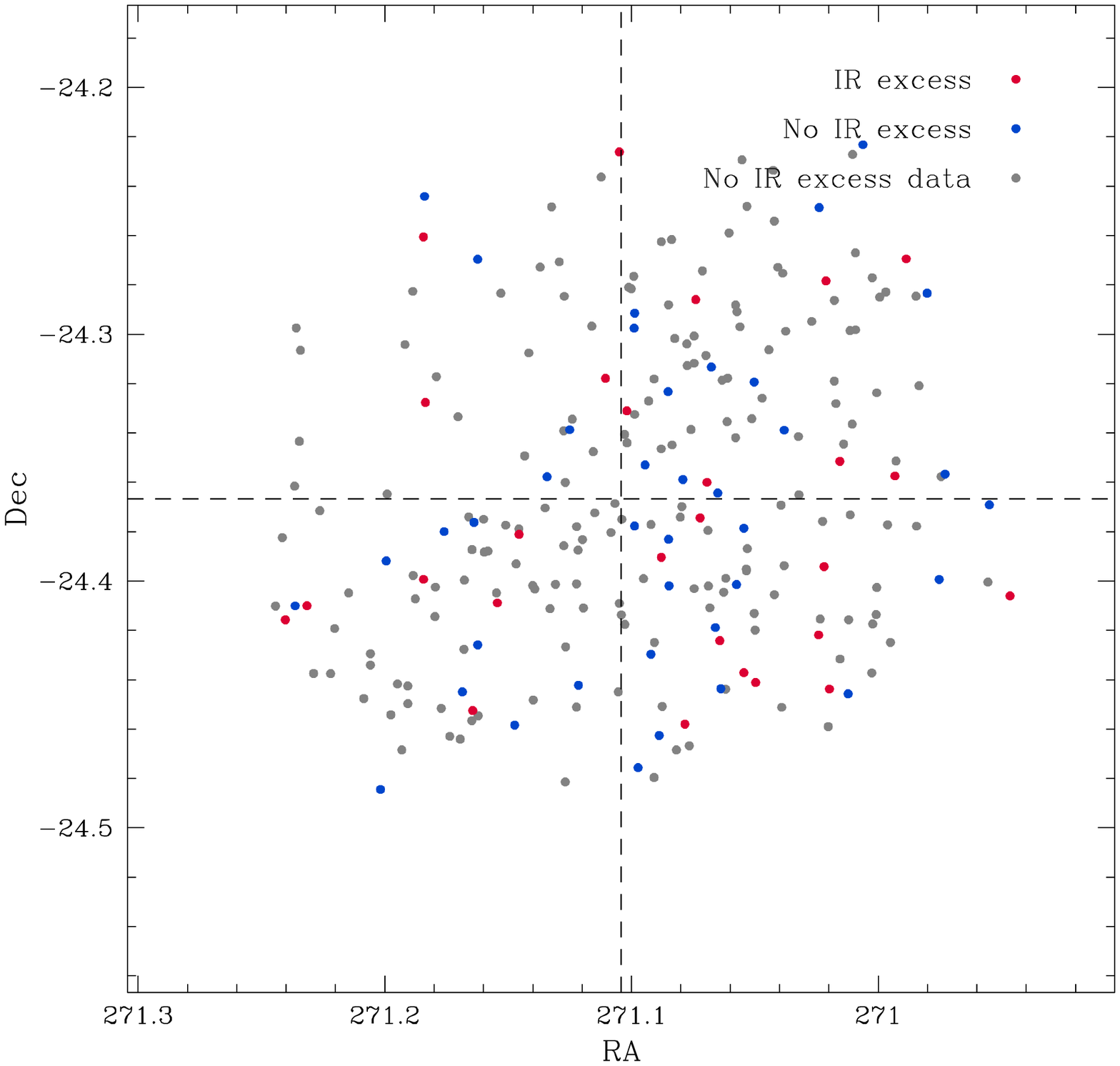}{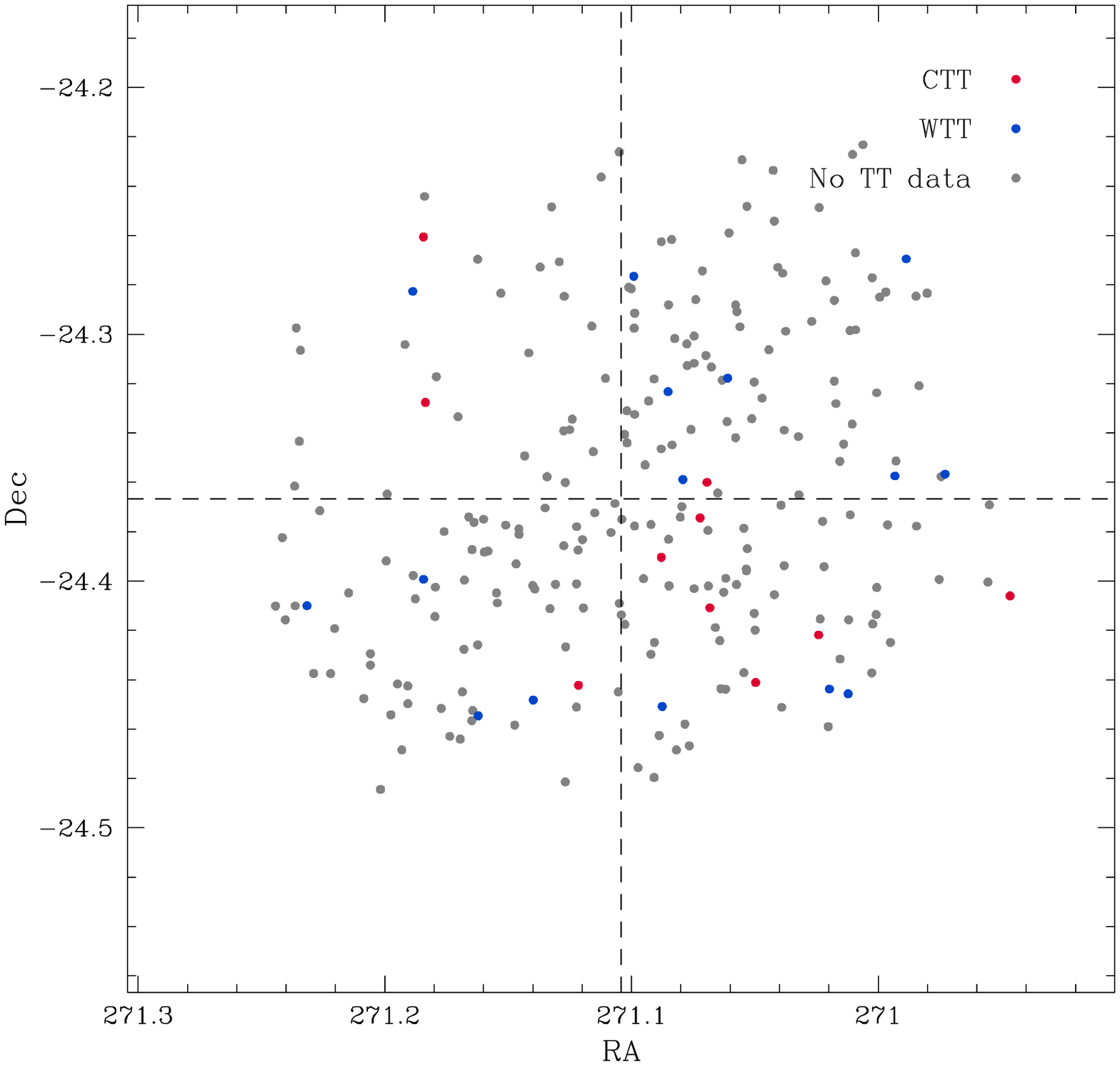}
\caption{The spatial distribution of our sample of NGC 6530 rotators. 
{\it (Left):}
Red points are those stars identified as having NIR excess in the $Q_{VIJK}$
index \citep{damiani06}, blue points are those with no NIR excess, and gray
points are those with no NIR classification. 
{\it (Right):}
Red points are those stars identified as CTTSs, blue points are WTTSs
\citep{prisinzano07, arias07}, and gray points are those with no 
H$\alpha$ spectrum.  In both panels, the dashed 
lines demarcate the quadrants that we use to investigate gradients
in age and rotation within the cluster \citep[see also][]{damiani04}.
A KS test suggests a potential southeast--northwest gradient in the rotation periods.
\label{fig:spatial.qvijk}}
\end{figure}

\begin{figure}[ht]
\plottwo{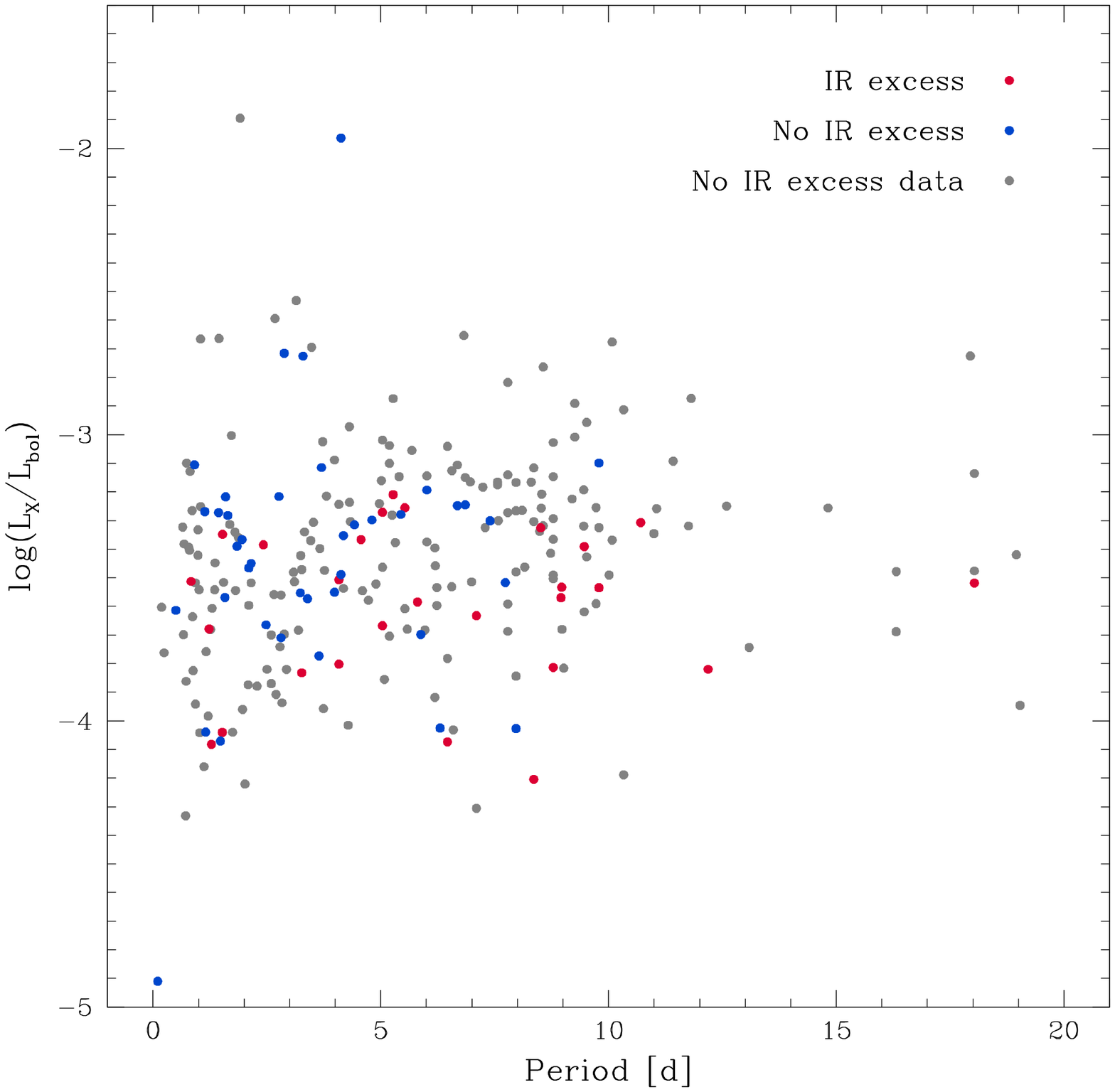}{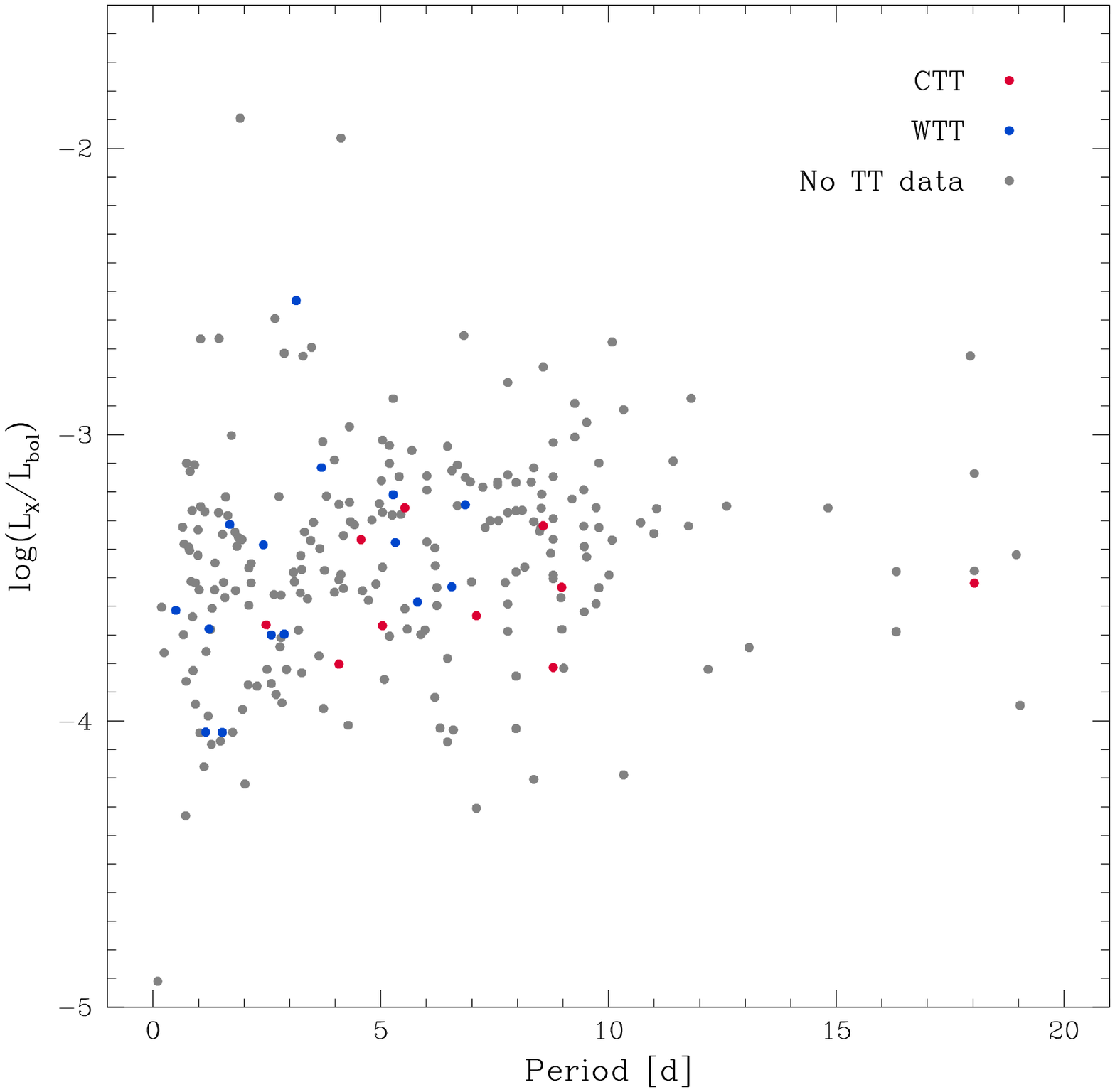}
\caption{$L_X / L_{\rm bol}$ vs.\ rotation period for our sample of rotators 
in NGC 6530.
{\it (Left):}
Red points are those stars identified as having NIR excess in the $Q_{VIJK}$
index \citep{damiani06}, blue points are those with no NIR excess, and gray
points are those with no NIR classification. 
{\it (Right):}
Red points are those stars identified as CTTSs, blue points are WTTSs
\citep{prisinzano07, arias07}, and gray points are those with no 
H$\alpha$ spectrum. 
A KS test and a Student's $t$ test both show a statistically significant
tendency for the most rapidly rotating stars to have lower 
$L_X / L_{\rm bol}$ (see Table~\ref{tab:xraystats}), 
suggestive of so-called ``super-saturation" \citep[e.g.][]{james00}.
\label{fig:lxlbol}}
\end{figure}

\begin{figure}[ht]
\plotone{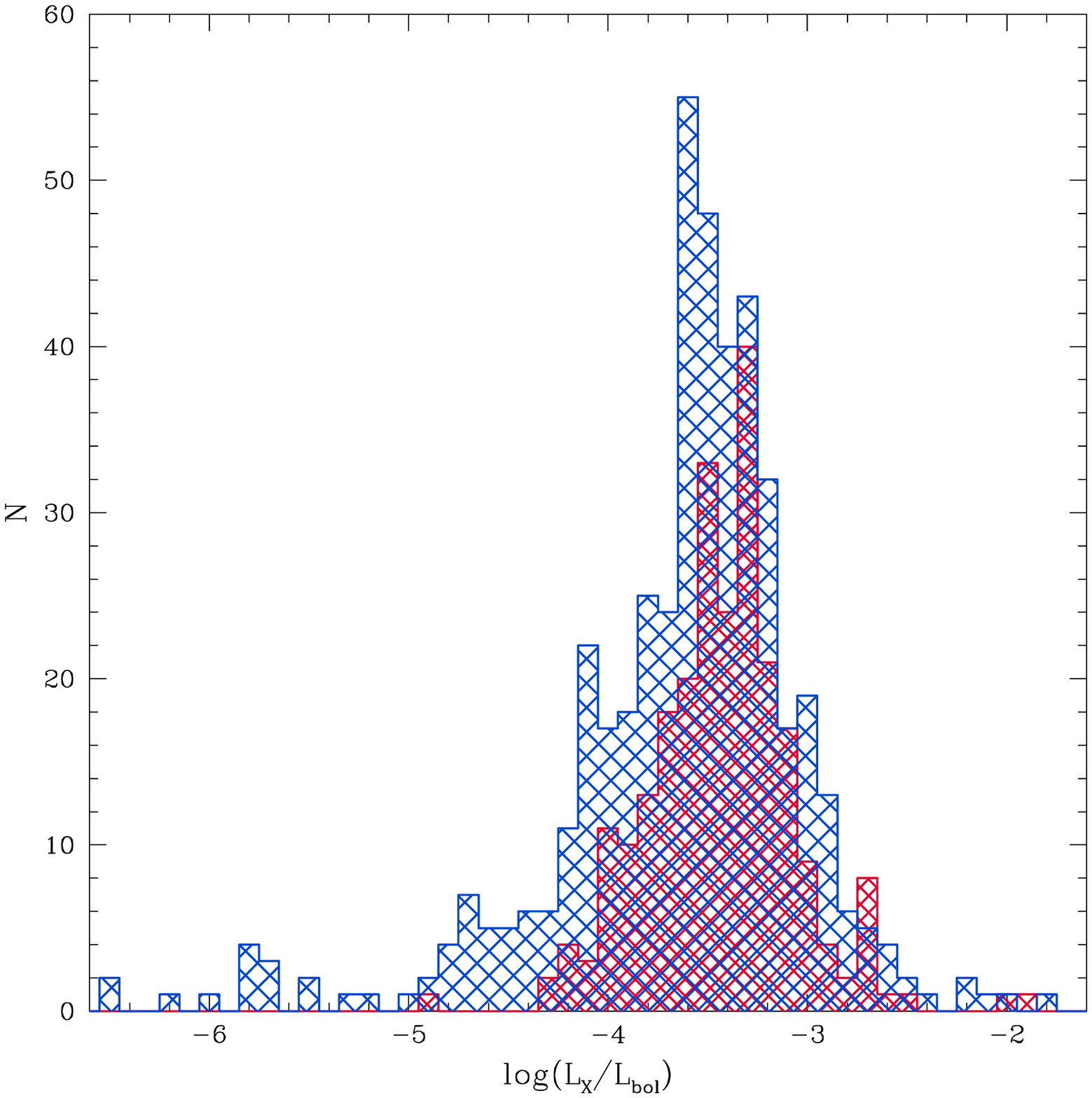}
\caption{The red shaded histogram represents $L_X / L_{\rm bol}$ for NGC
6530 stars with rotation periods, and the blue hatched histogram represents
$L_X / L_{\rm bol}$ for those without rotation periods.
Stars with rotation periods have significantly higher $L_X / L_{\rm bol}$
on average than those without detected periods (see Table~\ref{tab:xraystats}).
\label{fig:lxlbolhist}}
\end{figure}

\begin{figure}[ht]
\epsscale{0.7}
\plotone{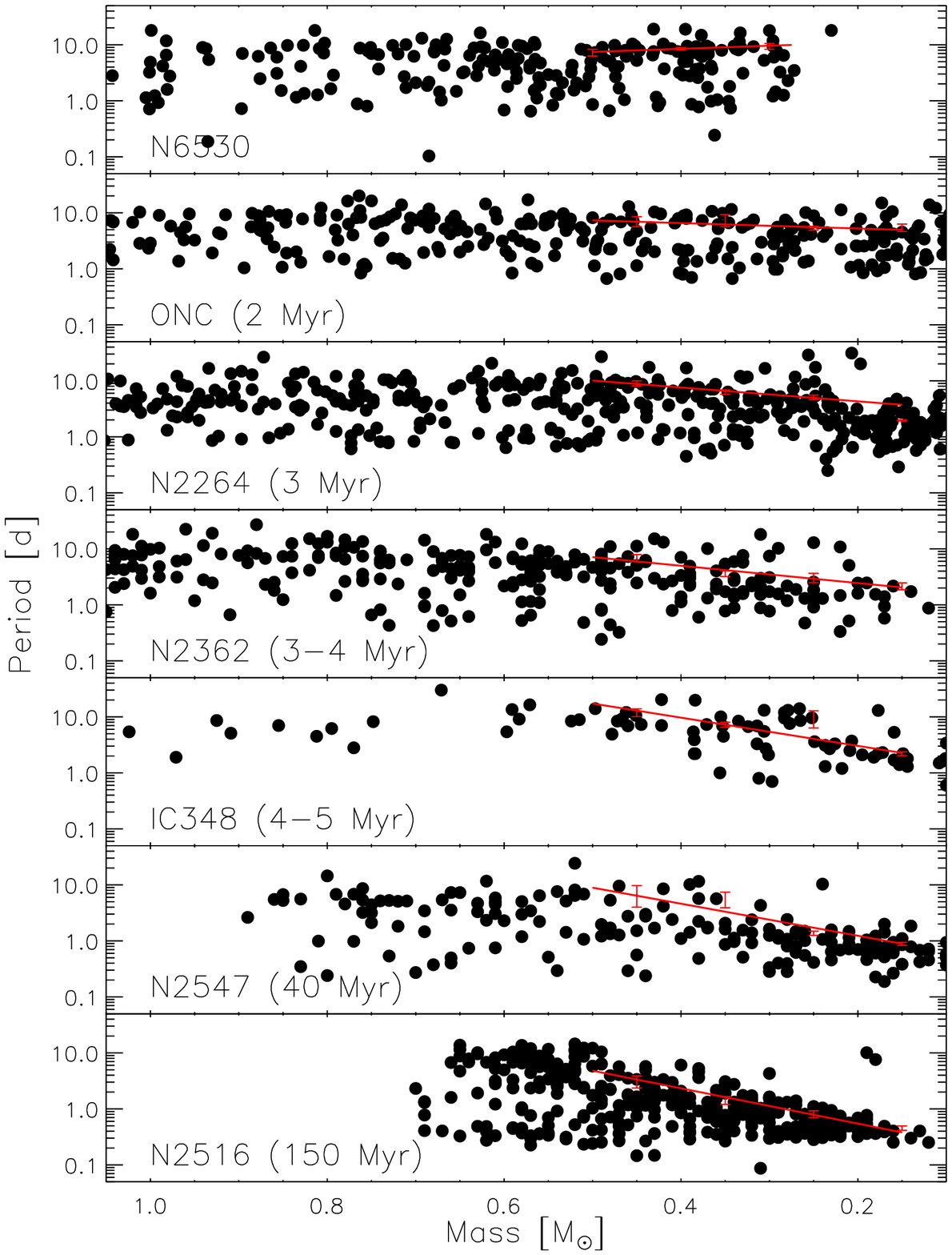}
\caption{Rotation period as a function of mass for (top to bottom):
NGC 6530, the ONC, NGC 2264, NGC 2362, IC 348, NGC 2547, and NGC 2516. 
Periods and masses for all clusters except NGC 6530 and IC 348 are from
the compilation of \citet{irwin08a}, with masses determined via interpolation
on the PMS evolutionary tracks of \citet{siess00}. Periods for IC 348 are from
\citet{cieza06} and we derived masses using data from that paper by
interpolating on the \citet{siess00} tracks. Periods and masses for NGC 6530
are from the present study.
Solid lines in each panel represent a least-squares fit to the 75\%-ile
upper envelope of periods in each 0.1-M$_\odot$ mass bin for masses in the
range 0.1--0.5 M$_\odot$, except for NGC 6530 which is limited to the range
0.2--0.5 M$_\odot$ (see \S\ref{sec:disc} for discussion of this figure).
\label{fig:permassclusters}}
\end{figure}

\begin{figure}[ht]
\plotone{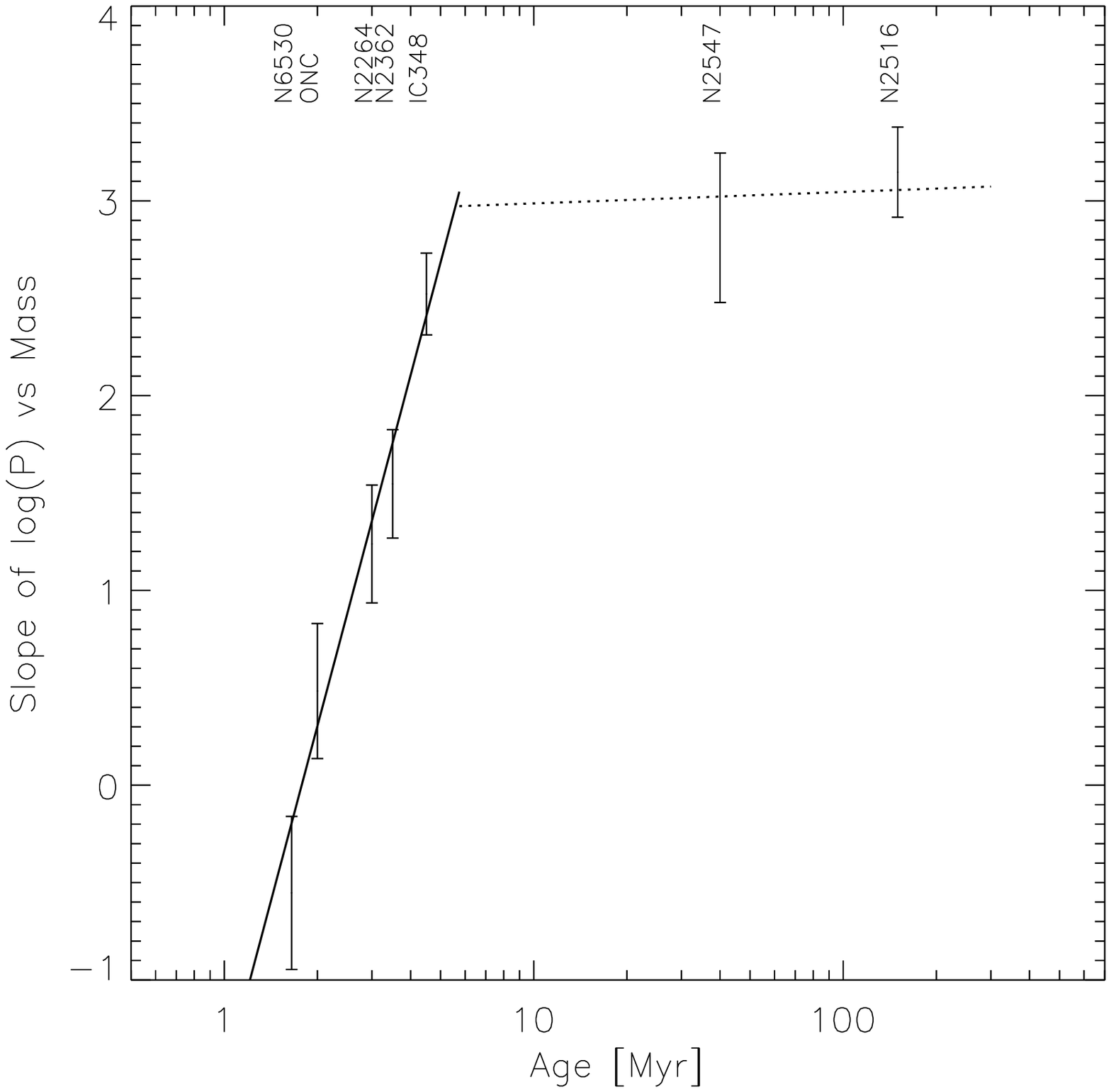}
\caption{
Slopes of the mass--period relationships from Fig.~\ref{fig:permassclusters}
versus cluster age. Cluster ages are from \citet{mayne07} and \citet{mayne08},
except for NGC 6530 whose age was adjusted here to be consistent with the
linear relationship (solid line) fitted to the ONC, NGC 2264, NGC 2362,
and IC 348. The maximum age inferred for NGC 6530 is 1.65~Myr, on an 
age scale where the ONC is 2~Myr. See \S\ref{sec:disc} for discussion of
this figure.
\label{fig:ageslopes}}
\end{figure}

\begin{figure}[ht]
\plotone{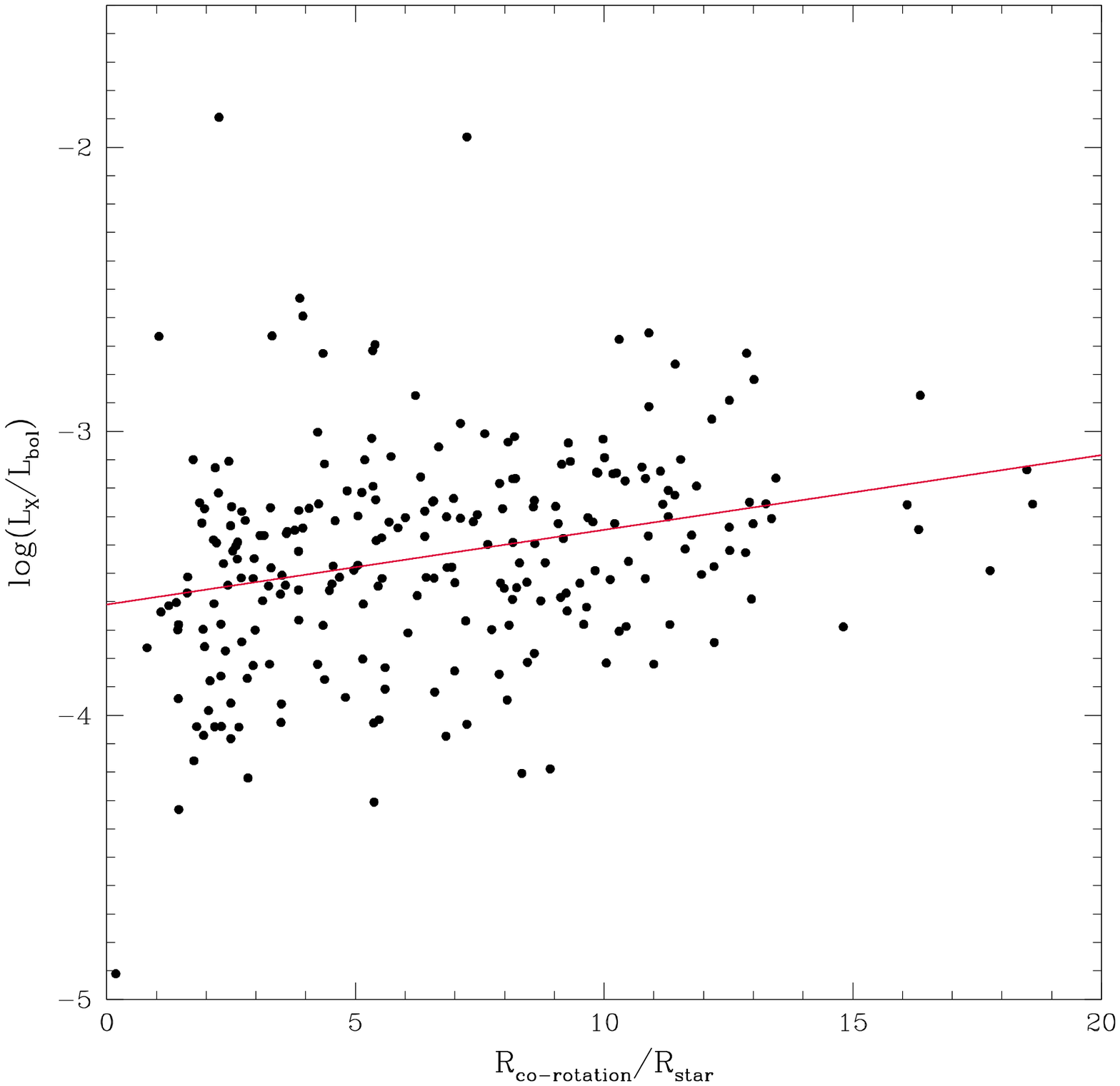}
\caption{$\log (L_X / L_{\rm bol})$ vs.\ $R_{\rm co} / R_\star$ for 
NGC 6530 stars with rotation periods. 
The line represents a linear best fit of the form 
$\log (L_X / L_{\rm bol}) = 0.026(\pm0.006) \times R_{\rm co}/R_\star
- 3.61(\pm0.05)$.
A Kendall's $\tau$ test shows the trend to be highly statistically 
significant. This suggests that centrifugal stripping of the stellar
coronae may be responsible for the super-saturation effect observed
in Fig.~\ref{fig:lxlbol} \citep[see also][]{wright11}.
\label{fig:corotation}}
\end{figure}

%==============================================================================
%                                TABLES
%==============================================================================

\clearpage
\begin{deluxetable}{ccc}
\tabletypesize{\scriptsize}
\tablecaption{All observational data \label{tab:obsdata}}
\tablewidth{0pt}
\tablehead{
\colhead{Season} &
\colhead{Telescope} &
\colhead{Nights of Data}
}
\startdata
June--July 2004   &   CTIO 0.9m   &   12   \\
July 2005         &   CTIO 1.0m   &   6    \\
June--July 2006   &   CTIO 1.0m   &   27   \\
June--July 2007   &   CTIO 1.0m   &   15   \\
April 2008        &   CTIO 1.0m   &   8    \\
May--June 2009    &   CTIO 1.0m   &   20
\enddata
\end{deluxetable}

\begin{deluxetable}{ccccc}
\tabletypesize{\scriptsize}
\tablecaption{2006 season data \label{tab:2006data}}
\tablewidth{0pt}
\tablehead{
\colhead{Field} &
\colhead{$\alpha_{2000}$} &
\colhead{$\delta_{2000}$} &
\colhead{Exp Time} &
\colhead{No. of Frames} \\
\colhead{} &
\colhead{[hh mm ss.s]} &
\colhead{[dd mm ss]} &
\colhead{[s]} &
\colhead{}
}
\startdata
1   &   18 05 01.6   &   -24 12 34   &   720   &   97   \\
2   &   18 05 01.6   &   -24 29 34   &   720   &   95   \\
3   &   18 03 46.9   &   -24 29 34   &   720   &   90   \\
4   &   18 03 46.9   &   -24 12 34   &   720   &   85
\enddata
\end{deluxetable}

\begin{deluxetable}{ccc}
\tabletypesize{\scriptsize}
\tablecaption{X-ray positional offsets \label{tab:xrayoffsets}}
\tablewidth{0pt}
\tablehead{
\colhead{Field} &
\colhead{$\Delta$RA} &
\colhead{$\Delta$DEC}
}
\startdata
1   &   0.20$\arcsec$    &   0.45$\arcsec$    \\
2   &   0.00$\arcsec$    &   0.15$\arcsec$    \\
3   &   -0.30$\arcsec$   &   -0.15$\arcsec$   \\
4   &   -1.05$\arcsec$   &   -0.20$\arcsec$
\enddata
\end{deluxetable}

\clearpage
\hoffset -0.6in
\LongTables
\begin{deluxetable}{ccccccccccccccc}
\label{tab:defparms}
\tabletypesize{\scriptsize}
\tablecaption{Parameters of the 244 cluster members with definite 
rotation periods \label{tab:defparms}}
\tablewidth{0pt}
\tablehead{
\colhead{ID} &
\colhead{ID} &
\colhead{RA} &
\colhead{DEC} &
\colhead{V} &
\colhead{I} &
\colhead{Period} &
\colhead{Mass} &
\colhead{Age} &
\colhead{L$_\mathrm{bol}$} &
\colhead{Radius} &
\colhead{log(L$_\mathrm{X}$)} &
\colhead{IR excess?} &
\colhead{H$\alpha$ class?$^{a}$} &
\colhead{SB2?} \\
\colhead{X-ray} &
\colhead{WFI} &
\colhead{[deg]} &
\colhead{[deg]} &
\colhead{[mag]} &
\colhead{[mag]} &
\colhead{[d]} &
\colhead{[M$_{\odot}$]} &
\colhead{[Myrs]} &
\colhead{[L$_{\odot}$]} &
\colhead{[R$_{\odot}$]} &
\colhead{[erg s$^{-1}$]} &
\colhead{} &
\colhead{} &
\colhead{}
}
\startdata
423   &   14403   &   271.09209   &   -24.42964   &   15.692   &   13.557   &    0.10352   &   0.69   &    0.35   &   5.12   &   4.49   &   29.382   &   N   &   -         &   N   \\
138   &   27470   &   271.02700   &   -24.29480   &   17.948   &   16.261   &    0.18743   &   0.94   &   22.23   &   0.39   &   0.96   &   29.566   &   -   &   -         &   N   \\
067   &   28743   &   270.99692   &   -24.28292   &   19.389   &   16.527   &    0.24286   &   0.36   &    2.02   &   0.32   &   1.44   &   29.331   &   -   &   -         &   -   \\
024   &   21464   &   270.97288   &   -24.35667   &   14.522   &   13.271   &    0.49948   &   1.85   &    6.22   &   7.02   &   2.61   &   30.815   &   N   &   W         &   N   \\
040   &   25374   &   270.98342   &   -24.32083   &   18.483   &   16.215   &    0.65205   &   0.57   &    3.00   &   0.43   &   1.37   &   29.895   &   -   &   -         &   -   \\
537   &   34984   &   271.11225   &   -24.23632   &   18.248   &   15.843   &    0.66582   &   0.48   &    1.44   &   0.62   &   1.76   &   29.675   &   -   &   -         &   -   \\
357   &   26832   &   271.08246   &   -24.30170   &   18.544   &   16.315   &    0.68347   &   0.60   &    3.81   &   0.39   &   1.28   &   29.794   &   -   &   -         &   -   \\
130   &   15487   &   271.02346   &   -24.41538   &   16.420   &   14.548   &    0.71534   &   1.00   &    1.33   &   1.85   &   2.32   &   29.518   &   -   &   -         &   -   \\
\enddata
\tablenotetext{a}{C denotes Classical T-Tauri stars while W identifies Weak-lined T-Tauri stars.}
\tablenotetext{b}{These H$\alpha$ classifications are taken from \citet{arias07}; all others come from \citet{prisinzano07}.}
\end{deluxetable}

\hoffset -0.65in
\clearpage
\LongTables
\begin{deluxetable}{ccccccccccccccc}
\tabletypesize{\scriptsize}
\tablecaption{Parameters of the 46 cluster members with possible 
rotation periods \label{tab:possparms}}
\tablewidth{0pt}
\tablehead{
\colhead{ID} &
\colhead{ID} &
\colhead{RA} &
\colhead{DEC} &
\colhead{V} &
\colhead{I} &
\colhead{Period} &
\colhead{Mass} &
\colhead{Age} &
\colhead{L$_\mathrm{bol}$} &
\colhead{Radius} &
\colhead{log(L$_\mathrm{X}$)} &
\colhead{IR excess?} &
\colhead{H$\alpha$ class?$^{a}$} &
\colhead{SB2?} \\
\colhead{X-ray} &
\colhead{WFI} &
\colhead{[deg]} &
\colhead{[deg]} &
\colhead{[mag]} &
\colhead{[mag]} &
\colhead{[d]} &
\colhead{[M$_{\odot}$]} &
\colhead{[Myrs]} &
\colhead{[L$_{\odot}$]} &
\colhead{[R$_{\odot}$]} &
\colhead{[erg s$^{-1}$]} &
\colhead{} &
\colhead{} &
\colhead{}
}
\startdata
217   &   25571   &   271.05200   &   -24.31766   &   17.959   &   15.489   &    0.3999630   &   0.45   &   1.01   &   0.84   &   2.12   &   29.890   &   -   &    -         &   -   \\
425   &   21654   &   271.09217   &   -24.35480   &   17.818   &   15.830   &    0.6353953   &   0.88   &   5.26   &   0.56   &   1.34   &   29.822   &   -   &    -         &   -   \\
581   &   13973   &   271.12150   &   -24.43629   &   19.875   &   16.704   &    0.6393026   &   0.32   &   1.95   &   0.30   &   1.45   &   29.406   &   -   &    -         &   -   \\
126   &   15028   &   271.02154   &   -24.42041   &   19.131   &   16.418   &    0.6978532   &   0.38   &   1.57   &   0.44   &   1.64   &   29.490   &   -   &    -         &   -   \\
732   &   28075   &   271.16463   &   -24.28881   &   17.645   &   15.552   &    0.7891157   &   0.73   &   2.56   &   0.73   &   1.64   &   29.601   &   Y   &    W         &   N   \\
307   &   30975   &   271.07117   &   -24.26600   &   19.945   &   17.015   &    0.8065604   &   0.35   &   3.05   &   0.22   &   1.18   &   29.465   &   -   &    -         &   -   \\
371   &   19029   &   271.08471   &   -24.38020   &   18.592   &   15.934   &    0.8771817   &   0.39   &   0.98   &   0.80   &   2.21   &   30.245   &   -   &    -         &   -   \\
758   &   21152   &   271.17558   &   -24.35950   &   18.223   &   15.931   &    0.9286347   &   0.55   &   1.99   &   0.56   &   1.59   &   29.923   &   -   &    -         &   -   \\
\enddata
\tablenotetext{a}{C denotes Classical T-Tauri stars while W identifies Weak-lined T-Tauri stars.}
\tablenotetext{b}{These H$\alpha$ classifications are taken from \citet{arias07}; all others come from \citet{prisinzano07}.}
\end{deluxetable}

\clearpage
\begin{deluxetable}{ccccc}
\tabletypesize{\scriptsize}
\tablecaption{Statistical comparisons of rotation periods
\label{tab:perstats}}
\tablewidth{0pt}
\tablehead{
\colhead{} &
\colhead{N} &
\colhead{$<$$P$$>$} &
\colhead{KS prob.} &
\colhead{$t$ prob.}
}
\startdata
NIR excess                 &   27    &    6.253   &              &              \\
No NIR excess              &   39    &    3.679   &   0.0125     &   0.00160    \\
\cline{1-5}
CTT                        &   10    &    7.316   &              &              \\
WTT                        &   15    &    3.376   &   0.0248     &   0.00590    \\
\cline{1-5} 
Low-mass                   &   98    &    6.357   &             &              \\
High-mass                  &   146   &    4.929   &   0.00104   &   0.00511    \\
\cline{1-5}
Young                      &   92    &    4.890   &              &              \\
Old                        &   152   &    5.873   &   0.0136     &   0.0575     \\
\cline{1-5}
Northwest                  &   79    &    5.199   &              &              \\
Southeast                  &   70    &    6.238   &   0.0115     &   0.117      \\
\cline{1-5}
Northeast                  &   30    &    5.982   &              &              \\
Southwest                  &   65    &    4.858   &   0.141      &   0.175
\enddata
\end{deluxetable}

\begin{deluxetable}{ccccc}
\tabletypesize{\scriptsize}
\tablecaption{Statistical comparisons of X-ray luminosities
\label{tab:xraystats}}
\tablewidth{0pt}
\tablehead{
\colhead{} &
\colhead{N} &
\colhead{$<$$\mathrm{log}(L_\mathrm{X}/L_\mathrm{bol})$$>$} &
\colhead{KS prob.} &
\colhead{$t$ prob.}
}
\startdata
Fast           &   79    &   -3.556   &              &                         \\
Slow           &   165   &   -3.372   &   0.000267   &   0.000338              \\
$P_{rot}$      &   244   &   -3.432   &              &                         \\
No $P_{rot}$   &   458   &   -3.632   &   0.000620   &   6.28$\times10^{-6}$
\enddata
\end{deluxetable}

\begin{deluxetable}{lcc}
\tabletypesize{\scriptsize}
\tablecaption{Slopes of the period--mass relationship for young clusters 
\label{tab:clusterages}}
\tablewidth{0pt}
\tablehead{\colhead{Cluster} & \colhead{Age [Myr]} & \colhead{slope}}
\startdata
N6530   &  1.65 &  $-0.55 \pm  0.39$ \\
ONC     &  2    &  $ 0.48 \pm  0.35$ \\
N2264   &  3    &  $ 1.24 \pm  0.30$ \\
N2362   &  3.5  &  $ 1.55 \pm  0.28$ \\
IC348   &  4.5  &  $ 2.52 \pm  0.21$ \\
N2547   &  40   &  $ 2.86 \pm  0.38$ \\
N2516   &  150  &  $ 3.15 \pm  0.23$
\enddata
\tablecomments{Slopes are from linear fits to the upper envelope of 
rotation periods versus mass, of the form $\log P = a \times M + b$,
where $P$ is in days and $M$ is in M$_\odot$. Ages are from
\citet{mayne07} and \citet{mayne08}, except for NGC 6530 whose age
here is determined from the linear relationship fitted to the other
clusters.  See \S\ref{sec:disc}.}
\end{deluxetable}


\clearpage
\begin{thebibliography}
\expandafter\ifx\csname natexlab\endcsname\relax\def\natexlab#1{#1}\fi

\bibitem[{{Aarnio} {et~al.}(2011){Aarnio}, {Stassun}, {Hughes}, \&
  {McGregor}}]{aarnio11}
{Aarnio}, A.~N., {Stassun}, K.~G., {Hughes}, W.~J., \& {McGregor}, S.~L. 2011,
  \solphys, 268, 195

\bibitem[{{Aarnio} {et~al.}(2009){Aarnio}, {Stassun}, \& {Matt}}]{aarnio09}
{Aarnio}, A.~N., {Stassun}, K.~G., \& {Matt}, S.~P. 2009, in American Institute
  of Physics Conference Series, Vol. 1094, American Institute of Physics
  Conference Series, ed. {E.~Stempels}, 337--340

\bibitem[{{Aarnio} {et~al.}(2010){Aarnio}, {Stassun}, \& {Matt}}]{aarnio10}
{Aarnio}, A.~N., {Stassun}, K.~G., \& {Matt}, S.~P. 2010, \apj, 717, 93

\bibitem[{{Allain} {et~al.}(1996){Allain}, {Bouvier}, {Prosser}, {Marschall},
  \& {Laaksonen}}]{allain96}
{Allain}, S., {Bouvier}, J., {Prosser}, C., {Marschall}, L.~A., \& {Laaksonen},
  B.~D. 1996, \aap, 305, 498

\bibitem[{{Arias} {et~al.}(2007){Arias}, {Barb{\'a}}, \& {Morrell}}]{arias07}
{Arias}, J.~I., {Barb{\'a}}, R.~H., \& {Morrell}, N.~I. 2007, \mnras, 374, 1253

\bibitem[{{Aspin}(2011)}]{aspin11}
{Aspin}, C. 2011, \aj, 142, 135

\bibitem[{{Aspin} {et~al.}(2006){Aspin}, {Barbieri}, {Boschi}, {Di Mille},
  {Rampazzi}, {Reipurth}, \& {Tsvetkov}}]{aspin06}
{Aspin}, C., {Barbieri}, C., {Boschi}, F., {Di Mille}, F., {Rampazzi}, F.,
  {Reipurth}, B., \& {Tsvetkov}, M. 2006, \aj, 132, 1298

\bibitem[{{Attridge} \& {Herbst}(1992)}]{attridge92}
{Attridge}, J.~M. \& {Herbst}, W. 1992, \apjl, 398, L61

\bibitem[{{Barnes}(2003)}]{barnes03}
{Barnes}, S.~A. 2003, \apj, 586, 464

\bibitem[{{Barnes}(2007)}]{barnes07}
---. 2007, \apj, 669, 1167

\bibitem[{{Barnes}(2010)}]{barnes10b}
---. 2010, \apj, 722, 222

\bibitem[{{Barnes} \& {Kim}(2010)}]{barnes10a}
{Barnes}, S.~A. \& {Kim}, Y.-C. 2010, \apj, 721, 675

\bibitem[{{Barnes} {et~al.}(1999){Barnes}, {Sofia}, {Prosser}, \&
  {Stauffer}}]{barnes99}
{Barnes}, S.~A., {Sofia}, S., {Prosser}, C.~F., \& {Stauffer}, J.~R. 1999,
  \apj, 516, 263

\bibitem[{{Basri} {et~al.}(1997){Basri}, {Johns-Krull}, \& {Mathieu}}]{basri97}
{Basri}, G., {Johns-Krull}, C.~M., \& {Mathieu}, R.~D. 1997, \aj, 114, 781

\bibitem[{{Bastien} {et~al.}(2011){Bastien}, {Stassun}, \&
  {Weintraub}}]{bastien11}
{Bastien}, F.~A., {Stassun}, K.~G., \& {Weintraub}, D.~A. 2011, \aj, 142, 141

\bibitem[{{Bouvier} {et~al.}(2007){Bouvier}, {Alencar}, {Harries},
  {Johns-Krull}, \& {Romanova}}]{bouvier07}
{Bouvier}, J., {Alencar}, S.~H.~P., {Harries}, T.~J., {Johns-Krull}, C.~M., \&
  {Romanova}, M.~M. 2007, Protostars and Planets V, 479

\bibitem[{{Bouvier} {et~al.}(1986){Bouvier}, {Bertout}, {Benz}, \&
  {Mayor}}]{bouvier86}
{Bouvier}, J., {Bertout}, C., {Benz}, W., \& {Mayor}, M. 1986, \aap, 165, 110

\bibitem[{{Bouvier} {et~al.}(1993){Bouvier}, {Cabrit}, {Fernandez}, {Martin},
  \& {Matthews}}]{bouvier93}
{Bouvier}, J., {Cabrit}, S., {Fernandez}, M., {Martin}, E.~L., \& {Matthews},
  J.~M. 1993, \aap, 272, 176

\bibitem[{{Bouvier} {et~al.}(1999){Bouvier}, {Chelli}, {Allain}, {Carrasco},
  {Costero}, {Cruz-Gonzalez}, {Dougados}, {Fern{\'a}ndez}, {Mart{\'{\i}}n},
  {M{\'e}nard}, {Mennessier}, {Mujica}, {Recillas}, {Salas}, {Schmidt}, \&
  {Wichmann}}]{bouvier99}
{Bouvier}, J., {Chelli}, A., {Allain}, S., {Carrasco}, L., {Costero}, R.,
  {Cruz-Gonzalez}, I., {Dougados}, C., {Fern{\'a}ndez}, M., {Mart{\'{\i}}n},
  E.~L., {M{\'e}nard}, F., {Mennessier}, C., {Mujica}, R., {Recillas}, E.,
  {Salas}, L., {Schmidt}, G., \& {Wichmann}, R. 1999, \aap, 349, 619

\bibitem[{{Bouvier} {et~al.}(2004){Bouvier}, {Dougados}, \&
  {Alencar}}]{bouvier04}
{Bouvier}, J., {Dougados}, C., \& {Alencar}, S.~H.~P. 2004, \apss, 292, 659

\bibitem[{{Bouvier} {et~al.}(2003){Bouvier}, {Grankin}, {Alencar}, {Dougados},
  {Fern{\'a}ndez}, {Basri}, {Batalha}, {Guenther}, {Ibrahimov}, {Magakian},
  {Melnikov}, {Petrov}, {Rud}, \& {Zapatero Osorio}}]{bouvier03}
{Bouvier}, J., {Grankin}, K.~N., {Alencar}, S.~H.~P., {Dougados}, C.,
  {Fern{\'a}ndez}, M., {Basri}, G., {Batalha}, C., {Guenther}, E., {Ibrahimov},
  M.~A., {Magakian}, T.~Y., {Melnikov}, S.~Y., {Petrov}, P.~P., {Rud}, M.~V.,
  \& {Zapatero Osorio}, M.~R. 2003, \aap, 409, 169

\bibitem[{{Bouvier} {et~al.}(1997){Bouvier}, {Wichmann}, {Grankin}, {Allain},
  {Covino}, {Fernandez}, {Martin}, {Terranegra}, {Catalano}, \&
  {Marilli}}]{bouvier97}
{Bouvier}, J., {Wichmann}, R., {Grankin}, K., {Allain}, S., {Covino}, E.,
  {Fernandez}, M., {Martin}, E.~L., {Terranegra}, L., {Catalano}, S., \&
  {Marilli}, E. 1997, \aap, 318, 495

\bibitem[{{Brice{\~n}o} {et~al.}(2004){Brice{\~n}o}, {Vivas}, {Hern{\'a}ndez},
  {Calvet}, {Hartmann}, {Megeath}, {Berlind}, {Calkins}, \&
  {Hoyer}}]{briceno04}
{Brice{\~n}o}, C., {Vivas}, A.~K., {Hern{\'a}ndez}, J., {Calvet}, N.,
  {Hartmann}, L., {Megeath}, T., {Berlind}, P., {Calkins}, M., \& {Hoyer}, S.
  2004, \apjl, 606, L123

\bibitem[{{Casey} {et~al.}(1998){Casey}, {Mathieu}, {Vaz}, {Andersen}, \&
  {Suntzeff}}]{casey98}
{Casey}, B.~W., {Mathieu}, R.~D., {Vaz}, L.~P.~R., {Andersen}, J., \&
  {Suntzeff}, N.~B. 1998, \aj, 115, 1617

\bibitem[{{Choi} \& {Herbst}(1996)}]{choi96}
{Choi}, P.~I. \& {Herbst}, W. 1996, \aj, 111, 283

\bibitem[{{Cieza} \& {Baliber}(2006)}]{cieza06}
{Cieza}, L. \& {Baliber}, N. 2006, \apj, 649, 862

\bibitem[{{Covey} {et~al.}(2011){Covey}, {Hillenbrand}, {Miller}, {Poznanski},
  {Cenko}, {Silverman}, {Bloom}, {Kasliwal}, {Fischer}, {Rayner}, {Rebull},
  {Butler}, {Filippenko}, {Law}, {Ofek}, {Ag{\"u}eros}, {Dekany}, {Rahmer},
  {Hale}, {Smith}, {Quimby}, {Nugent}, {Jacobsen}, {Zolkower}, {Velur},
  {Walters}, {Henning}, {Bui}, {McKenna}, {Kulkarni}, \& {Klein}}]{covey11}
{Covey}, K.~R., {Hillenbrand}, L.~A., {Miller}, A.~A., {Poznanski}, D.,
  {Cenko}, S.~B., {Silverman}, J.~M., {Bloom}, J.~S., {Kasliwal}, M.~M.,
  {Fischer}, W., {Rayner}, J., {Rebull}, L.~M., {Butler}, N.~R., {Filippenko},
  A.~V., {Law}, N.~M., {Ofek}, E.~O., {Ag{\"u}eros}, M., {Dekany}, R.~G.,
  {Rahmer}, G., {Hale}, D., {Smith}, R., {Quimby}, R.~M., {Nugent}, P.,
  {Jacobsen}, J., {Zolkower}, J., {Velur}, V., {Walters}, R., {Henning}, J.,
  {Bui}, K., {McKenna}, D., {Kulkarni}, S.~R., \& {Klein}, C. 2011, \aj, 141,
  40

\bibitem[{{Covino} {et~al.}(2000){Covino}, {Catalano}, {Frasca}, {Marilli},
  {Fern{\'a}ndez}, {Alcal{\'a}}, {Melo}, {Paladino}, {Sterzik}, \&
  {Stelzer}}]{covino00}
{Covino}, E., {Catalano}, S., {Frasca}, A., {Marilli}, E., {Fern{\'a}ndez}, M.,
  {Alcal{\'a}}, J.~M., {Melo}, C., {Paladino}, R., {Sterzik}, M.~F., \&
  {Stelzer}, B. 2000, \aap, 361, L49

\bibitem[{{Da Rio} {et~al.}(2010){Da Rio}, {Robberto}, {Soderblom}, {Panagia},
  {Hillenbrand}, {Palla}, \& {Stassun}}]{dario10}
{Da Rio}, N., {Robberto}, M., {Soderblom}, D.~R., {Panagia}, N., {Hillenbrand},
  L.~A., {Palla}, F., \& {Stassun}, K.~G. 2010, \apj, 722, 1092

\bibitem[{{Damiani} {et~al.}(2004){Damiani}, {Flaccomio}, {Micela},
  {Sciortino}, {Harnden}, \& {Murray}}]{damiani04}
{Damiani}, F., {Flaccomio}, E., {Micela}, G., {Sciortino}, S., {Harnden}, Jr.,
  F.~R., \& {Murray}, S.~S. 2004, \apj, 608, 781

\bibitem[{{Damiani} {et~al.}(2006){Damiani}, {Prisinzano}, {Micela}, \&
  {Sciortino}}]{damiani06}
{Damiani}, F., {Prisinzano}, L., {Micela}, G., \& {Sciortino}, S. 2006, \aap,
  459, 477

\bibitem[{{Favata} {et~al.}(2005){Favata}, {Flaccomio}, {Reale}, {Micela},
  {Sciortino}, {Shang}, {Stassun}, \& {Feigelson}}]{favata05}
{Favata}, F., {Flaccomio}, E., {Reale}, F., {Micela}, G., {Sciortino}, S.,
  {Shang}, H., {Stassun}, K.~G., \& {Feigelson}, E.~D. 2005, \apjs, 160, 469

\bibitem[{{Feigelson} {et~al.}(2007){Feigelson}, {Townsley}, {G{\"u}del}, \&
  {Stassun}}]{feigelson07}
{Feigelson}, E., {Townsley}, L., {G{\"u}del}, M., \& {Stassun}, K. 2007,
  Protostars and Planets V, 313

\bibitem[{{Feigelson} {et~al.}(2002){Feigelson}, {Garmire}, \&
  {Pravdo}}]{feigelson02}
{Feigelson}, E.~D., {Garmire}, G.~P., \& {Pravdo}, S.~H. 2002, \apj, 572, 335

\bibitem[{{Feigelson} {et~al.}(2005){Feigelson}, {Getman}, {Townsley},
  {Garmire}, {Preibisch}, {Grosso}, {Montmerle}, {Muench}, \&
  {McCaughrean}}]{feigelson05}
{Feigelson}, E.~D., {Getman}, K., {Townsley}, L., {Garmire}, G., {Preibisch},
  T., {Grosso}, N., {Montmerle}, T., {Muench}, A., \& {McCaughrean}, M. 2005,
  \apjs, 160, 379

\bibitem[{{Feigelson} {et~al.}(2011){Feigelson}, {Getman}, {Townsley}, {Broos},
  {Povich}, {Garmire}, {King}, {Montmerle}, {Preibisch}, {Smith}, {Stassun},
  {Wang}, {Wolk}, \& {Zinnecker}}]{feigelson11}
{Feigelson}, E.~D., {Getman}, K.~V., {Townsley}, L.~K., {Broos}, P.~S.,
  {Povich}, M.~S., {Garmire}, G.~P., {King}, R.~R., {Montmerle}, T.,
  {Preibisch}, T., {Smith}, N., {Stassun}, K.~G., {Wang}, J., {Wolk}, S., \&
  {Zinnecker}, H. 2011, \apjs, 194, 9

\bibitem[{{Getman} {et~al.}(2005{\natexlab{a}}){Getman}, {Feigelson}, {Grosso},
  {McCaughrean}, {Micela}, {Broos}, {Garmire}, \& {Townsley}}]{getman05b}
{Getman}, K.~V., {Feigelson}, E.~D., {Grosso}, N., {McCaughrean}, M.~J.,
  {Micela}, G., {Broos}, P., {Garmire}, G., \& {Townsley}, L.
  2005{\natexlab{a}}, \apjs, 160, 353

\bibitem[{{Getman} {et~al.}(2005{\natexlab{b}}){Getman}, {Flaccomio}, {Broos},
  {Grosso}, {Tsujimoto}, {Townsley}, {Garmire}, {Kastner}, {Li}, {Harnden},
  {Wolk}, {Murray}, {Lada}, {Muench}, {McCaughrean}, {Meeus}, {Damiani},
  {Micela}, {Sciortino}, {Bally}, {Hillenbrand}, {Herbst}, {Preibisch}, \&
  {Feigelson}}]{getman05a}
{Getman}, K.~V., {Flaccomio}, E., {Broos}, P.~S., {Grosso}, N., {Tsujimoto},
  M., {Townsley}, L., {Garmire}, G.~P., {Kastner}, J., {Li}, J., {Harnden},
  Jr., F.~R., {Wolk}, S., {Murray}, S.~S., {Lada}, C.~J., {Muench}, A.~A.,
  {McCaughrean}, M.~J., {Meeus}, G., {Damiani}, F., {Micela}, G., {Sciortino},
  S., {Bally}, J., {Hillenbrand}, L.~A., {Herbst}, W., {Preibisch}, T., \&
  {Feigelson}, E.~D. 2005{\natexlab{b}}, \apjs, 160, 319

\bibitem[{{G{\'o}mez Maqueo Chew} {et~al.}(2009){G{\'o}mez Maqueo Chew},
  {Stassun}, {Pr{\v s}a}, \& {Mathieu}}]{gomez09}
{G{\'o}mez Maqueo Chew}, Y., {Stassun}, K.~G., {Pr{\v s}a}, A., \& {Mathieu},
  R.~D. 2009, \apj, 699, 1196

\bibitem[{{Grosso} {et~al.}(2005){Grosso}, {Feigelson}, {Getman}, {Townsley},
  {Broos}, {Flaccomio}, {McCaughrean}, {Micela}, {Sciortino}, {Bally}, {Smith},
  {Muench}, {Garmire}, \& {Palla}}]{grosso05}
{Grosso}, N., {Feigelson}, E.~D., {Getman}, K.~V., {Townsley}, L., {Broos}, P.,
  {Flaccomio}, E., {McCaughrean}, M.~J., {Micela}, G., {Sciortino}, S.,
  {Bally}, J., {Smith}, N., {Muench}, A.~A., {Garmire}, G.~P., \& {Palla}, F.
  2005, \apjs, 160, 530

\bibitem[{{G{\"u}del} {et~al.}(2007){G{\"u}del}, {Briggs}, {Arzner}, {Audard},
  {Bouvier}, {Feigelson}, {Franciosini}, {Glauser}, {Grosso}, {Micela},
  {Monin}, {Montmerle}, {Padgett}, {Palla}, {Pillitteri}, {Rebull}, {Scelsi},
  {Silva}, {Skinner}, {Stelzer}, \& {Telleschi}}]{guedel07}
{G{\"u}del}, M., {Briggs}, K.~R., {Arzner}, K., {Audard}, M., {Bouvier}, J.,
  {Feigelson}, E.~D., {Franciosini}, E., {Glauser}, A., {Grosso}, N., {Micela},
  G., {Monin}, J.-L., {Montmerle}, T., {Padgett}, D.~L., {Palla}, F.,
  {Pillitteri}, I., {Rebull}, L., {Scelsi}, L., {Silva}, B., {Skinner}, S.~L.,
  {Stelzer}, B., \& {Telleschi}, A. 2007, \aap, 468, 353

\bibitem[{{Guenther} {et~al.}(2007){Guenther}, {Kallinger}, {Zwintz}, {Weiss},
  \& {Tanner}}]{guenther07}
{Guenther}, D.~B., {Kallinger}, T., {Zwintz}, K., {Weiss}, W.~W., \& {Tanner},
  J. 2007, \apj, 671, 581

\bibitem[{{Gullbring} \& {Gahm}(1996)}]{gullbring96}
{Gullbring}, E. \& {Gahm}, G.~F. 1996, \aap, 308, 821

\bibitem[{{G{\"u}ver} \& {{\"O}zel}(2009)}]{guver09}
{G{\"u}ver}, T. \& {{\"O}zel}, F. 2009, \mnras, 400, 2050

\bibitem[{{Hartman} {et~al.}(2008){Hartman}, {Gaudi}, {Holman}, {McLeod},
  {Stanek}, {Barranco}, {Pinsonneault}, \& {Kalirai}}]{hartman08}
{Hartman}, J.~D., {Gaudi}, B.~S., {Holman}, M.~J., {McLeod}, B.~A., {Stanek},
  K.~Z., {Barranco}, J.~A., {Pinsonneault}, M.~H., \& {Kalirai}, J.~S. 2008,
  \apj, 675, 1254

\bibitem[{{Hartmann} {et~al.}(1986){Hartmann}, {Hewett}, {Stahler}, \&
  {Mathieu}}]{hartmann86}
{Hartmann}, L., {Hewett}, R., {Stahler}, S., \& {Mathieu}, R.~D. 1986, \apj,
  309, 275

\bibitem[{{Hebb} {et~al.}(2010){Hebb}, {Stempels}, {Aigrain},
  {Collier-Cameron}, {Hodgkin}, {Irwin}, {Maxted}, {Pollacco}, {Street},
  {Wilson}, \& {Stassun}}]{hebb10}
{Hebb}, L., {Stempels}, H.~C., {Aigrain}, S., {Collier-Cameron}, A., {Hodgkin},
  S.~T., {Irwin}, J.~M., {Maxted}, P.~F.~L., {Pollacco}, D., {Street}, R.~A.,
  {Wilson}, D.~M., \& {Stassun}, K.~G. 2010, \aap, 522, A37+

\bibitem[{{Herbst} {et~al.}(2001){Herbst}, {Bailer-Jones}, \&
  {Mundt}}]{herbst01}
{Herbst}, W., {Bailer-Jones}, C.~A.~L., \& {Mundt}, R. 2001, \apjl, 554, L197

\bibitem[{{Herbst} {et~al.}(2002){Herbst}, {Bailer-Jones}, {Mundt},
  {Meisenheimer}, \& {Wackermann}}]{herbst02}
{Herbst}, W., {Bailer-Jones}, C.~A.~L., {Mundt}, R., {Meisenheimer}, K., \&
  {Wackermann}, R. 2002, \aap, 396, 513

\bibitem[{{Herbst} {et~al.}(2007){Herbst}, {Eisl{\"o}ffel}, {Mundt}, \&
  {Scholz}}]{herbst07}
{Herbst}, W., {Eisl{\"o}ffel}, J., {Mundt}, R., \& {Scholz}, A. 2007,
  Protostars and Planets V, 297

\bibitem[{{Herbst} {et~al.}(2008){Herbst}, {Hamilton}, {Leduc}, {Winn},
  {Johns-Krull}, {Mundt}, \& {Ibrahimov}}]{herbst08}
{Herbst}, W., {Hamilton}, C.~M., {Leduc}, K., {Winn}, J.~N., {Johns-Krull},
  C.~M., {Mundt}, R., \& {Ibrahimov}, M. 2008, \nat, 452, 194

\bibitem[{{Herbst} {et~al.}(1994){Herbst}, {Herbst}, {Grossman}, \&
  {Weinstein}}]{herbst94}
{Herbst}, W., {Herbst}, D.~K., {Grossman}, E.~J., \& {Weinstein}, D. 1994, \aj,
  108, 1906

\bibitem[{{Herbst} {et~al.}(2010){Herbst}, {LeDuc}, {Hamilton}, {Winn},
  {Ibrahimov}, {Mundt}, \& {Johns-Krull}}]{herbst10}
{Herbst}, W., {LeDuc}, K., {Hamilton}, C.~M., {Winn}, J.~N., {Ibrahimov}, M.,
  {Mundt}, R., \& {Johns-Krull}, C.~M. 2010, \aj, 140, 2025

\bibitem[{{Herbst} {et~al.}(2000){Herbst}, {Maley}, \& {Williams}}]{herbst00b}
{Herbst}, W., {Maley}, J.~A., \& {Williams}, E.~C. 2000, \aj, 120, 349

\bibitem[{{Hillenbrand}(1997)}]{hillenbrand97}
{Hillenbrand}, L.~A. 1997, \aj, 113, 1733

\bibitem[{{Honeycutt}(1992)}]{honeycutt92}
{Honeycutt}, R.~K. 1992, \pasp, 104, 435

\bibitem[{{Irwin} {et~al.}(2007){Irwin}, {Aigrain}, {Hodgkin}, {Stassun},
  {Hebb}, {Irwin}, {Moraux}, {Bouvier}, {Alapini}, {Alexander}, {Bramich},
  {Holtzman}, {Mart{\'{\i}}n}, {McCaughrean}, {Pont}, {Verrier}, \& {Zapatero
  Osorio}}]{irwin07}
{Irwin}, J., {Aigrain}, S., {Hodgkin}, S., {Stassun}, K.~G., {Hebb}, L.,
  {Irwin}, M., {Moraux}, E., {Bouvier}, J., {Alapini}, A., {Alexander}, R.,
  {Bramich}, D.~M., {Holtzman}, J., {Mart{\'{\i}}n}, E.~L., {McCaughrean},
  M.~J., {Pont}, F., {Verrier}, P.~E., \& {Zapatero Osorio}, M.~R. 2007,
  \mnras, 380, 541

\bibitem[{{Irwin} {et~al.}(2008{\natexlab{a}}){Irwin}, {Hodgkin}, {Aigrain},
  {Bouvier}, {Hebb}, {Irwin}, \& {Moraux}}]{irwin08a}
{Irwin}, J., {Hodgkin}, S., {Aigrain}, S., {Bouvier}, J., {Hebb}, L., {Irwin},
  M., \& {Moraux}, E. 2008{\natexlab{a}}, \mnras, 384, 675

\bibitem[{{Irwin} {et~al.}(2008{\natexlab{b}}){Irwin}, {Hodgkin}, {Aigrain},
  {Bouvier}, {Hebb}, \& {Moraux}}]{irwin08b}
{Irwin}, J., {Hodgkin}, S., {Aigrain}, S., {Bouvier}, J., {Hebb}, L., \&
  {Moraux}, E. 2008{\natexlab{b}}, \mnras, 383, 1588

\bibitem[{{James} {et~al.}(2000){James}, {Jardine}, {Jeffries}, {Randich},
  {Collier Cameron}, \& {Ferreira}}]{james00}
{James}, D.~J., {Jardine}, M.~M., {Jeffries}, R.~D., {Randich}, S., {Collier
  Cameron}, A., \& {Ferreira}, M. 2000, \mnras, 318, 1217

\bibitem[{{Jeffries} {et~al.}(2011){Jeffries}, {Jackson}, {Briggs}, {Evans},
  \& {Pye}}]{jeffries11}
{Jeffries}, R.~D., {Jackson}, R.~J., {Briggs}, K.~R., {Evans}, P.~A., \&
  {Pye}, J.~P. 2011, \mnras, 411, 2099

\bibitem[{{Jensen} {et~al.}(2007){Jensen}, {Dhital}, {Stassun}, {Patience},
  {Herbst}, {Walter}, {Simon}, \& {Basri}}]{jensen07}
{Jensen}, E.~L.~N., {Dhital}, S., {Stassun}, K.~G., {Patience}, J., {Herbst},
  W., {Walter}, F.~M., {Simon}, M., \& {Basri}, G. 2007, \aj, 134, 241

\bibitem[{{Kastner} {et~al.}(2004){Kastner}, {Richmond}, {Grosso}, {Weintraub},
  {Simon}, {Frank}, {Hamaguchi}, {Ozawa}, \& {Henden}}]{kastner04}
{Kastner}, J.~H., {Richmond}, M., {Grosso}, N., {Weintraub}, D.~A., {Simon},
  T., {Frank}, A., {Hamaguchi}, K., {Ozawa}, H., \& {Henden}, A. 2004, \nat,
  430, 429

\bibitem[{{Kearns} {et~al.}(1997){Kearns}, {Eaton}, {Herbst}, \&
  {Mazzurco}}]{kearns97}
{Kearns}, K.~E., {Eaton}, N.~L., {Herbst}, W., \& {Mazzurco}, C.~J. 1997, \aj,
  114, 1098

\bibitem[{{Kearns} \& {Herbst}(1998)}]{kearns98}
{Kearns}, K.~E. \& {Herbst}, W. 1998, \aj, 116, 261

\bibitem[{{Kenyon} \& {Hartmann}(1995)}]{kenyon95}
{Kenyon}, S.~J. \& {Hartmann}, L. 1995, \apjs, 101, 117

\bibitem[{{Koenigl}(1991)}]{konigl91}
{Koenigl}, A. 1991, \apjl, 370, L39

\bibitem[{{Krishnamurthi} {et~al.}(1998){Krishnamurthi}, {Terndrup},
  {Pinsonneault}, {Sellgren}, {Stauffer}, {Schild}, {Backman}, {Beisser},
  {Dahari}, {Dasgupta}, {Hagelgans}, {Seeds}, {Anand}, {Laaksonen},
  {Marschall}, \& {Ramseyer}}]{krishnamurthi98}
{Krishnamurthi}, A., {Terndrup}, D.~M., {Pinsonneault}, M.~H., {Sellgren}, K.,
  {Stauffer}, J.~R., {Schild}, R., {Backman}, D.~E., {Beisser}, K.~B.,
  {Dahari}, D.~B., {Dasgupta}, A., {Hagelgans}, J.~T., {Seeds}, M.~A., {Anand},
  R., {Laaksonen}, B.~D., {Marschall}, L.~A., \& {Ramseyer}, T. 1998, \apj,
  493, 914

\bibitem[{{Lada} {et~al.}(1976){Lada}, {Gottlieb}, {Gottlieb}, \&
  {Gull}}]{lada76}
{Lada}, C.~J., {Gottlieb}, C.~A., {Gottlieb}, E.~W., \& {Gull}, T.~R. 1976,
  \apj, 203, 159

\bibitem[{{Lamm} {et~al.}(2005){Lamm}, {Mundt}, {Bailer-Jones}, \&
  {Herbst}}]{lamm05}
{Lamm}, M.~H., {Mundt}, R., {Bailer-Jones}, C.~A.~L., \& {Herbst}, W. 2005,
  \aap, 430, 1005

\bibitem[{{Lang} {et~al.}(2010){Lang}, {Hogg}, {Mierle}, {Blanton}, \&
  {Roweis}}]{lang10}
{Lang}, D., {Hogg}, D.~W., {Mierle}, K., {Blanton}, M., \& {Roweis}, S. 2010,
  \aj, 139, 1782

\bibitem[{{Le Blanc} {et~al.}(2011){Le Blanc}, {Covey}, \&
  {Stassun}}]{leblanc11}
{Le Blanc}, T.~S., {Covey}, K.~R., \& {Stassun}, K.~G. 2011, ArXiv e-prints

\bibitem[{{Loktin} {et~al.}(1997){Loktin}, {Zakharova}, {Gerasimenko}, \&
  {Malisheva}}]{loktin97}
{Loktin}, A., {Zakharova}, P., {Gerasimenko}, T., \& {Malisheva}, L. 1997,
  Baltic Astronomy, 6, 316

\bibitem[{{Lomb}(1976)}]{lomb76}
{Lomb}, N.~R. 1976, \apss, 39, 447

\bibitem[{{Makidon} {et~al.}(2004){Makidon}, {Rebull}, {Strom}, {Adams}, \&
  {Patten}}]{makidon04}
{Makidon}, R.~B., {Rebull}, L.~M., {Strom}, S.~E., {Adams}, M.~T., \& {Patten},
  B.~M. 2004, \aj, 127, 2228

\bibitem[{{Mathieu} {et~al.}(1997){Mathieu}, {Stassun}, {Basri}, {Jensen},
  {Johns-Krull}, {Valenti}, \& {Hartmann}}]{mathieu97}
{Mathieu}, R.~D., {Stassun}, K., {Basri}, G., {Jensen}, E.~L.~N.,
  {Johns-Krull}, C.~M., {Valenti}, J.~A., \& {Hartmann}, L.~W. 1997, \aj, 113,
  1841

\bibitem[{{Matt} \& {Pudritz}(2004)}]{matt04}
{Matt}, S. \& {Pudritz}, R.~E. 2004, \apjl, 607, L43

\bibitem[{{Matt} \& {Pudritz}(2005{\natexlab{a}})}]{matt05b}
---. 2005{\natexlab{a}}, \apjl, 632, L135

\bibitem[{{Matt} \& {Pudritz}(2005{\natexlab{b}})}]{matt05a}
---. 2005{\natexlab{b}}, \mnras, 356, 167

\bibitem[{{Matt} \& {Pudritz}(2008{\natexlab{a}})}]{matt08a}
---. 2008{\natexlab{a}}, \apj, 678, 1109

\bibitem[{{Matt} \& {Pudritz}(2008{\natexlab{b}})}]{matt08b}
---. 2008{\natexlab{b}}, \apj, 681, 391

\bibitem[{{Mayne} \& {Naylor}(2008)}]{mayne08}
{Mayne}, N.~J. \& {Naylor}, T. 2008, \mnras, 386, 261

\bibitem[{{Mayne} {et~al.}(2007){Mayne}, {Naylor}, {Littlefair}, {Saunders}, \&
  {Jeffries}}]{mayne07}
{Mayne}, N.~J., {Naylor}, T., {Littlefair}, S.~P., {Saunders}, E.~S., \&
  {Jeffries}, R.~D. 2007, \mnras, 375, 1220

\bibitem[{{Menten} {et~al.}(2007){Menten}, {Reid}, {Forbrich}, \&
  {Brunthaler}}]{menten07}
{Menten}, K.~M., {Reid}, M.~J., {Forbrich}, J., \& {Brunthaler}, A. 2007, \aap,
  474, 515

\bibitem[{{Mohanty} {et~al.}(2010){Mohanty}, {Stassun}, \&
  {Doppmann}}]{mohanty10}
{Mohanty}, S., {Stassun}, K.~G., \& {Doppmann}, G.~W. 2010, \apj, 722, 1138

\bibitem[{{Mohanty} {et~al.}(2009){Mohanty}, {Stassun}, \&
  {Mathieu}}]{mohanty09}
{Mohanty}, S., {Stassun}, K.~G., \& {Mathieu}, R.~D. 2009, \apj, 697, 713

\bibitem[{{Najita}(1995)}]{najita95}
{Najita}, J. 1995, in Revista Mexicana de Astronomia y Astrofisica, vol. 27,
  Vol.~1, Revista Mexicana de Astronomia y Astrofisica Conference Series, ed.
  {S.~Lizano \& J.~M.~Torrelles}, 293--+

\bibitem[{{Ostriker} \& {Shu}(1995)}]{ostriker95}
{Ostriker}, E.~C. \& {Shu}, F.~H. 1995, \apj, 447, 813

\bibitem[{{Patten} \& {Simon}(1996)}]{patten96}
{Patten}, B.~M. \& {Simon}, T. 1996, \apjs, 106, 489

\bibitem[{{Pizzolato} {et~al.}(2003){Pizzolato}, {Maggio}, {Micela},
  {Sciortino}, \& {Ventura}}]{pizzolato03}
{Pizzolato}, N., {Maggio}, A., {Micela}, G., {Sciortino}, S., \& {Ventura}, P.
  2003, \aap, 397, 147

\bibitem[{{Press} \& {Rybicki}(1989)}]{press89}
{Press}, W.~H. \& {Rybicki}, G.~B. 1989, \apj, 338, 277

\bibitem[{{Press} {et~al.}(1992){Press}, {Teukolsky}, {Vetterling}, \&
  {Flannery}}]{press92}
{Press}, W.~H., {Teukolsky}, S.~A., {Vetterling}, W.~T., \& {Flannery}, B.~P.
  1992, {Numerical recipes in C. The art of scientific computing}, ed. {Press,
  W.~H., Teukolsky, S.~A., Vetterling, W.~T., \& Flannery, B.~P. }

\bibitem[{{Prisinzano} {et~al.}(2007){Prisinzano}, {Damiani}, {Micela}, \&
  {Pillitteri}}]{prisinzano07}
{Prisinzano}, L., {Damiani}, F., {Micela}, G., \& {Pillitteri}, I. 2007, \aap,
  462, 123

\bibitem[{{Prisinzano} {et~al.}(2005){Prisinzano}, {Damiani}, {Micela}, \&
  {Sciortino}}]{prisinzano05}
{Prisinzano}, L., {Damiani}, F., {Micela}, G., \& {Sciortino}, S. 2005, \aap,
  430, 941

\bibitem[{{Radick} {et~al.}(1990){Radick}, {Skiff}, \& {Lockwood}}]{radick90}
{Radick}, R.~R., {Skiff}, B.~A., \& {Lockwood}, G.~W. 1990, \apj, 353, 524

\bibitem[{{Radick} {et~al.}(1987){Radick}, {Thompson}, {Lockwood}, {Duncan}, \&
  {Baggett}}]{radick87}
{Radick}, R.~R., {Thompson}, D.~T., {Lockwood}, G.~W., {Duncan}, D.~K., \&
  {Baggett}, W.~E. 1987, \apj, 321, 459

\bibitem[{{Rebull}(2001)}]{rebull01}
{Rebull}, L.~M. 2001, \aj, 121, 1676

\bibitem[{{Scargle}(1982)}]{scargle82}
{Scargle}, J.~D. 1982, \apj, 263, 835

\bibitem[{{Shu} {et~al.}(1994){Shu}, {Najita}, {Ostriker}, {Wilkin}, {Ruden},
  \& {Lizano}}]{shu94}
{Shu}, F., {Najita}, J., {Ostriker}, E., {Wilkin}, F., {Ruden}, S., \&
  {Lizano}, S. 1994, \apj, 429, 781

\bibitem[{{Siess} {et~al.}(2000){Siess}, {Dufour}, \& {Forestini}}]{siess00}
{Siess}, L., {Dufour}, E., \& {Forestini}, M. 2000, \aap, 358, 593

\bibitem[{{Stassun} \& {Wood}(1999)}]{stassun99b}
{Stassun}, K. \& {Wood}, K. 1999, \apj, 510, 892

\bibitem[{{Stassun} {et~al.}(2004{\natexlab{a}}){Stassun}, {Ardila}, {Barsony},
  {Basri}, \& {Mathieu}}]{stassun04b}
{Stassun}, K.~G., {Ardila}, D.~R., {Barsony}, M., {Basri}, G., \& {Mathieu},
  R.~D. 2004{\natexlab{a}}, \aj, 127, 3537

\bibitem[{{Stassun} {et~al.}(2009){Stassun}, {Hebb}, {L{\'o}pez-Morales}, \&
  {Pr{\v s}a}}]{stassun09}
{Stassun}, K.~G., {Hebb}, L., {L{\'o}pez-Morales}, M., \& {Pr{\v s}a}, A. 2009,
  in IAU Symposium, Vol. 258, IAU Symposium, ed. {E.~E.~Mamajek,
  D.~R.~Soderblom, \& R.~F.~G.~Wyse}, 161--170

\bibitem[{{Stassun} {et~al.}(2008){Stassun}, {Mathieu}, {Cargile}, {Aarnio},
  {Stempels}, \& {Geller}}]{stassun08}
{Stassun}, K.~G., {Mathieu}, R.~D., {Cargile}, P.~A., {Aarnio}, A.~N.,
  {Stempels}, E., \& {Geller}, A. 2008, \nat, 453, 1079

\bibitem[{{Stassun} {et~al.}(1999){Stassun}, {Mathieu}, {Mazeh}, \&
  {Vrba}}]{stassun99a}
{Stassun}, K.~G., {Mathieu}, R.~D., {Mazeh}, T., \& {Vrba}, F.~J. 1999, \aj,
  117, 2941

\bibitem[{{Stassun} {et~al.}(2006{\natexlab{a}}){Stassun}, {Mathieu}, \&
  {Valenti}}]{stassun06a}
{Stassun}, K.~G., {Mathieu}, R.~D., \& {Valenti}, J.~A. 2006{\natexlab{a}},
  \nat, 440, 311

\bibitem[{{Stassun} {et~al.}(2007{\natexlab{a}}){Stassun}, {Mathieu}, \&
  {Valenti}}]{stassun07a}
---. 2007{\natexlab{a}}, \apj, 664, 1154

\bibitem[{{Stassun} {et~al.}(2004{\natexlab{b}}){Stassun}, {Mathieu}, {Vaz},
  {Stroud}, \& {Vrba}}]{stassun04a}
{Stassun}, K.~G., {Mathieu}, R.~D., {Vaz}, L.~P.~R., {Stroud}, N., \& {Vrba},
  F.~J. 2004{\natexlab{b}}, \apjs, 151, 357

\bibitem[{{Stassun} {et~al.}(2001){Stassun}, {Mathieu}, {Vrba}, {Mazeh}, \&
  {Henden}}]{stassun01}
{Stassun}, K.~G., {Mathieu}, R.~D., {Vrba}, F.~J., {Mazeh}, T., \& {Henden}, A.
  2001, \aj, 121, 1003

\bibitem[{{Stassun} \& {Terndrup}(2003)}]{stassun03}
{Stassun}, K.~G. \& {Terndrup}, D. 2003, \pasp, 115, 505

\bibitem[{{Stassun} {et~al.}(2007{\natexlab{b}}){Stassun}, {van den Berg}, \&
  {Feigelson}}]{stassun07b}
{Stassun}, K.~G., {van den Berg}, M., \& {Feigelson}, E. 2007{\natexlab{b}},
  \apj, 660, 704

\bibitem[{{Stassun} {et~al.}(2006{\natexlab{b}}){Stassun}, {van den Berg},
  {Feigelson}, \& {Flaccomio}}]{stassun06}
{Stassun}, K.~G., {van den Berg}, M., {Feigelson}, E., \& {Flaccomio}, E.
  2006{\natexlab{b}}, \apj, 649, 914

\bibitem[{{Stassun} {et~al.}(2002){Stassun}, {van den Berg}, {Mathieu}, \&
  {Verbunt}}]{stassun02}
{Stassun}, K.~G., {van den Berg}, M., {Mathieu}, R.~D., \& {Verbunt}, F. 2002,
  \aap, 382, 899

\bibitem[{{Stauffer} \& {Hartmann}(1987)}]{stauffer87}
{Stauffer}, J.~R. \& {Hartmann}, L.~W. 1987, \apj, 318, 337

\bibitem[{{Stempels} {et~al.}(2008){Stempels}, {Hebb}, {Stassun}, {Holtzman},
  {Dunstone}, {Glowienka}, \& {Frandsen}}]{stempels08}
{Stempels}, H.~C., {Hebb}, L., {Stassun}, K.~G., {Holtzman}, J., {Dunstone},
  N., {Glowienka}, L., \& {Frandsen}, S. 2008, \aap, 481, 747

\bibitem[{{Sung} {et~al.}(2000){Sung}, {Chun}, \& {Bessell}}]{sung00}
{Sung}, H., {Chun}, M.-Y., \& {Bessell}, M.~S. 2000, \aj, 120, 333

\bibitem[{{van den Ancker} {et~al.}(1997){van den Ancker}, {The}, {Feinstein},
  {Vazquez}, {de Winter}, \& {Perez}}]{vandenancker97}
{van den Ancker}, M.~E., {The}, P.~S., {Feinstein}, A., {Vazquez}, R.~A., {de
  Winter}, D., \& {Perez}, M.~R. 1997, \aaps, 123, 63

\bibitem[{{van Eyken} {et~al.}(2011){van Eyken}, {Ciardi}, {Rebull},
  {Stauffer}, {Akeson}, {Beichman}, {Boden}, {von Braun}, {Gelino}, {Hoard},
  {Howell}, {Kane}, {Plavchan}, {Ram{\'{\i}}rez}, {Bloom}, {Cenko}, {Kasliwal},
  {Kulkarni}, {Law}, {Nugent}, {Ofek}, {Poznanski}, {Quimby}, {Grillmair},
  {Laher}, {Levitan}, {Mattingly}, \& {Surace}}]{vaneyken11}
{van Eyken}, J.~C., {Ciardi}, D.~R., {Rebull}, L.~M., {Stauffer}, J.~R.,
  {Akeson}, R.~L., {Beichman}, C.~A., {Boden}, A.~F., {von Braun}, K.,
  {Gelino}, D.~M., {Hoard}, D.~W., {Howell}, S.~B., {Kane}, S.~R., {Plavchan},
  P., {Ram{\'{\i}}rez}, S.~V., {Bloom}, J.~S., {Cenko}, S.~B., {Kasliwal},
  M.~M., {Kulkarni}, S.~R., {Law}, N.~M., {Nugent}, P.~E., {Ofek}, E.~O.,
  {Poznanski}, D., {Quimby}, R.~M., {Grillmair}, C.~J., {Laher}, R., {Levitan},
  D., {Mattingly}, S., \& {Surace}, J.~A. 2011, \aj, 142, 60

\bibitem[{{Vogel} \& {Kuhi}(1981)}]{vogel81}
{Vogel}, S.~N. \& {Kuhi}, L.~V. 1981, \apj, 245, 960

\bibitem[{{Walter} {et~al.}(1987){Walter}, {Neff}, {Gibson}, {Linsky},
  {Rodono}, {Gary}, \& {Butler}}]{walter87}
{Walter}, F.~M., {Neff}, J.~E., {Gibson}, D.~M., {Linsky}, J.~L., {Rodono}, M.,
  {Gary}, D.~E., \& {Butler}, C.~J. 1987, \aap, 186, 241

\bibitem[{{Walter} {et~al.}(2004){Walter}, {Stringfellow}, {Sherry}, \&
  {Field-Pollatou}}]{walter04}
{Walter}, F.~M., {Stringfellow}, G.~S., {Sherry}, W.~H., \& {Field-Pollatou},
  A. 2004, \aj, 128, 1872

\bibitem[{{Winn} {et~al.}(2006){Winn}, {Hamilton}, {Herbst}, {Hoffman},
  {Holman}, {Johnson}, \& {Kuchner}}]{winn06}
{Winn}, J.~N., {Hamilton}, C.~M., {Herbst}, W.~J., {Hoffman}, J.~L., {Holman},
  M.~J., {Johnson}, J.~A., \& {Kuchner}, M.~J. 2006, \apj, 644, 510

\bibitem[{{Wright} {et~al.}(2011){Wright}, {Drake}, {Mamajek}, \&
  {Henry}}]{wright11}
{Wright}, N.~J., {Drake}, J.~J., {Mamajek}, E.~E., \& {Henry}, G.~W. 2011,
  ArXiv e-prints

\bibitem[{{Zwintz} \& {Weiss}(2006)}]{zwintz06}
{Zwintz}, K. \& {Weiss}, W.~W. 2006, \aap, 457, 237

\end{thebibliography}
\end{document}